\documentclass[12pt,a4paper]{article}
\usepackage{amsmath,amsfonts,amsthm,graphicx,lscape}
\usepackage[dvips]{psfrag}
\usepackage{cite}
\usepackage[normalem]{ulem}

\usepackage{textcomp}
\usepackage{amsmath,amsthm,amssymb,mathrsfs,cases}
\usepackage{amsfonts,graphicx,lscape}

\usepackage{accents}
\makeatletter
\def\widebar{\accentset{{\cc@style\underline{\mskip10mu}}}}
\makeatother

\numberwithin{equation}{section}

\theoremstyle{definition}
\newtheorem{theorem}{Theorem}[section]

\newtheorem{proposition}[theorem]{Proposition}

\newtheorem{lemma}[theorem]{Lemma}

\newcommand{\vt}[1]{\mbox{\boldmath$#1$}}
\newcommand{\dprod}[2]{\prod^{\stackrel{#1}{#2}}}
\newcommand{\sca}[2]{\langle #1, #2 \rangle}

\begin{document}
\title{
\vspace{-9mm}
A 
refined and 
unified
version of 
the 
inverse scattering method
for 
the Ablowitz--Ladik 
lattice
and 
derivative 
NLS
lattices
}
\author{Takayuki \textsc{Tsuchida}
}
\maketitle
\begin{abstract} 
We 
refine and 
develop the inverse scattering 
theory 
on a lattice
in such a way that the 
Ablowitz--Ladik
lattice and 
derivative 
NLS 
lattices 
as well as 
their matrix analogs can be solved 
in a unified way. 
The inverse scattering method 
for the (matrix analog of the) 
Ablowitz--Ladik lattice 
is
simplified  
to the same level as 
that 
for the continuous NLS system. 
Using the 
linear 
eigenfunctions 
of the 
Lax 
pair 
for 
the Ablowitz--Ladik lattice, 
we 
can 
construct 
solutions 
of 
the derivative NLS lattices
such as the 
discrete Gerdjikov--Ivanov 
(also known as 
Ablowitz--Ramani--Segur)
system 
and the 
discrete  
Kaup--Newell system. 
Thus, explicit solutions such as the 
multisoliton 
solutions for 
these systems 
can be obtained 
by 
solving linear summation equations 
of the 
Gel'fand--Levitan--Marchenko type. 
The derivation of 
the 
discrete 
Kaup--Newell system 
from the Ablowitz--Ladik lattice 
is based on a new 
method 
that 
allows us to generate 
new integrable systems
from known 
systems 
in a systematic manner. 
In an appendix, 
we 
describe 
the reduction of the matrix Ablowitz--Ladik lattice to 
a vector analog of the 
modified Volterra lattice 
from the point of view of the inverse scattering method. 
%
%
%
\end{abstract}
%
\vspace{15mm}
\begin{minipage}{13cm}
\end{minipage}

%
%
%
\newpage
\noindent
\tableofcontents

\newpage
\section{Introduction}
The cubic nonlinear Schr\"odinger (NLS) equation~\cite{ZS1,ZS2} 
is probably the most prominent 
example of an 
integrable 
partial differential equation 
in 
\mbox{$1+1$} 
space-time 
dimensions. 
The 
inverse scattering 
method 
for the 
NLS 
equation 
devised 
by Zakharov and Shabat~\cite{ZS1}
was 
reformulated by 
Ablowitz, Kaup, Newell and Segur~\cite{AKNS73,AKNS74}
in a 
user-friendly 
and broadly-applicable 
manner.
Since then
various extensions of the NLS equation 
have been 
obtained 
within the framework of the inverse scattering method. 
%
Among them,
we mention 
two kinds 
of 
extensions:\
(i) space-discrete 
NLS
systems\footnote{The 
problem of how to discretize the continuous time variable in 
integrable 
space-discrete systems 
has a long history, see~\cite{Tsu2010JPA} and references therein.}
wherein 
the spatial variable 
is discretized~\cite{AL76,GI82}
and (ii) derivative NLS
systems 
wherein the nonlinear terms 
involve 
differentiation with respect to
the spatial variable~\cite{KN,CLL,ARS,GI,Kun}.
For these extensions, 
the inverse scattering method 
can be 
applied 
on a case-by-case basis, 
but 
its 
application 
requires 
more steps 
and 
is 
apparently 
more complicated 
than 
that 
for 
the 
original 
NLS 
system~\cite{AKNS73,AKNS74}. 

%
%
The main objective 
of this paper 
is twofold.
First, 
we 
develop 
the inverse scattering method 
on a lattice and 
correct 
the widespread impression that 
it is essentially 
more complicated 
than 
the inverse scattering method 
on the line. 
Second, we 
show 
that the derivative NLS systems 
can be solved 
using 
the inverse scattering method for 
the NLS system; however, 
this paper focuses on 
the space-discrete case.\footnote{Relevant 
results 
on 
continuous
derivative NLS systems  
can be found in~\cite{talk07,talk08}.}
To be  
specific, 
we consider 
(a matrix generalization
of) the Ablowitz--Ladik
lattice~\cite{AL76} that is an integrable space discretization 
of the NLS system. 
The inverse scattering 
method 
for the Ablowitz--Ladik lattice 
reported in the 
existing 
literature 
involves 
some onerous 
processes 
peculiar to the discrete case; 
in fact, they are redundant and  
can be 
avoided. 
For this purpose, 
we only need to 
start with an 
eigenvalue problem
that 
is 
trivially 
equivalent to the 
Ablowitz--Ladik eigenvalue problem 
up to a 
similarity transformation 
and 
inversion of the spatial coordinate.
Then, all the key quantities such as the 
scattering data 
become even functions of 
the conventional spectral parameter;
thus, we can use 
its square as
a new (and more 
essential) 
parameter.
This considerably simplifies the subsequent 
computations. 
Moreover, 
by the inversion of the spatial coordinate, 
we no longer need to
normalize 
the ``integral kernels" of 
the 
linear eigenfunctions 
(Jost solutions) 
to express the 
potentials 
in 
the Ablowitz--Ladik eigenvalue problem 
explicitly. 
This is in contrast to the conventional approach 
wherein 
one has to first introduce 
the ``integral kernels" and 
then normalize them;  
in the literature (see, {\em e.g.}, \cite{Ab78,AS81}), 
they are usually denoted as 
$K(n,m)$, $\widebar{K}(n,m)$ 
and 
$\kappa (n,m)$, $\widebar{\kappa}(n,m)$, 
respectively. 
%
Thus, we can successfully refine 
the inverse scattering method 
associated with the (matrix) Ablowitz--Ladik eigenvalue problem; 
both the potentials 
and the linear eigenfunctions 
are 
determined from the scattering data 
through 
a set of linear summation equations 
in the most transparent manner.  
%
%

In 
our 
previous paper~\cite{TsuJMP11}, 
we proposed a systematic method of
generating 
new integrable systems 
from known systems 
through 
inverse Miura maps.\footnote{
The original 
Miura map 
transforms 
%
the modified KdV 
equation 
(or, more generally, 
its one-parameter 
generalization called the Gardner equation) 
to 
the KdV equation~\cite{Miura68}.}
As a result, 
two derivative NLS systems, namely, 
the Gerdjikov--Ivanov 
(also known as 
Ablowitz--Ramani--Segur) 
system~\cite{ARS,GI}
and the Chen--Lee--Liu system~\cite{CLL}, 
were
constructed from
the Lax representation\footnote{The term 
was coined after 
Lax's 
work on 
the KdV hierarchy~\cite{Lax}.}
for 
the NLS system;  
the same 
prescription 
applies 
to the space-discrete case. 
Thus, the inverse scattering method for 
the Ablowitz--Ladik lattice can also 
provide the solutions of the 
space-discrete Gerdjikov--Ivanov 
system 
and 
the space-discrete 
Chen--Lee--Liu system
in a unified way. 
%
%
In this paper, we propose yet another 
method of
generating new integrable systems 
from known systems. 
In particular, 
by applying this new method 
to the Ablowitz--Ladik lattice, we 
obtain a
lattice system that is 
essentially equivalent to 
the space-discrete Kaup--Newell system studied 
in~\cite{Tsuchi02,TsuJMP10}; 
its solutions can 
be 
expressed 
in terms of 
the 
linear 
eigenfunctions 
associated with 
the Ablowitz--Ladik lattice. 
Thus, 
the space-discrete Kaup--Newell 
system can also 
be solved using the inverse scattering method 
for the Ablowitz--Ladik lattice; 
note that among the derivative NLS systems,  
the Kaup--Newell system~\cite{KN}
is the most important 
for physical applications.  

The main body 
of 
this paper 
is organized as follows. 
In section 2, we start with the Lax representation for  
the Ablowitz--Ladik lattice;
using 
its linear eigenfunctions, 
we derive 
two 
derivative NLS 
lattices. 
First, we 
apply 
the method proposed in~\cite{TsuJMP11} 
and derive 
the space-discrete Gerdjikov--Ivanov system. 
Second, we propose a new systematic 
method and 
obtain the space-discrete Kaup--Newell system. 
We can also derive and solve the space-discrete 
Chen--Lee--Liu system, 
but 
we do not present it in this paper;  
the interested reader is referred to 
appendix B of~\cite{TsuJMP11} 
and section 3 of~\cite{Tsuchi02}. 
In section 3, we present a streamlined version 
of the inverse scattering method for the Ablowitz--Ladik lattice. 
In section 4, we combine 
the results of sections 2 and 3 
and 
show 
that the 
derivative NLS lattices 
can be solved by the inverse scattering method 
associated with the Ablowitz--Ladik eigenvalue problem. 
Their multisoliton solutions can be derived 
from the linear summation equations in a straightforward 
manner. 
The last section, section 5, is devoted to concluding remarks. 

In this 
paper, 
we consider 
the general 
case 
where
the dependent variables 
take their values in matrices~\cite{GI82,ZS3}
in such a way that 
the operations such as 
addition and 
multiplication 
in the equations of motion 
make sense. 
In 
appendix~\ref{app2}, 
we 
consider 
the 
reduction of 
the matrix 
Ablowitz--Ladik lattice 
to 
a vector analog of the modified Volterra
lattice 
and discuss 
the effect of the reduction on the scattering data; 
this 
considerably refines 
our previous 
results 
given 
in~\cite{TUW98,TUW99}. 
%
%
%
%
%
%

\section{Ablowitz--Ladik lattice and derivative NLS lattices}
\label{sect2}


\subsection{Lax
representation 
for the Ablowitz--Ladik lattice}

We 
start with 
the matrix Ablowitz--Ladik eigenvalue problem 
written 
in the nonstandard 
form:\footnote{Actually, in the simplest \mbox{$2 \times 2$} matrix case, 
(\ref{mAL1}) 
can be identified with 
a special 
case of the eigenvalue problem 
studied in~\cite{AL75}.}
\begin{align}
&
\left[
\begin{array}{c}
 \Psi_{1, n} \\
 \Psi_{2, n} \\
\end{array}
\right]
= 
 \left[
 \begin{array}{cc}
  z I & z Q_n \\
 z^{-1} R_n & z^{-1} I \\
 \end{array}
 \right]
\left[
\begin{array}{c}
 \Psi_{1, n+1} \\
 \Psi_{2, n+1} \\
\end{array}
\right],
\label{mAL1}
\end{align}
where 
$z$ is 
a constant 
spectral parameter and $I$ 
is the 
identity matrix of arbitrary size. 
%
For simplicity, 
we assume that 
all the entries 
in (\ref{mAL1}), 
such as the potentials 
$Q_n$ and $R_n$, are \mbox{$\l \times l$} 
square matrices. 
It is also possible 
to consider 
the more general case 
where 
$Q_n$ is an \mbox{$\l_1 \times l_2$} 
matrix 
and $R_n$ is an \mbox{$\l_2 \times l_1$} 
matrix; 
however, the results for 
that 
case 
can be easily 
obtained 
by setting
some rows and columns in 
$Q_n$ and $R_n$
as 
identically 
zero. 

The 
time evolution 
of the linear eigenfunction can be 
introduced 
in such a way that it is 
compatible with the 
eigenvalue problem (\ref{mAL1}). 
The most illustrative 
example
is 
given by 
%
\begin{align}
& 
\left[
\begin{array}{c}
 \Psi_{1, n} \\
 \Psi_{2, n} \\
\end{array}
\right]_t
= 
 \left[
 \begin{array}{cc}
  (-z^2+1)bI + b Q_{n-1}R_n & -z^2 b Q_{n-1} -a Q_n \\
  -b R_n - z^{-2} a R_{n-1} & (1-z^{-2})a I + a R_{n-1} Q_n \\
 \end{array}
 \right]
\left[
\begin{array}{c}
 \Psi_{1, n} \\
 \Psi_{2, n} \\
\end{array}
\right].
\label{mAL-time}
\end{align}
%
Here, $a$ and $b$ are 
arbitrary scalar 
constants; 
in fact, they may
depend 
on the time variable $t$ in an arbitrary manner
(see, {\it e.g.}, 
\cite{Calo76,Vakh02}), 
but we do not discuss it in this paper.
The compatibility condition for 
the overdetermined 
linear system,
(\ref{mAL1}) and (\ref{mAL-time}), 
with the isospectral condition \mbox{$z_t=0$}
implies 
the 
time evolution equations for $Q_n$ and 
$R_n$: 
\begin{subnumcases}{\label{mALsys}}
{}
\label{AL-Qt}
 Q_{n,t} - a Q_{n+1} + b Q_{n-1} + (a-b) Q_n 
	+a Q_n R_n Q_{n+1} -b Q_{n-1} R_n Q_n = O, 
\hspace{13mm}
\\[1mm]
\label{AL-Rt}
 R_{n,t} -b R_{n+1} +a R_{n-1} + (b-a) R_n 
	+b R_n Q_n R_{n+1} -a R_{n-1} Q_n R_n =O.
\hspace{13mm}
\end{subnumcases}
We call (\ref{mALsys}) the (matrix) Ablowitz--Ladik 
lattice/system~\cite{AL76,GI82}; 
(\ref{mAL1}) and (\ref{mAL-time}) 
comprise 
its Lax
representation. 
The symbol
$O$
on the right-hand side
of the
equations
implies that the dependent variables can take
their
values in
matrices. 
In fact, 
there exist 
infinitely 
many ways to 
define such an 
isospectral
time evolution (cf.~\cite{Chiu77,Kako}), 
%
depending on 
the choice of the
temporal Lax matrix in (\ref{mAL-time})
as
a Laurent polynomial in $z^2$; 
they 
provide
the positive flows 
of the Ablowitz--Ladik hierarchy 
(cf.~appendix A of~\cite{Tsuchi02}
for the negative flows)
and each of them is uniquely 
determined by its linear part
or, equivalently, the dispersion relation.

\subsection{Space-discrete Gerdjikov--Ivanov system}
\label{subs2.2}

In this subsection, using 
the 
method 
proposed in~\cite{TsuJMP11}, 
we
derive the space-discrete Gerdjikov--Ivanov system
from the Lax representation for the Ablowitz--Ladik lattice. 
The result is essentially the same as that given 
in~\cite{TsuJMP11}, 
but we 
restate
it here 
for the self-containedness and readability 
of the paper.

%
%
We consider 
a \mbox{$2l \times l$}
matrix-valued solution
to the pair of linear equations 
(\ref{mAL1}) and (\ref{mAL-time}) 
such that $\Psi_{1,n}$ is an \mbox{$l \times l$}
invertible
matrix.
Then,
in terms of
the \mbox{$l \times l$} matrix 
\mbox{$P_n := \Psi_{2,n} \Psi_{1,n}^{-1}$},
(\ref{mAL1}) and (\ref{mAL-time}) can be
rewritten as a pair of discrete and continuous 
matrix Riccati equations for $P_n$, 
\begin{subequations}
\label{AL-R}
\begin{align}
& R_n = \mu P_{n} - P_{n+1} + \mu P_{n}Q_n P_{n+1},
\label{AL-R1}
\\[2mm]
& 
P_{n,t} = -b R_{n} -\mu^{-1} a R_{n-1} 
 +(1-\mu^{-1})a P_n +( \mu -1 ) b P_n 
\nonumber \\
& \hphantom{P_{n,t} =}\hspace{1pt}
+a R_{n-1} Q_{n} P_n - b P_n Q_{n-1} R_{n} 
+ \mu b P_n Q_{n-1} P_n + a P_n Q_n P_n,
\label{AL-R2}
\end{align}
\end{subequations}
where \mbox{$\mu := z^2$}. 
The first relation (\ref{AL-R1}) defines
the Miura map \mbox{$(Q_n,P_n) \mapsto (Q_n,R_n)$}. 
Using (\ref{AL-R1}), 
we can 
eliminate $R_n$ and $R_{n-1}$ in 
(\ref{AL-Qt}) and (\ref{AL-R2}) 
to obtain a closed 
system for \mbox{$(Q_n,P_n)$}, 
i.e., 
the space-discrete Gerdjikov--Ivanov system~\cite{Tsuchi02}:
\begin{subnumcases}{\label{sdGI}}
{}
 Q_{n,t}- a Q_{n+1} + b Q_{n-1} + (a-b)Q_n 
	+ a Q_{n} 
\left( \mu P_{n} - P_{n+1} \right) Q_{n+1} 
\nonumber \\
\mbox{}- b Q_{n-1} \left( \mu  P_{n} 
 - P_{n+1} \right) Q_{n}
+ a \mu Q_{n} P_{n}Q_n P_{n+1} Q_{n+1} 
- b \mu Q_{n-1}  P_{n}Q_n P_{n+1} Q_{n}
=O, \hspace{12mm}
\label{}
\\[2mm]
P_{n,t} -b P_{n+1} + a P_{n-1} +(b-a)P_n -b P_{n}
\left( Q_{n-1} - \mu Q_n \right) P_{n+1}
\nonumber \\
\mbox{}+a P_{n-1}\left( Q_{n-1} -\mu Q_n \right) P_{n}
+ b \mu P_{n} Q_{n-1} P_{n}Q_{n} P_{n+1} 
- a \mu P_{n-1} Q_{n-1} P_{n}Q_{n} P_{n}
=O.
\label{}
\end{subnumcases}
%

\subsection{Space-discrete Kaup--Newell system}
\label{subs2.3}

In this subsection, we 
propose yet 
another 
method of
generating new integrable systems from known 
systems  
using 
a fundamental 
set of 
linear eigenfunctions. 
The method is 
applicable to 
a 
matrix 
Lax representation of arbitrary size, 
but for brevity we describe it
in 
the simplest case of a \mbox{$2 \times 2$} 
Lax representation 
as well as 
its block-matrix generalization 
involving two potentials. 
In this paper, we consider 
the space-discrete 
case
(see~\cite{talk08} for the 
continuous
case). 

Suppose that 
a 
lattice system
admits the 
Lax representation
\begin{subequations}
\label{gen-Lax}
\begin{align}
\label{gen-Lax1}
& 
\left[
\begin{array}{c}
 \Psi_{1, n}^{(j)} \\
 \Psi_{2, n}^{(j)} \\
\end{array}
\right]
= 
 \left[
 \begin{array}{cc}
  L_{11,n} & L_{12,n} \\
  L_{21,n} & L_{22,n} \\
 \end{array}
 \right]
\left[
\begin{array}{c}
 \Psi_{1, n+1}^{(j)} \\
 \Psi_{2, n+1}^{(j)} \\
\end{array}
\right],
\\[2.5mm]
\label{gen-Lax2}
& \left[
\begin{array}{c}
 \Psi_{1, n}^{(j)} \\
 \Psi_{2, n}^{(j)} \\
\end{array}
\right]_t
= 
 \left[
 \begin{array}{cc}
  M_{11,n} & M_{12,n} \\
  M_{21,n} & M_{22,n} \\
 \end{array}
 \right]
\left[
\begin{array}{c}
 \Psi_{1, n}^{(j)} \\
 \Psi_{2, n}^{(j)} \\
\end{array}
\right],
\end{align}
\end{subequations}
%
where all the entries 
are assumed to be square matrices of the same size. 
%
Because 
the method requires 
a full
set of 
linearly independent 
eigenfunctions, 
the superscript ${}^{(j)}$ with $j=1$ or $2$
is used to designate 
the two 
eigenfunctions. 
%
%
%
We apply
a gauge transformation 
defined using 
one eigenfunction 
to 
the other 
eigenfunction
as 
\[
\left[
\begin{array}{c}
 \Psi_{1, n}^{(2)} \\
 \Psi_{2, n}^{(2)} \\
\end{array}
\right]
\mapsto 
\left[
\begin{array}{cc}
 \Psi_{1, n}^{(1)\, -1} & O \\
 O & \Psi_{2, n}^{(1)\, -1} \\
\end{array}
\right]
\left[
\begin{array}{c}
 \Psi_{1, n}^{(2)} \\
 \Psi_{2, n}^{(2)} \\
\end{array}
\right]
= \left[
\begin{array}{c}
 \Psi_{1, n}^{(1)\, -1} \Psi_{1, n}^{(2)} \\
 \Psi_{2, n}^{(1)\, -1} \Psi_{2, n}^{(2)} \\
\end{array}
\right]. 
\]
Then, the Lax representation (\ref{gen-Lax}) 
is transformed 
to 
the 
degenerate 
form: 
\begin{align}
&
\left[
\begin{array}{c}
 \Psi_{1, n}^{(1)\, -1} \Psi_{1, n}^{(2)} \\
 \Psi_{2, n}^{(1)\, -1} \Psi_{2, n}^{(2)} \\
\end{array}
\right] =
\left[
\begin{array}{cc}
 I- \Psi_{1, n}^{(1)\, -1} L_{12,n} \Psi_{2, n+1}^{(1)} & 
 \Psi_{1, n}^{(1)\, -1} L_{12,n} \Psi_{2, n+1}^{(1)} \\
 \Psi_{2, n}^{(1)\, -1} L_{21,n} \Psi_{1, n+1}^{(1)} &
 I- \Psi_{2, n}^{(1)\, -1} L_{21,n} \Psi_{1, n+1}^{(1)} \\
\end{array}
\right]
\left[
\begin{array}{c}
 \Psi_{1, n+1}^{(1)\, -1} \Psi_{1, n+1}^{(2)} \\
 \Psi_{2, n+1}^{(1)\, -1} \Psi_{2, n+1}^{(2)} \\
\end{array}
\right],
\nonumber \\[2.5mm]
&
\left[
\begin{array}{c}
 \Psi_{1, n}^{(1)\, -1} \Psi_{1, n}^{(2)} \\
 \Psi_{2, n}^{(1)\, -1} \Psi_{2, n}^{(2)} \\
\end{array}
\right]_t = 
\left[
\begin{array}{cc}
 -\Psi_{1, n}^{(1)\, -1} M_{12,n} \Psi_{2, n}^{(1)} & 
 \Psi_{1, n}^{(1)\, -1} M_{12,n} \Psi_{2, n}^{(1)} \\
 \Psi_{2, n}^{(1)\, -1} M_{21,n} \Psi_{1, n}^{(1)} &
 -\Psi_{2, n}^{(1)\, -1} M_{21,n} \Psi_{1, n}^{(1)} \\
\end{array}
\right]
\left[
\begin{array}{c}
 \Psi_{1, n}^{(1)\, -1} \Psi_{1, n}^{(2)} \\
 \Psi_{2, n}^{(1)\, -1} \Psi_{2, n}^{(2)} \\
\end{array}
\right]. 
\nonumber
\end{align}
%
Indeed, 
there are only \mbox{$2+2$}
independent quantities  
in 
these 
Lax matrices. 
Thus, 
we can express them 
in terms of the components of the linear eigenfunction 
as 
\begin{align}
&
\Psi_{1,n}^{(1)\,-1} L_{12,n} \Psi_{2,n+1}^{(1)} = 
\left( 
\Psi_{1,n}^{(1)\,-1} \Psi_{1,n}^{(2)} 
- \Psi_{1,n+1}^{(1)\,-1} \Psi_{1,n+1}^{(2)} \right)
\left( 
\Psi_{2,n+1}^{(1)\,-1} \Psi_{2,n+1}^{(2)} 
- \Psi_{1,n+1}^{(1)\,-1} \Psi_{1,n+1}^{(2)} \right)^{-1},
\label{psi1-uv} \\
&
\Psi_{2,n}^{(1)\,-1} L_{21,n} \Psi_{1,n+1}^{(1)} = 
\left( 
\Psi_{2,n}^{(1)\,-1} \Psi_{2,n}^{(2)} 
- \Psi_{2,n+1}^{(1)\,-1} \Psi_{2,n+1}^{(2)} \right)
\left( 
\Psi_{1,n+1}^{(1)\,-1} \Psi_{1,n+1}^{(2)} 
- \Psi_{2,n+1}^{(1)\,-1} \Psi_{2,n+1}^{(2)} \right)^{-1},
\label{psi2-uv}
\end{align}
and 
\begin{align}
& \Psi_{1,n}^{(1)\,-1} M_{12,n} \Psi_{2,n}^{(1)} 
= \left( \Psi_{1,n}^{(1)\,-1} \Psi_{1,n}^{(2)} \right)_t 
 \left( 
 \Psi_{2,n}^{(1)\,-1} \Psi_{2,n}^{(2)} 
 -\Psi_{1,n}^{(1)\,-1} \Psi_{1,n}^{(2)} 
\right)^{-1},
\label{M12-uv}
\\
& \Psi_{2,n}^{(1)\,-1} M_{21,n} \Psi_{1,n}^{(1)} 
= \left( \Psi_{2,n}^{(1)\,-1} \Psi_{2,n}^{(2)} \right)_t 
 \left( 
 \Psi_{1,n}^{(1)\,-1} \Psi_{1,n}^{(2)} 
 -\Psi_{2,n}^{(1)\,-1} \Psi_{2,n}^{(2)} 
\right)^{-1}.
\label{M21-uv}
\end{align}
Note that 
(\ref{gen-Lax1}) also implies the two important relations:
\begin{align}
\label{L11-12}
& \Psi_{1,n}^{(1)\,-1} L_{11,n} \Psi_{1,n+1}^{(1)}
= I - \Psi_{1,n}^{(1)\,-1} L_{12,n} \Psi_{2,n+1}^{(1)}, 
\\[1mm]
\label{L22-21}
& \Psi_{2,n}^{(1)\,-1} L_{22,n} \Psi_{2,n+1}^{(1)} = 
  I - \Psi_{2,n}^{(1)\,-1} L_{21,n} \Psi_{1,n+1}^{(1)}.
\end{align}
Typically, 
$L_{11,n}$ and $L_{22,n}$ are 
ultralocal functions 
of $L_{12,n}$ and $L_{21,n}$, 
such as 
\[
L_{11,n} = \alpha I + \beta L_{12,n} L_{21,n}, \hspace{5mm} 
L_{22,n} = \gamma I + \delta L_{21,n} L_{12,n}, 
\]
where $\alpha$, $\beta$, $\gamma$ and $\delta$ are 
scalar functions of the spectral parameter. 
Thus, we can try to
solve (\ref{L11-12}) and (\ref{L22-21}) 
to express $\Psi_{1,n}^{(1)\,-1} \Psi_{1,n+1}^{(1)}$
and $\Psi_{2,n}^{(1)\,-1} \Psi_{2,n+1}^{(1)}$ 
in terms of $\Psi_{1,n}^{(1)\,-1} L_{12,n} \Psi_{2,n+1}^{(1)}$ 
and $\Psi_{2,n}^{(1)\,-1} L_{21,n} \Psi_{1,n+1}^{(1)}$. 
This is relatively 
easy if either 
$L_{11,n}$ or $L_{22,n}$ is a constant scalar
matrix,  
{\it e.g.}, 
$\beta$ or $\delta$ vanishes in the above example. 
Then, 
using (\ref{psi1-uv}) and (\ref{psi2-uv}), 
we can also 
express 
\begin{align}
& \Psi_{1,n}^{(1)\,-1} L_{12,n} \Psi_{2,n}^{(1)}
= \Psi_{1,n}^{(1)\,-1} L_{12,n} \Psi_{2,n+1}^{(1)}
  \left( \Psi_{2,n}^{(1)\,-1} \Psi_{2,n+1}^{(1)} \right)^{-1},
\nonumber \\[1mm]
& \Psi_{1,n+1}^{(1)\,-1} L_{12,n} \Psi_{2,n+1}^{(1)}
= \left( \Psi_{1,n}^{(1)\,-1} \Psi_{1,n+1}^{(1)} \right)^{-1} 
 \Psi_{1,n}^{(1)\,-1} L_{12,n} \Psi_{2,n+1}^{(1)}, 
\nonumber \\[1mm]
& \Psi_{2,n}^{(1)\,-1} L_{21,n} \Psi_{1,n}^{(1)}
 = \Psi_{2,n}^{(1)\,-1} L_{21,n} \Psi_{1,n+1}^{(1)}
  \left( \Psi_{1,n}^{(1)\,-1} \Psi_{1,n+1}^{(1)} \right)^{-1}, 
\nonumber \\[1mm]
&  \Psi_{2,n+1}^{(1)\,-1} L_{21,n} \Psi_{1,n+1}^{(1)}
= \left( \Psi_{2,n}^{(1)\,-1} \Psi_{2,n+1}^{(1)} \right)^{-1}
 \Psi_{2,n}^{(1)\,-1} L_{21,n} \Psi_{1,n+1}^{(1)}, \hspace{5mm} 
\mathrm{etc.}
\nonumber 
\end{align}
recursively 
in terms of 
\begin{equation}
\label{uv-def}
u_n := \Psi_{1,n}^{(1)\,-1} \Psi_{1,n}^{(2)}, 
\hspace{5mm}
v_n := \Psi_{2,n}^{(1)\,-1} \Psi_{2,n}^{(2)}. 
\end{equation}
%
In general, 
$M_{12,n}$ and $M_{21,n}$ are local (but not ultralocal) 
functions of $L_{12,n}$ and $L_{21,n}$. 
Therefore, 
with the aid of the above relations, 
(\ref{M12-uv}) and (\ref{M21-uv}) 
%
can be rewritten as 
a 
closed lattice 
system 
for 
$u_n$ 
and 
$v_n$. 

Let us illustrate 
the 
method 
using the matrix 
Ablowitz--Ladik lattice (\ref{mALsys}) as an example. 
By setting 
\[
L_{11,n} = z I, \hspace{5mm} L_{22,n} = z^{-1} I, 
\]
(\ref{L11-12}) and (\ref{L22-21}) provide the useful 
relations
\begin{align}
\label{aux1}
& \Psi_{1,n+1}^{(1)\,-1} \Psi_{1,n}^{(1)} 
= z \left( I - \Psi_{1,n}^{(1)\,-1} L_{12,n} \Psi_{2,n+1}^{(1)} 
 \right)^{-1}, 
\\[1mm]
\label{aux2}
& \Psi_{2,n+1}^{(1)\, -1} \Psi_{2,n}^{(1)} 
 = z^{-1} \left( I - \Psi_{2,n}^{(1)\,-1} L_{21,n} \Psi_{1,n+1}^{(1)} 
 \right)^{-1}.
\end{align}
The Lax representation, 
(\ref{mAL1}) and (\ref{mAL-time}),  
implies
simple relations between off-diagonal elements of the 
temporal and spatial Lax matrices, i.e.
\begin{align}
\nonumber 
& M_{12,n} + z^{-1}a L_{12,n} +z b L_{12,n-1} =O, 
\\[1mm]
& M_{21,n} + z b L_{21,n} +z^{-1} a L_{21,n-1} =O, 
\nonumber
\end{align}
which can be rewritten as 
\begin{align}
\nonumber 
 \Psi_{1,n}^{(1)\,-1} M_{12,n} \Psi_{2,n}^{(1)} 
& + z^{-1} a \Psi_{1,n}^{(1)\,-1} L_{12,n} \Psi_{2,n+1}^{(1)}
  \Psi_{2,n+1}^{(1)\, -1} \Psi_{2,n}^{(1)} 
\\ & 
+ z b  \Psi_{1,n}^{(1)\, -1} \Psi_{1,n-1}^{(1)}
 \Psi_{1,n-1}^{(1)\,-1} L_{12,n-1} \Psi_{2,n}^{(1)} =O,
\nonumber 
\\[2mm]
\Psi_{2,n}^{(1)\,-1} M_{21,n} \Psi_{1,n}^{(1)} 
& + z b \Psi_{2,n}^{(1)\,-1} L_{21,n} \Psi_{1,n+1}^{(1)}
  \Psi_{1,n+1}^{(1)\, -1}  \Psi_{1,n}^{(1)} 
\nonumber \\ & 
+ z^{-1} a \Psi_{2,n}^{(1)\, -1} \Psi_{2,n-1}^{(1)} 
 \Psi_{2,n-1}^{(1)\,-1} L_{21,n-1} \Psi_{1,n}^{(1)} = O. 
\nonumber 
\end{align}
Substituting 
(\ref{aux1}) and (\ref{aux2}) 
and subsequently 
(\ref{psi1-uv})--(\ref{M21-uv}) into the above 
relations,
we arrive at a closed system for 
$\Psi_{1,n}^{(1)\,-1} \Psi_{1,n}^{(2)}$ and 
$\Psi_{2,n}^{(1)\,-1} \Psi_{2,n}^{(2)}$.
\\
\begin{proposition}
Consider two linearly independent 
solutions of 
(\ref{mAL1}) and (\ref{mAL-time}) 
and write them as in (\ref{gen-Lax}) 
with $j=1$ or $2$. 
Then, 
$u_n$ and $v_n$ defined in
(\ref{uv-def})
satisfy the following system:
\begin{subnumcases}{\label{lHF1}}
{}
 u_{n,t} + \frac{a}{\mu
} \left( u_n - u_{n+1} \right) 
	\left[ I + (v_n-u_n)^{-1}(u_n-u_{n+1}) \right]^{-1} 
\nonumber \\ \hphantom{u_{n,t}}
	+\mu
b \left[ I - (u_{n-1}-u_n) (v_n-u_n)^{-1}\right]^{-1}
	(u_{n-1}-u_n) = O, \hspace{10mm}
 \\[2mm]
 v_{n,t} + \mu
b \left( v_n -v_{n+1} \right) 
	\left[ I + (u_n-v_n)^{-1} (v_n -v_{n+1})\right]^{-1} 
\nonumber \\ \hphantom{v_{n,t}}
	+\frac{a}{\mu
} \left[ I - (v_{n-1} -v_n ) (u_n -v_n)^{-1}\right]^{-1}
 (v_{n-1} - v_n ) = O,
\end{subnumcases}
where \mbox{$\mu = z^2$}.
\label{prop1}
\end{proposition}
\noindent
{\it Remarks:}
\\
(i) 
The system 
(\ref{lHF1}) 
with 
\mbox{$\mu
b = (a/\mu
)^\ast$}
allows 
both 
the complex conjugation reduction 
\mbox{$v_n = \sigma u_n^\ast$}
and the Hermitian conjugation reduction 
\mbox{$v_n = \sigma u_n^\dagger$}, where 
\mbox{$\sigma = \pm 1$}. 
In addition, by setting \mbox{$v_n = -u_n$}, 
(\ref{lHF1}) 
with 
\mbox{$\mu
b = a/\mu
$} 
reduces to a single matrix equation, 
\[
u_{n,t} = u_n 
\left[ \left( u_{n+1} + u_n \right)^{-1} 
 - 
\left( u_n + u_{n-1} \right)^{-1} \right] u_n, 
\]
up to a rescaling of 
$t$. 
In the scalar case, this 
belongs to 
Yamilov's 
list of 
Volterra-type lattices 
in~\cite{Yamilov83};
in the matrix case, it 
allows further 
reductions to 
various multicomponent 
systems 
(cf.~\cite{AdSviYam99}).  
\\
\\
(ii) 
The system 
(\ref{lHF1}) 
provides 
a space-discrete 
analog 
of the 
system studied by Svinolupov and Sokolov~\cite{SviSok94}; 
their system 
gives 
a matrix generalization of 
the Heisenberg ferromagnet model 
written in 
a two-component 
form~\cite{MS2,MikShYam87}. 
Indeed,
(\ref{lHF1}) in the scalar case 
is 
closely 
related to 
the lattice Heisenberg ferromagnet model~\cite{Ishi82}
and its 
simplest 
higher 
symmetry~\cite{GIV86,Papanico}. 
\\
\\
(iii) The system (\ref{lHF1}) in the scalar case 
appeared in the recent paper~\cite{Veks11}
(also see~\cite{AdSha06}). 
In this context, it is natural to rewrite 
(\ref{lHF1}) as a two-component system for 
the pair of variables $u_n$ and $v_n^{-1}$
(cf.~(4.8) in~\cite{MS2}). 
\\

In (\ref{lHF1}), 
the two variables $u_n$ and $v_n$ 
interact with each other 
through the quantity \mbox{$\left( v_n - u_n \right)^{-1}$},  
which can be 
used as a new 
dependent 
variable. 
Indeed, a direct calculation 
shows the following two 
propositions.
\\
\begin{proposition}
Let 
$u_n$ and $v_n$
satisfy the system (\ref{lHF1}). 
Then, the new pair of
variables $q_n$ and $r_n$, 
\[
q_n :=\boldsymbol{\Delta}_n^+ u_n 
	\left( = u_{n+1} - u_{n} \right) , \hspace{5mm}
r_n := \left( v_n - u_n \right)^{-1}, 
\]
satisfies the space-discrete Kaup--Newell system~\cite{Tsuchi02}: 
\begin{subnumcases}{\label{sdKN}}
{}
 q_{n,t} - \boldsymbol{\Delta}_n^+ 
 \left[ \frac{a}{\mu
} \left( I - q_{n}r_{n} \right)^{-1} q_{n}
 + \mu
b \left( I + q_{n-1} r_{n} \right)^{-1} q_{n-1} \right] = O,
\\[1mm]
 r_{n,t} - \boldsymbol{\Delta}_n^+ 
 \left[ \mu
b  \left( I + r_{n} q_{n-1} \right)^{-1} r_{n}
 + \frac{a}{\mu
} \left( I - r_{n-1} q_{n-1} \right)^{-1} r_{n-1} \right] 
 = O, 
\hspace{15mm} 
\end{subnumcases}
where $\boldsymbol{\Delta}_n^+$ 
denotes the forward difference operator. 
\\
\label{prop2}
\end{proposition}
%
\begin{proposition}
Let 
$u_n$ and $v_n$
satisfy the system (\ref{lHF1}). 
Then, the new pair of
variables $\widetilde{q}_n$ and $\widetilde{r}_n$, 
\[
\widetilde{q}_n := \left( v_n - u_n \right)^{-1}, \hspace{5mm}
\widetilde{r}_n := \boldsymbol{\Delta}_n^+ v_{n-1} 
	\left( = v_{n} - v_{n-1} \right), 
\]
also 
satisfies the space-discrete Kaup--Newell
system (\ref{sdKN}) for $\widetilde{q}_n$ and $\widetilde{r}_n$.
\\
%
\label{prop3}
\end{proposition}
%

\section{Inverse scattering method for the 
Ablowitz--Ladik lattice}
\label{sec3}

\subsection{Revisiting the 
Ablowitz--Ladik eigenvalue problem}

In this section, we describe 
the 
inverse scattering method 
associated with the matrix Ablowitz--Ladik eigenvalue 
problem 
(\ref{mAL1}), 
\begin{equation}
\left[
\begin{array}{c}
 \Psi_{1, n} \\
 \Psi_{2, n} \\
\end{array}
\right]
= 
 \left[
 \begin{array}{cc}
  z I & z Q_n \\
 z^{-1} R_n & z^{-1} I \\
 \end{array}
 \right]
\left[
\begin{array}{c}
 \Psi_{1, n+1} \\
 \Psi_{2, n+1} \\     
\end{array}
\right].
\label{mAL0}
\end{equation}
Here, 
the potentials $Q_n$ and $R_n$ are 
assumed to decay 
sufficiently rapidly at spatial infinity: 
\begin{equation}
\label{zero-bc}
\lim_{n \to \pm \infty} Q_n = \lim_{n \to \pm \infty} R_n =O.
\end{equation}
The 
matrix generalization of the 
Ablowitz--Ladik 
lattice~\cite{AL76} 
first considered 
in the early 1980s~\cite{GI82}
is still 
a 
topic of interest in 
discrete 
integrable systems 
(see~\cite{Tsuchi02,APT, DM2010} and references therein). 
In our previous papers~\cite{TUW98,TUW99},
we 
presented 
the inverse scattering method 
for the matrix Ablowitz--Ladik 
lattice
while assuming 
some 
symmetry 
conditions on 
the potentials $Q_n$ and $R_n$.
Here, we 
remove 
such assumptions 
and 
consider the 
general 
case of 
\mbox{$\l \times l$} 
square 
matrices $Q_n$ and $R_n$; 
recall that 
the 
results on rectangular 
matrix potentials 
can 
be 
obtained 
by setting 
some 
rows/columns in 
$Q_n$ and $R_n$ as 
zero. 
The inverse scattering method 
reported 
here 
bypasses
some redundant computation and consideration 
contained in the existing literature 
on the same 
subject, 
so 
we believe that 
this 
is the most streamlined 
version. 
The results 
in 
the previous work~\cite{TUW98,TUW99} 
can be 
reproduced 
by 
imposing 
some reduction conditions 
on the scattering data; 
this 
is briefly sketched 
in 
appendix~\ref{app2}\@. 
In addition, 
we fix 
some minor inconsistencies 
in~\cite{TUW98,TUW99}, 
though 
they do not 
affect the 
main 
results 
of 
these 
papers.  
%
%

All the flows of 
the matrix 
Ablowitz--Ladik hierarchy 
are associated with 
the same 
eigenvalue problem (\ref{mAL0}), 
so they can be solved 
together 
by the 
inverse scattering method. 
%
However, in the following, 
we concentrate on 
the matrix Ablowitz--Ladik lattice 
(\ref{mALsys})
to illustrate the method 
in an easy-to-read manner. 
We stress that in contrast to 
other methods of obtaining 
special 
solutions, 
the inverse scattering method 
can 
provide 
the general solution 
formulas.
Moreover, the method 
can determine not only 
the potentials 
$Q_n$ and $R_n$ but also 
a fundamental 
set of 
linear 
eigenfunctions, 
which will be used in 
section~4.

\subsection{Jost solutions and relevant quantities}
\label{subs3.2}

To analyze 
the general case 
of the matrix 
potentials $Q_n$ and $R_n$, 
we 
consider 
the 
adjoint equation,\footnote{Note 
that if we consider a square matrix solution 
$\Psi_n$ to the eigenvalue problem \mbox{$\Psi_{n} = L_n(z) \Psi_{n+1}$}, 
then its 
inverse 
\mbox{$\Phi_n := \Psi_n^{-1}$} satisfies the 
eigenvalue problem \mbox{$\Phi_{n+1} = \Phi_{n} L_n (z)$}.}
%
\begin{equation}
\left[
\begin{array}{cc}
\!
\Phi_{1,n+1}
 \! & \!  
\Phi_{2,n+1} \!
\end{array}
\right]
= 
\left[
\begin{array}{cc}
\! \Phi_{1,n} \! & \! \Phi_{2,n} \!
\end{array}
\right]
\left[
 \begin{array}{cc}
  z I & z Q_n \\
  z^{-1} R_n & z^{-1} I \\
 \end{array}
 \right].
\label{mAL2}
\end{equation}
Indeed, 
a discrete analog of 
Lagrange's identity, 
\begin{align}
& \left[
\begin{array}{cc}
\! \Phi_{1,n} \! & \! \Phi_{2,n} \!
\end{array}
\right] 
\left\{
\left[
\begin{array}{c}
 \Psi_{1, n} \\
 \Psi_{2, n} \\
\end{array}
\right]
-
 \left[
 \begin{array}{cc}
  z I & z Q_n \\
 z^{-1} R_n & z^{-1} I \\
 \end{array}
 \right]
\left[
\begin{array}{c}
 \Psi_{1, n+1} \\
 \Psi_{2, n+1} \\
\end{array}
\right] \right\}
\nonumber \\
& \mbox{} - \left\{
\left[
\begin{array}{cc}
\!
\Phi_{1,n+1}
 \! & \!  
\Phi_{2,n+1} \!
\end{array}
\right]
- 
\left[
\begin{array}{cc}
\! \Phi_{1,n} \! & \! \Phi_{2,n} \!
\end{array}
\right]
\left[
 \begin{array}{cc}
  z I & z Q_n \\
  z^{-1} R_n & z^{-1} I \\
 \end{array}
 \right]
\right\} \left[
\begin{array}{c}
 \Psi_{1, n+1} \\
 \Psi_{2, n+1} \\
\end{array}
\right]
\nonumber \\[1mm]
& =
- \boldsymbol{\Delta}_n^+ 
\left\{ 
\left[
\begin{array}{cc}
\! \Phi_{1,n} \! & \! \Phi_{2,n} \!
\end{array}
\right]
\left[
\begin{array}{c}
 \Psi_{1, n} \\
 \Psi_{2, n} \\
\end{array}
\right]
\right\},
\nonumber 
\end{align}
implies 
that (\ref{mAL0}) and (\ref{mAL2}) 
can be 
said to be 
adjoint to 
each other. 
%
Thus, 
we can introduce an $l \times l$ matrix function 
\mbox{$W[ \, \cdot\, , \, \cdot \,]$}
for a pair of 
solutions to (\ref{mAL0}) and 
(\ref{mAL2}) as
%
\[
 W [ \Phi_n, 
\Psi_n ] := 
\left[
\begin{array}{cc}
\! \Phi_{1,n} \! & \! \Phi_{2,n} \! 
\end{array}
\right]
\left[
\begin{array}{c}
 \Psi_{1,n} \\
 \Psi_{2,n} \\
\end{array}
\right] 
= \Phi_{1,n} \Psi_{1,n} + \Phi_{2,n} \Psi_{2,n}, 
\]
which is $n$-independent:
%
 \[
W [\Phi_{n}, \Psi_{n}] 
	= W [\Phi_{n+1}, \Psi_{n+1}].
 \]
 
In addition to the 
rapidly decaying boundary conditions (\ref{zero-bc}), 
we 
assume 
that the spatial Lax matrix defining the eigenvalue problem 
is invertible: 
\begin{equation}
\det \left( I-Q_n R_n \right) 
\neq 0, \hspace{3mm} 
\forall\hspace{1pt}
n\in {\mathbb Z}. 
\label{nonzero-det}
\end{equation}
Because 
\mbox{$\log \left[ \det \left( I-Q_n R_n \right) \right]$} 
is a conserved density
for 
the matrix Ablowitz--Ladik hierarchy~\cite{GI82}, 
this assumption is preserved under the time evolution. 
Thus, 
a 
set of column-vector (or row-vector) solutions 
to the eigenvalue problem (\ref{mAL0}) 
(or (\ref{mAL2})) 
that are linearly independent 
at some lattice 
site, 
say 
\mbox{$n=n_0$},  
remain 
independent for all \mbox{$n \in {\mathbb Z}$}. 
%

We introduce 
Jost solutions 
$\phi_n (z)$, $\widebar{\phi}_n (z)$ 
and $\psi_n (z)$,
$\widebar{\psi}_n (z)$ at a fixed time
that satisfy 
(\ref{mAL0}) 
and the boundary conditions,
\begin{subequations}
\label{leftJost}
 \begin{equation}
\left.
\begin{array}{l}
 z^{n} \phi_n  \to
 \left[
 \begin{array}{c}
   I  \\
   O  \\
 \end{array}
 \right] 
\vspace{2mm}
\\
z^{-n} \widebar{\phi}_n  \to 
 \left[
 \begin{array}{c}
   O  \\
   -I  \\
 \end{array}
 \right]
\end{array}
\right\}
 \hspace{4mm}
 {\rm as}~~~ n \rightarrow -\infty
\label{phi_bar}
\end{equation}
and
\begin{equation}
\left.
\begin{array}{l}
 z^{-n} \psi_n  \to 
 \left[
 \begin{array}{c}
   O  \\
   I  \\
 \end{array}
 \right] 
\vspace{2mm}
\\
 z^{n} \widebar{\psi}_n  \to 
 \left[
 \begin{array}{c}
   I  \\
   O  \\
 \end{array}
 \right] 
\end{array}
\right\}
\hspace{4mm}
 {\rm as}~~~ n \rightarrow +\infty.
 \label{psi_bar}
\end{equation}
\end{subequations}
The time evolution of the Jost solutions 
will be considered 
in subsection~\ref{sTDsd}. 
Note that the overbar 
does {\it not} mean 
complex conjugation in this paper. 
%
Similarly, 
we 
introduce 
adjoint Jost solutions 
$\phi_n^{\mathrm{ad}} (z)$, $\widebar{\phi}_n^{\mathrm{ad}} (z)$ 
and $\psi_n^{\mathrm{ad}}(z)$,
$\widebar{\psi}_n^{\mathrm{ad}} (z)$ 
that satisfy (\ref{mAL2}) and the boundary conditions, 
%
\begin{subequations}
\label{rightJost}
 \begin{equation}
\left.
\begin{array}{l}
 z^{n} \phi_n^{\mathrm{ad}}   \to
\left[
\begin{array}{cc}
 \! O \! & \! -I \!
\end{array}
\right]
\vspace{2mm}
\\
  z^{-n} \widebar{\phi}_n^{\mathrm{ad}}  \to 
 \left[
 \begin{array}{cc}
 \!  I \! & \! O \!
 \end{array}
 \right]
\end{array}
\right\}
 \hspace{4mm}
 {\rm as}~~~ n \rightarrow -\infty
 \end{equation}
and
 \begin{equation}
\left.
\begin{array}{l}
 z^{-n} \psi_n^{\mathrm{ad}} \to 
 \left[
 \begin{array}{cc}
  \! I  \! & \! O \!
 \end{array}
 \right] 
\vspace{2mm}
\\
 z^{n} \widebar{\psi}_n^{\mathrm{ad}}   \to 
 \left[
 \begin{array}{cc}
 \!  O  \! & \! I  \!
 \end{array}
 \right] 
\end{array}
\right\}
\hspace{4mm}
 {\rm as}~~~ n \rightarrow +\infty.
 \label{psi_ad}
 \end{equation}
\end{subequations}
Because 
the \mbox{$l+l\hspace{2pt}(=2l)$} columns 
of the Jost solutions 
$\psi_n$ and $\widebar{\psi}_n$ 
form a fundamental 
set of 
solutions to 
the eigenvalue problem (\ref{mAL0}), we can set 
on the unit circle \mbox{$|z|=1$} as 
\begin{subequations}
\label{ref102}
 \begin{align}
 \phi_n (z) & = \widebar{\psi}_n (z) A
        + \psi_n (z)B,
 \label{phi_relation} \\[0.5mm]
 \widebar{\phi}_n (z) &= \widebar{\psi}_n (z) \widebar{B}
        - \psi_n (z) \widebar{A}.
 \label{phi_bar_relation}
 \end{align}
\end{subequations}
Here, 
$A$, $B$, $\widebar{B}$ and $\widebar{A}$ 
are $n$-independent \mbox{$l \times l$} matrices, 
which 
depend on the spectral parameter $z$ 
and are called scattering data. 
%
%
According to the asymptotic behaviors of the Jost solutions 
(\ref{leftJost})--(\ref{rightJost}), we can express them 
on 
\mbox{$|z|=1$} as
\begin{subequations}
\label{W_rel}
\begin{align}
A &= W[ \psi_n^{\mathrm{ad}}, \phi_n ],
\label{rep1}
\\[1mm]
B &= W[ \widebar{\psi}_n^{\mathrm{ad}}, \phi_n ],
 \label{rep3} 
\\[1mm]
 \widebar{B} &= 
	W [ \psi_n^{\mathrm{ad}}, \widebar{\phi}_n ],
 \label{rep4}
 \\[1mm]
%
 \widebar{A} &= 
	- W [ \widebar{\psi}_n^{\mathrm{ad}}, \widebar{\phi}_n ].
 \label{rep2}
\end{align}
\end{subequations}

We can rewrite the eigenvalue problem (\ref{mAL0}) 
in the following 
equivalent 
forms:\footnote{In the 
area 
of orthogonal polynomials, 
%
the Ablowitz--Ladik eigenvalue problem 
in a similar 
rewritten form was studied by G.~Baxter in the early 1960s
after the seminal work of G.~Szeg\"o, see~\cite{Bax1,Bax2}.
}
%
\begin{subequations}
\label{Jost-1}
\begin{align}
%
\left[
\begin{array}{c}
 z^{-n} \Psi_{1, n} \\
 z^{-n} \Psi_{2, n} \\
\end{array}
\right]
&=
 \left[
 \begin{array}{cc}
  z^2 I & z^2 Q_n \\
  R_n & I \\
 \end{array}
 \right]
\left[
\begin{array}{c}
 z^{-(n+1)} \Psi_{1, n+1} \\
 z^{-(n+1)} \Psi_{2, n+1} \\
\end{array}
\right]_{\vphantom \int},
\label{mAL3}
\\[1.5mm]
\left[
\begin{array}{c}
 z^{n} \Psi_{1, n} \\
 z^{n} \Psi_{2, n} \\
\end{array}
\right]
&=
 \left[
 \begin{array}{cc}
  I & Q_n \\
  z^{-2} R_n & z^{-2} I \\
 \end{array}
 \right]
\left[
\begin{array}{c}
 z^{n+1} \Psi_{1, n+1} \\
 z^{n+1} \Psi_{2, n+1} \\
\end{array}
\right].
\label{mAL4}
\end{align}
\end{subequations}
%
Thus, in view of the boundary conditions (\ref{leftJost}), 
$z^{n} \phi_n$, 
$z^{-n} \widebar{\phi}_n$, 
$z^{-n} \psi_n$ and 
$z^{n} \widebar{\psi}_n$ 
depend on 
$z$ only through \mbox{$z^2$}.
%
Similarly, 
(\ref{mAL2}) can be rewritten 
as 
%
\begin{subequations}
\label{Jost-2}
\begin{align}
\left[
\begin{array}{cc}
 \! z^{n+1} \Phi_{1, n+1} \! & \! 
 z^{n+1} \Phi_{2, n+1} \!
\end{array}
\right]
&=
\left[
\begin{array}{cc}
 \! z^{n} \Phi_{1, n} \! & \! 
 z^{n} \Phi_{2, n} \!
\end{array}
\right]
\left[
 \begin{array}{cc}
  z^2 I & z^2 Q_n \\
  R_n & I \\
 \end{array}
 \right], 
\label{mAL5}
\\[1.5mm]
\left[
\begin{array}{cc}
 \! z^{-(n+1)} \Phi_{1, n+1} \! & \! 
 z^{-(n+1)} \Phi_{2, n+1} \!
\end{array}
\right]
&=
\left[
\begin{array}{cc}
\! z^{-n} \Phi_{1, n} \! & \! 
 z^{-n} \Phi_{2, n} \!
\end{array}
\right]
\left[
 \begin{array}{cc}
  I & Q_n \\
  z^{-2} R_n & z^{-2} I \\
 \end{array}
 \right], 
\label{mAL6}
\end{align}
\end{subequations}
%
so 
$z^{-n} \psi_n^{\mathrm{ad}}$ and 
$z^{n} \widebar{\psi}_n^{\mathrm{ad}}$ 
depend on $z$ only through \mbox{$z^2$}. 
Therefore, 
relations (\ref{W_rel}) imply 
that the scattering data 
$A$, $B$, $\widebar{B}$ and $\widebar{A}$ are even 
functions of $z$ 
(cf.~\cite{Papanico});  
they 
can be denoted as 
$A(\mu)$, $B(\mu)$, $\widebar{B}(\mu)$ and $\widebar{A}(\mu)$, 
%
where 
\[
\mu 
=z^2. 
\]
%
%
%

We introduce 
the following 
representations of 
the Jost solutions 
$\psi_n$ and $\widebar{\psi}_n$:
 \begin{subequations}
 \label{psi_form}
\begin{align}
&  z^{-n} \psi_n
=
\dprod{\displaystyle \curvearrowright}{\infty}_{i=n}
\left[
 \begin{array}{cc}
  \mu I & \mu Q_i \\
  R_i & I \\
 \end{array}
 \right]
\left[
\begin{array}{c}
 O \\
 I \\
\end{array}
\right]
=:
\left[
\begin{array}{c}
 O \\
 I \\
\end{array}
\right]
+ 
\sum_{k=0}^{\infty} \mu^{k+1} K (n, n+k), 
\label{Kdef}
\\[0.5mm]
& z^{n} \widebar{\psi}_n
=
\dprod{\displaystyle \curvearrowright}{\infty}_{i=n}
 \left[
 \begin{array}{cc}
  I & Q_i \\
  \mu^{-1} R_i & \mu^{-1} I \\
 \end{array}
 \right]
\left[
\begin{array}{c}
 I \\
 O \\
\end{array}
\right]
=: 
\left[
\begin{array}{c}
 I \\
 O \\
\end{array}
\right] +
\sum_{k=0}^{\infty} \mu^{-k-1} \widebar{K} (n,n+k),
\label{5Kdef}
\end{align}
\end{subequations}
which are 
assumed 
to be uniformly convergent 
in \mbox{$|\mu|\le 1$} and \mbox{$|\mu|\ge 1$}, respectively 
(cf.~(\ref{zero-bc})). 
Here and hereafter, 
the order of the matrix product 
is 
defined 
as 
\[
\dprod{\displaystyle \curvearrowright}{m
}_{i=n} 
X_i := X_n X_{n+1} 
	\cdots X_m, \hspace{5mm} 
\dprod{\displaystyle \curvearrowleft}{m
}_{i=n} 
X_i := X_m X_{m-1} \cdots 
	X_n, \hspace{5mm} 
m \ge n
,
\]
and the ``integral kernels"
$K(n,m)$ and $\widebar{K}(n,m)$ 
are $\mu$-independent 
\mbox{$2l \times l$} matrices denoted in terms 
of \mbox{$l \times l$} 
matrices as 
 \[
 K(n,m) =
 \left[
\begin{array}{c}
  K_1(n,m)  \\
  K_2(n,m)  \\
 \end{array}
 \right],
 \hspace{5mm}
 \widebar{K}(n,m) =
 \left[
 \begin{array}{c}
  \widebar{K}_1(n,m)  \\
  \widebar{K}_2(n,m)  \\
 \end{array}
 \right], \hspace{5mm} m \ge n.
 \]
We substitute (\ref{Kdef}) and (\ref{5Kdef}) 
into 
(\ref{mAL3}) and (\ref{mAL4}), respectively. 
Noting that they are identities in $\mu$, we 
can express the ``integral kernels" recursively 
in terms of the potentials $Q_n$ and $R_n$; 
the 
most 
important relations are
%
%
%
\begin{align}
& K_1(n,n) =Q_n,
\label{K2}
\\[1mm]
%
&  K_2 (n,n) = \sum_{j=n}^\infty R_j Q_{j+1},
\nonumber
%
\end{align}
%
and 
%
%
\begin{align}
%
& \widebar{K}_2 (n,n)=R_n,
\label{K_bar2}
\\[1mm]
%
& \widebar{K}_1 (n,n) = \sum_{j=n}^\infty Q_j R_{j+1}. 
\nonumber
\end{align}

In a way similar to (\ref{psi_form}), we can also 
express 
the other (adjoint) Jost solutions as
power series in
either 
$\mu$ or $\mu^{-1}$. 
Indeed, noting 
the identity, 
\[
\left[
 \begin{array}{cc}
  \mu I & \mu Q_n \\
  R_n & I \\
 \end{array}
 \right] 
  \left[
 \begin{array}{cc}
  \mu^{-1} ( I-Q_n R_n )^{-1} & -Q_n ( I-R_n Q_n )^{-1} \\
  -\mu^{-1} R_n ( I-Q_n R_n )^{-1} & ( I-R_n Q_n )^{-1} \\
 \end{array}
 \right]
 =
 \left[
 \begin{array}{cc}
  I & O \\
  O  & I \\
 \end{array}
 \right], 
\]
we obtain
\begin{subequations}
\begin{align}
&  z^{n} \phi_n
=
\dprod{\displaystyle \curvearrowleft}{n-1}_{i=-\infty}
 \left[
 \begin{array}{cc}
 \left( I-Q_i R_i \right)^{-1} 
	& -\mu Q_i \left( I-R_i Q_i \right)^{-1} \\
  - R_i \left( I-Q_i R_i \right)^{-1} 
	& \mu \left( I-R_i Q_i \right)^{-1} \\
 \end{array}
 \right]
\left[
\begin{array}{c}
 I \\
 O \\
\end{array}
\right],
\label{phi-ex}
\\[0.5mm]
& z^{-n} \widebar{\phi}_n
=
\dprod{\displaystyle \curvearrowleft}{n-1}_{i=-\infty}
 \left[
 \begin{array}{cc}
  \mu^{-1} \left( I-Q_i R_i \right)^{-1} 
	& -Q_i \left( I-R_i Q_i \right)^{-1} \\
  -\mu^{-1} R_i \left( I-Q_i R_i \right)^{-1} 
	& \left( I-R_i Q_i \right)^{-1} \\
 \end{array}
 \right]
\left[
\begin{array}{c}
 O \\
 -I \\
\end{array}
\right],
\label{phi_bar-ex}
\\
&  z^{-n} \psi_n^{\mathrm{ad}} =
 \left[
 \begin{array}{cc}
  \! I  \! & \! O \!
 \end{array}
 \right] 
\dprod{\displaystyle \curvearrowleft}{\infty}_{i=n}
 \left[
 \begin{array}{cc}
 \left( I-Q_i R_i \right)^{-1} 
	& -\mu Q_i \left( I-R_i Q_i \right)^{-1} \\
  - R_i \left( I-Q_i R_i \right)^{-1} 
	& \mu \left( I-R_i Q_i \right)^{-1} \\
 \end{array}
 \right],
\nonumber
\\[0.5mm]
& z^{n} \widebar{\psi}_n^{\mathrm{ad}} =
  \left[
 \begin{array}{cc}
  \! O \! & \! I \!
 \end{array}
 \right] 
\dprod{\displaystyle \curvearrowleft}{\infty}_{i=n}
 \left[
 \begin{array}{cc}
  \mu^{-1} \left( I-Q_i R_i \right)^{-1} 
	& -Q_i \left( I-R_i Q_i \right)^{-1} \\
  -\mu^{-1} R_i \left( I-Q_i R_i \right)^{-1} 
	& \left( I-R_i Q_i \right)^{-1} \\
 \end{array}
 \right], \hspace{3mm} \mathrm{etc.}
\nonumber
\end{align}
\end{subequations}
Thus, 
(\ref{rep1}) and (\ref{rep2}) 
imply that $A(\mu)$ and $\widebar{A}(\mu)$ can be 
written 
explicitly 
as 
\begin{subequations}
\label{AAbar-ex}
\begin{align}
&  A(\mu)
= \left[
 \begin{array}{cc}
  \! I  \! & \! O \!
 \end{array}
 \right] 
\dprod{\displaystyle \curvearrowleft}{\infty}_{i=-\infty}
 \left[
 \begin{array}{cc}
 \left( I-Q_i R_i \right)^{-1} 
	& -\mu Q_i \left( I-R_i Q_i \right)^{-1} \\
  - R_i \left( I-Q_i R_i \right)^{-1} 
	& \mu \left( I-R_i Q_i \right)^{-1} \\
 \end{array}
 \right]
\left[
\begin{array}{c}
 I \\
 O \\
\end{array}
\right],
\label{A-ex}
\\
& \widebar{A}(\mu)
=  \left[
 \begin{array}{cc}
  \! O \! & \! I \!
 \end{array}
 \right] 
\dprod{\displaystyle \curvearrowleft}{\infty}_{i=-\infty}
 \left[
 \begin{array}{cc}
  \mu^{-1} \left( I-Q_i R_i \right)^{-1} 
	& -Q_i \left( I-R_i Q_i \right)^{-1} \\
  -\mu^{-1} R_i \left( I-Q_i R_i \right)^{-1} 
	& \left( I-R_i Q_i \right)^{-1} \\
 \end{array}
 \right]
\left[
\begin{array}{c}
 O \\
 I \\
\end{array}
\right]. 
\label{Abar-ex}
\end{align}
\end{subequations}
%

Therefore, 
as long as $Q_n$ and $R_n$ decay 
sufficiently rapidly 
as \mbox{$n \to \pm \infty$}, 
$z^{n}\phi_n$ 
and $z^{-n} \psi_n^{\mathrm{ad}}$ 
are analytic 
on and inside the unit circle (\mbox{$|\mu|\le 1$}), 
and 
$z^{-n} \widebar{\phi}_n$ and $z^{n} \widebar{\psi}_n^{\mathrm{ad}}$ 
are analytic 
on and outside the unit circle (\mbox{$|\mu|\ge 1$}). 
Consequently, $A(\mu)$ and $\widebar{A}(\mu)$ 
can be analytically continued 
for 
\mbox{$|\mu|\le1$}
and 
\mbox{$|\mu| \ge 1$},
respectively.\footnote{Here, we use the term 
``analytic continuation" 
loosely. 
See appendix B of~\cite{Vakh10} 
for a 
rigorous treatment
of ``analytic continuation" 
in the delicate 
case, 
{\it e.g.}, when 
$Q_n$ and $R_n$ do not decay 
exponentially 
fast 
as \mbox{$n \to \pm \infty$}.}
%
A more precise discussion on the 
analytical properties of the Jost solutions 
can be made 
using 
a discrete 
analog 
of
the approach 
in~\cite{AKNS74}; that is, 
we can rewrite the eigenvalue problem in the form of 
linear summation equations 
and 
discuss 
the convergence of 
their Liouville--Neumann-type series solutions. 
However, 
we 
omit 
such a discussion in this paper. 

\subsection{Gel'fand--Levitan--Marchenko equations}
\label{GLM_eq}

%
We 
multiply 
(\ref{phi_relation}) and (\ref{phi_bar_relation}) from the right 
by $z^{n} A(\mu)^{-1}$ and $z^{-n} \widebar{A}(\mu)^{-1}$, 
respectively, 
to obtain
\begin{subequations}
\begin{align}
 [ z^{n} \phi_n ] (\mu) A(\mu)^{-1} 
	&= [z^{n} \widebar{\psi}_n] (\mu) 
	+ [z^{-n} \psi_n] (\mu) B(\mu)A(\mu)^{-1} \mu^{n},
 \label{prep1}
\\[1mm]
 %
 [z^{-n} \widebar{\phi}_n] (\mu) \widebar{A}(\mu)^{-1} 
 &= - [z^{-n} \psi_n] (\mu) 
	+ [z^{n} \widebar{\psi}_n ] (\mu) 
	\widebar{B}(\mu) \widebar{A}(\mu)^{-1} \mu^{-n}.
 \label{prep2}
\end{align}
\end{subequations}
Here, ``$(\mu)$'' 
emphasizes that the argument of 
the functions 
is \mbox{$\mu \hspace{1pt}(=z^2)$} rather than $z$. 

Then, 
we substitute the 
summation representations 
(\ref{psi_form}) into the right-hand side of 
(\ref{prep1}) and operate 
with 
\begin{align}
 \frac{1}{2\pi \mathrm{i}} \oint_{C} 
\mathrm{d} \mu \, \mu^{m-n} \hspace{3mm} (m \ge n)
\label{oper1}
\end{align}
on both sides.
Here, $C$ denotes the counterclockwise
contour along the 
unit circle \mbox{$|\mu|=1$}. 
Thus, 
we obtain
\begin{equation}
J(n,m) = \widebar{K} (n,m) +
\left[
\begin{array}{c}
 O \\
 F_{\mathrm C}(m) \\
\end{array}
\right]
+ \sum_{k=0}^\infty K(n,n+k) F_{\mathrm C} (m+k+1),
\label{GLM-pre1}
\end{equation}
where 
%
\begin{align}
J(n,m) 
&:=
\frac{1}{2\pi \mathrm{i}} \oint_{C_{\vphantom \int}} [z^{n} \phi_n ]
	(\mu) A(\mu)^{-1} \mu^{m-n}\mathrm{d} \mu,
\label{eq_ref6}
\\[1mm]
F_{\mathrm C} (m) &:=
\frac{1}{2 \pi \mathrm{i}} \oint_C 
         B(\mu)A(\mu)^{-1} \mu^{m} \mathrm{d} \mu.
\nonumber
\end{align}
Because of the analyticity of 
$[z^{n} \phi_n ](\mu)$ and $A(\mu)$ 
in 
\mbox{$|\mu|\le 1$}, we can 
evaluate 
$J(n,m)$ using the residue theorem. 
Recall that 
the inverse of the 
matrix 
$A(\mu)$
is given by
 \[
 A(\mu)^{-1} = \frac{1}{\det A(\mu)} \widetilde{A}(\mu),
 \]
where 
the tilde denotes the adjugate 
(i.e., transposed cofactor)
matrix. 
Thus, 
the singularities of the integrand 
in (\ref{eq_ref6}) 
are determined by 
the zeros of $\det A(\mu)$. 
For 
simplicity, 
we assume that 
the matrix function 
$A(\mu)^{-1}$ only has 
isolated simple poles in \mbox{$|\mu| < 1$}, 
denoted as 
\mbox{$\{\mu_1, \mu_2, \ldots, \mu_{N} \}$}, 
and is regular on 
\mbox{$|\mu|=1$}.\footnote{We
should 
not 
assume 
such 
a strong condition 
as 
$\det A(\mu)$ 
has 
only simple zeros, 
which 
was 
assumed 
in 
our previous papers~\cite{TUW98,TUW99}. 
Indeed, a 
zero of multiplicity 
$k$ 
of $\det A(\mu)$ may 
be cancelled by 
a zero of multiplicity 
\mbox{$k-1$}  
of $\widetilde{A}(\mu)$ to give a simple 
pole of $A(\mu)^{-1}$. 
However, 
this correction
does not 
affect the validity of 
the 
formulas 
in~\cite{TUW98,TUW99}.  
} 
In fact, the more general case where
$A(\mu)^{-1}$ 
also 
has 
higher order poles
can be 
recovered 
by taking a suitable 
coalescence limit of 
two or more 
simple poles afterward. 

In the neighborhood 
of \mbox{$\mu = \mu_j$}, we can expand $A(\mu)$ and 
$A(\mu)^{-1}$ as 
(cf.~\cite{Kamijo,Olme85})
\begin{eqnarray}
&& A(\mu) = A(\mu_j) + (\mu-\mu_j) A'(\mu_j) 
 + \mathrm{O} ((\mu-\mu_j)^2 ), 
\hspace{5mm} \det A(\mu_j) =0_{\vphantom \sum}, 
\nonumber \\[1mm]
&& A(\mu)^{-1} = \frac{1}{\mu-\mu_j} A_j^{(-1)} + A_j^{(0)} 
	+ \mathrm{O}( \mu-\mu_j ), \hspace{5mm} A_j^{(-1)} \neq O, 
\label{A-inv}	
\end{eqnarray}
where
\[
A(\mu_j) A_j^{(-1)} 
=O,  \hspace{5mm}
A(\mu_j) A_j^{(0)} + A'(\mu_j) A_j^{(-1)} 
= I.
\]	
Thus, using (\ref{psi_bar}), (\ref{psi_ad}) and 
(\ref{rep1}), we obtain 
\begin{align}
%
z^{-n} \psi_n^{\mathrm{ad}} 
\left[
\begin{array}{cc}
\! z^{n} \phi_n A_j^{(-1)}  \! & \!  
z^{n} \psi_n \!
\end{array}
\right]
&=
\left[
\begin{array}{cc}
\! A(\mu) A_j^{(-1)} \! & \! O \! 
\end{array}
\right]
\nonumber \\[1mm]
&=
\left[
\begin{array}{cc}
\! O \! & \! O \!
\end{array}
\right]
\;\; {\rm at} \;\; \mu = \mu_j. 
\nonumber
\end{align}
%
Because 
$z^{-n}\psi_n^{\mathrm{ad}}$ 
consisting of $l$ rows 
satisfies 
the boundary condition in (\ref{psi_ad}) 
and
the eigenvalue problem (\ref{mAL6}),   
the rank of $[z^{-n} \psi_n^{\mathrm{ad}}](\mu_j)$ 
is equal to $l$ for all \mbox{$n \in {\mathbb Z}$}. 
Similarly, the rank of $[z^{n} \psi_n ](\mu_j)$ is 
$l$ for all \mbox{$n \in {\mathbb Z}$}. Therefore, 
there 
exists an 
\mbox{$\l \times l$} matrix $C_j$ such that 
%
%
\begin{eqnarray}
[z^{n} \phi_n ] (\mu_j) A_j^{(-1)} 
	= [z^{-n} \psi_n ](\mu_j) C_j \mu_j^{n}.
\label{connect3}
\end{eqnarray}
Here, 
$C_j$ must be 
$n$-independent, 
because 
both 
$z^{n} \phi_n$ 
and $z^{n} \psi_n$
satisfy 
the same eigenvalue problem 
(\ref{mAL4}). 
The matrix $C_j$ can 
be 
intuitively 
considered as 
\mbox{$B(\mu_j) \lim_{\mu \to \mu_j} (\mu -\mu_j) A(\mu)^{-1}$}, 
but it is, in general, 
different from 
the naive residue
$\lim_{\mu \to \mu_j} (\mu -\mu_j) B(\mu) A(\mu)^{-1}$.
Indeed, 
$B(\mu)$ 
can have a {\it discontinuity} 
at \mbox{$\mu = \mu_j$}. 
Because 
$\phi_n A_j^{(-1)}$ 
and $\psi_n C_j$ 
vanish exponentially 
for \mbox{$n \to -\infty$} and 
\mbox{$n \to +\infty$}
respectively, 
each nonzero column vector of 
the \mbox{$2l \times l$} matrix 
\mbox{$\phi_n A_j^{(-1)}=\psi_n C_j$} 
at \mbox{$z^2 = \mu_j$} 
gives 
a bound state in the 
potentials $Q_n$ and $R_n$. 

Therefore, 
using the residue theorem 
with the aid of 
(\ref{A-inv}) and (\ref{connect3}), 
we can 
compute the right-hand side of 
(\ref{eq_ref6})
as
\begin{align}
J(n,m) &= \sum_{j=1}^{N} 
	[z^{-n} \psi_n ](\mu_j) C_j \mu_j^{m} 
\nonumber \\[1mm]
&= \sum_{j=1}^{N} 
\left\{ 
\left[
\begin{array}{c}
 O \\
 I \\
\end{array}
\right] +
\sum_{k=0}^\infty \mu_j^{k+1} K(n,n+k) 
\right\} 
C_j \mu_j^{m}
\nonumber \\[1mm]
&= - \left[
\begin{array}{c}
 O \\
 F_{\mathrm D} (m) \\
\end{array}
\right] -
\sum_{k=0}^\infty K(n,n+k)F_{\mathrm D} (m+k+1), 
\nonumber
\end{align}
%
where 
\[
F_{\mathrm D} (m) := -\sum_{j=1}^{N} C_j \mu_j^{m}. 
\]
Substituting this expression for $J(n,m)$
into (\ref{GLM-pre1}), 
we obtain 
a linear summation equation of the 
Gel'fand--Levitan--Marchenko type,
 \begin{equation}
 \widebar{K}(n,m) +  
\left[
\begin{array}{c}
 O \\
 F(m) \\
\end{array}
\right]+
\sum_{k=0}^{\infty} K(n,n+k) F(m+k+1) = 
\left[
\begin{array}{c}
 O \\
 O \\
\end{array}
\right], \hspace{5mm} m \geq n.
 \label{GLM_1}
 \end{equation}
Here, $F(m)$ is defined as
\begin{align}
 F(m) &:= F_{\mathrm C} (m) + F_{\mathrm D} (m) \nonumber \\
& \hphantom{:} 
 = \frac{1}{2\pi \mathrm{i}} \oint_{C} B(\mu)A(\mu)^{-1} 
	\mu^{m} \mathrm{d} \mu - \sum_{j=1}^{N} C_j \mu_j^{m}.
\label{F_form}
\end{align}
Note that 
$F_{\mathrm C}$ and $F_{\mathrm D}$ 
correspond to the 
contributions 
of the continuous and discrete
spectra, 
respectively.\footnote{In this section, we often follow 
the notation of Ablowitz
{\it et al.}~\cite{AL76,Ab78,
AS81}.} 
\\
\\
{\it Remark.}
Using the expressions 
(\ref{phi-ex}) and (\ref{A-ex}), we can 
evaluate 
$[z^{n} \phi_n](\mu)$ 
and $A(\mu)$ 
in the 
limit 
\mbox{$\mu \to 0$} 
as (cf.~\cite{AL75,Ab78,APT})
\begin{align}
& \lim_{\mu \to 0} [z^{n} \phi_n](\mu) = 
\left[
\begin{array}{c}
 I \\
 -R_{n-1} \\
\end{array}
\right] 
\dprod{\displaystyle \curvearrowleft}{n-1}_{i=-\infty}
 \left( I-Q_i R_i \right)^{-1}, 
\nonumber \\[1mm] & 
\lim_{\mu \to 0} A(\mu) = \dprod{\displaystyle 
\curvearrowleft}{\infty}_{i=-\infty}
 \left( I-Q_i R_i \right)^{-1}.
\nonumber
\end{align}
Because \mbox{$\lim_{\mu \to 0} A(\mu)$} is 
invertible 
(cf.~(\ref{nonzero-det})), 
we 
have 
\mbox{$\mu_j \neq 0$},  
\mbox{$j=1, 2, \ldots, N$}. 
Note that 
instead of (\ref{oper1}), 
we can 
operate 
with 
$ \frac{1}{2\pi \mathrm{i}} \oint_{C} \mathrm{d} \mu \, \mu^{-1}$ 
on (\ref{prep1}) 
with (\ref{psi_form}).
Thus, 
we 
can express 
\mbox{$\lim_{\mu \to 0} [z^{n} \phi_n](\mu) A(\mu)^{-1}$} as
\[
\lim_{\mu \to 0} [z^{n} \phi_n](\mu) A(\mu)^{-1} 
 = \left[
\begin{array}{c}
 I \\
 F(n-1) \\
\end{array}
\right] 
+\sum_{k=0}^\infty 
\left[
\begin{array}{c}
  K_1(n,n+k)  \\
  K_2(n,n+k)  \\
 \end{array}
 \right] F(n+k).
\]
From 
the above 
three 
relations, 
we 
obtain 
%
\[
\left[
\begin{array}{c}
 I \\
 -R_{n-1} \\
\end{array}
\right] 
\dprod{\displaystyle \curvearrowright}{\infty}_{i=n}
 \left( I-Q_i R_i \right) 
 = \left[
\begin{array}{c}
 I \\
 F(n-1) \\
\end{array}
\right] 
+\sum_{k=0}^\infty 
\left[
\begin{array}{c}
  K_1(n,n+k)  \\
  K_2(n,n+k)  \\
 \end{array}
 \right] F(n+k).
\]
The 
nonlocal quantities on the left-hand side 
can be used 
to transform 
the matrix 
Ablowitz--Ladik lattice (\ref{mALsys}) 
to 
other 
systems~\cite{GI82}. 
\\
%

Next, we 
substitute the 
summation representations 
(\ref{psi_form}) into the right-hand side of 
(\ref{prep2}) and 
operate with 
 \begin{equation}
 \frac{1}{2\pi \mathrm{i}}\oint_{C} \mathrm{d} \mu \, \mu^{n-m-2} \hspace{5mm} (m \ge n)
\label{oper2} 
\end{equation}
on both sides. 
Thus, 
we obtain
\begin{equation}
\widebar{J} (n,m) = - K(n,m) 
+\left[
\begin{array}{c}
 \widebar{F}_{\mathrm C} (m)\\
 O \\
\end{array}
\right]+ \sum_{k=0}^\infty \widebar{K}(n,n+k) 
\widebar{F}_{\mathrm C} (m+k+1),
\label{GLM-pre2}
\end{equation}
where 
\begin{align}
\widebar{J}(n,m) &:= \frac{1}{2\pi \mathrm{i}} \oint_C [z^{-n} \widebar{\phi}_n ](\mu) 
	\widebar{A}(\mu)^{-1} \mu^{n-m-2} \mathrm{d} \mu,
\label{J_bar} 
\\[1mm]
\widebar{F}_{\mathrm C} (m) &:= \frac{1}{2 \pi \mathrm{i}} \oint_C 
         \widebar{B}(\mu)\widebar{A}(\mu)^{-1} \mu^{-m-2} 
	\mathrm{d} \mu.
\nonumber
\end{align}
%
Because of the analyticity of 
$[z^{-n} \widebar{\phi}_n](\mu)$ 
and $\widebar{A}(\mu)$ 
in \mbox{$|\mu| \ge 1$}, 
we can evaluate \mbox{$\widebar{J}(n,m)$} using 
the residue theorem. 
We assume that $\widebar{A}(\mu)^{-1}$ only 
has 
isolated simple poles in \mbox{$|\mu| > 1$},
denoted as $\{\widebar{\mu}_1, \widebar{\mu}_2, \ldots, 
\widebar{\mu}_{\widebar{N}} \}$, 
and is regular on \mbox{$|\mu| = 1$}.

In the neighborhood 
of $\mu = \widebar{\mu}_j$, we expand $\widebar{A}(\mu)$ and 
$\widebar{A}(\mu)^{-1}$ as 
\begin{eqnarray}
&& \widebar{A}(\mu) = \widebar{A}(\widebar{\mu}_j) 
 + (\mu-\widebar{\mu}_j) \widebar{A}\hspace{1pt}'(\widebar{\mu}_j) 
	+ \mathrm{O}( (\mu-\widebar{\mu}_j)^2 ), 
\hspace{5mm} \det \widebar{A}(\widebar{\mu}_j) =0_{\vphantom \sum}, 
\nonumber \\[1mm]
&& \widebar{A} (\mu)^{-1} = \frac{1}{\mu-\widebar{\mu}_j} 
\widebar{A}_j^{\hspace{1pt}(-1)} + \widebar{A}_j^{\hspace{1pt}(0)} 
+ \mathrm{O}( \mu-\widebar{\mu}_j ), 
\hspace{5mm} \widebar{A}_j^{\hspace{1pt}(-1)} \neq O, 
\label{A_bar-inv}
\end{eqnarray}
where
\[
\widebar{A}(\widebar{\mu}_j) 
 \widebar{A}_j^{\hspace{1pt}(-1)} 
=O,  \hspace{5mm}
\widebar{A}(\widebar{\mu}_j) \widebar{A}_j^{\hspace{1pt}(0)} 
+ \widebar{A}\hspace{1pt}'(\widebar{\mu}_j) 
 \widebar{A}_j^{\hspace{1pt}(-1)} 
= I.
\]	
%
Thus, 
using 
(\ref{psi_bar}), (\ref{psi_ad}) and 
(\ref{rep2}), 
we obtain 
\begin{align}
%
z^{n} \widebar{\psi}_n^{\mathrm{ad}} 
\left[
\begin{array}{cc}
\! z^{-n} \widebar{\phi}_n \widebar{A}_j^{\hspace{1pt}(-1)}  \! & \! 
z^{-n} \widebar{\psi}_n \!
\end{array}
\right]
&=
\left[
\begin{array}{cc}
\! -\widebar{A} (\mu) \widebar{A}_j^{\hspace{1pt}(-1)} \! & \! O \!
\end{array}
\right]
\nonumber \\[1mm]
 &= \left[
\begin{array}{cc}
\! O \! & \! O \!
\end{array}
\right] \;\; {\rm at} \;\; \mu = \widebar{\mu}_j. 
\nonumber
\end{align}
In the same way as 
the derivation of (\ref{connect3}), 
there 
exists an $n$-independent \mbox{$l \times l$} 
matrix $\widebar{C}_j$ such that 
\begin{eqnarray}
[z^{-n} \widebar{\phi}_n ] (\widebar{\mu}_j) 
 \widebar{A}_j^{\hspace{1pt}(-1)} 
  = [z^{n} \widebar{\psi}_n ](\widebar{\mu}_j) 
 \widebar{C}_j \widebar{\mu}_j^{\hspace{1pt}-n}.
\label{connect4}
\end{eqnarray}

Therefore, 
using the residue theorem with the aid of 
(\ref{A_bar-inv}) and (\ref{connect4}), 
we can compute 
the right-hand side of 
(\ref{J_bar}) 
as 
\begin{align}
\widebar{J} (n,m)
&= - 
\sum_{j=1}^{\widebar{N}} [z^{n} \widebar{\psi}_n ](\widebar{\mu}_j) \widebar{C}_j
\widebar{\mu}_j^{\hspace{1pt}-m-2} 
\nonumber \\[1mm]
&= -
\sum_{j=1}^{\widebar{N}} 
\left\{
\left[
\begin{array}{c}
 I \\
 O \\
\end{array}
\right] +
\sum_{k=0}^\infty \widebar{\mu}_j^{\hspace{1pt}-k-1} 
 \widebar{K}(n,n+k)
\right\}
\widebar{C}_j \widebar{\mu}_j^{\hspace{1pt}-m-2}
\nonumber 
\\[1mm]
&= -
\left[
\begin{array}{c}
 \widebar{F}_{\mathrm D} (m)  \\
 O \\
\end{array}
\right] -
\sum_{k=0}^\infty 
 \widebar{K}(n,n+k) \widebar{F}_{\mathrm D} (m+k+1),
\nonumber
\end{align}
%
where 
\[
\widebar{F}_{\mathrm D} (m) := \sum_{j=1}^{\widebar{N}} \widebar{C}_j \widebar{\mu}_j^{\hspace{1pt}-m-2}.
\]
Substituting this expression for \mbox{$\widebar{J} (n,m)$} 
into (\ref{GLM-pre2}), 
we obtain another linear summation equation of 
the Gel'fand--Levitan--Marchenko 
type,
 \begin{equation}
 -K(n,m) +
\left[
\begin{array}{c}
 \widebar{F}(m) \\
 O \\
\end{array}
\right] + \sum_{k=0}^{\infty} \widebar{K}(n,n+k) 
 \widebar{F}(m+k+1) = 
\left[
\begin{array}{c}
 O \\
 O \\
\end{array}
\right], 
\hspace{5mm} m \geq n.
 \label{GLM_2}
 \end{equation}
Here, $\widebar{F}(m)$ is defined as
 \begin{align}
 \widebar{F}(m) &:= \widebar{F}_{\mathrm C} (m) 
 + \widebar{F}_{\mathrm D} (m) \nonumber \\
& \hphantom{:} = 
\frac{1}{2\pi \mathrm{i}} \oint_{C} 
\widebar{B}(\mu)\widebar{A}(\mu)^{-1} \mu^{-m-2}\mathrm{d} \mu
 + \sum_{j=1}^{\widebar{N}} \widebar{C}_j 
 \widebar{\mu}_j^{\hspace{1pt}-m-2}.
\label{F_bar_form}
\end{align}
\\
{\it Remark.}
%
Using the expressions 
(\ref{phi_bar-ex}) and (\ref{Abar-ex}), we can 
evaluate 
$[z^{-n} \widebar{\phi}_n ](\mu)$ 
and $\widebar{A}(\mu)$ 
in the 
limit 
\mbox{$|\mu| \to \infty$} 
as (cf.~\cite{APT})
\begin{align}
& \lim_{|\mu| \to \infty}
[z^{-n} \widebar{\phi}_n ](\mu) = 
\left[
\begin{array}{c}
 Q_{n-1} \\
 -I \\
\end{array}
\right] 
\dprod{\displaystyle \curvearrowleft}{n-1}_{i=-\infty}
 \left( I-R_i Q_i \right)^{-1}, 
\nonumber \\[1mm] & 
\lim_{|\mu| \to \infty} 
	 \widebar{A}(\mu) = \dprod{\displaystyle 
\curvearrowleft}{\infty}_{i=-\infty}
 \left( I-R_i Q_i \right)^{-1}.
\nonumber
\end{align}
Because 
\mbox{$\lim_{|\mu| \to \infty} \widebar{A}(\mu)$} is 
invertible 
(cf.~(\ref{nonzero-det})), 
$\widebar{A}(\mu)^{-1}$ does not 
have
poles at infinity, 
which 
is indeed 
consistent with 
the 
computation 
for $\widebar{J} (n,m)$. 
Note that 
instead of (\ref{oper2}), 
we can 
operate 
with 
\mbox{$ \frac{1}{2\pi \mathrm{i}} \oint_{C} \mathrm{d} 
\mu \, \mu^{-1}$} 
on (\ref{prep2}) 
with (\ref{psi_form}).
Thus, 
we 
can express  
\mbox{$\lim_{|\mu| \to \infty}
[z^{-n} \widebar{\phi}_n ](\mu) 
	 \widebar{A}(\mu)^{-1}$} as
\begin{align}
\lim_{|\mu| \to \infty}
[z^{-n} \widebar{\phi}_n ](\mu) 
	 \widebar{A}(\mu)^{-1}
 = -\left[
\begin{array}{c}
 -\widebar{F}(n-1) \\
 I \\
\end{array}
\right] 
 +\sum_{k=0}^\infty 
\left[
\begin{array}{c}
  \widebar{K}_1(n,n+k)  \\
  \widebar{K}_2(n,n+k)  \\
 \end{array}
 \right] \widebar{F}(n+k).
\nonumber
\end{align}
From 
the above 
three 
relations, 
we obtain  
\begin{align}
\left[
\begin{array}{c}
 -Q_{n-1} \\
 I \\
\end{array}
\right] 
\dprod{\displaystyle \curvearrowright}{\infty}_{i=n}
 \left( I-R_i Q_i \right) 
 = \left[
\begin{array}{c}
 -\widebar{F} (n-1) \\
 I \\
\end{array}
\right] 
-\sum_{k=0}^\infty 
\left[
\begin{array}{c}
  \widebar{K}_1(n,n+k)  \\
  \widebar{K}_2(n,n+k)  \\
 \end{array}
 \right] \widebar{F}(n+k).
\nonumber
\end{align}
The 
nonlocal quantities on the left-hand side can be used 
to transform
the matrix 
Ablowitz--Ladik lattice (\ref{mALsys}) 
to other 
systems~\cite{GI82}.
\\

The Gel'fand--Levitan--Marchenko equations 
(\ref{GLM_1}) and (\ref{GLM_2}) 
relate 
the scattering data to the potentials $Q_n$ and $R_n$ 
through 
(\ref{K2}) and (\ref{K_bar2}). 
The required 
set of the scattering data is given by 
$B(\mu)A(\mu)^{-1}$
and 
$\widebar{B}(\mu)\widebar{A}(\mu)^{-1}$ for \mbox{$|\mu|=1$},
\mbox{$\{ \mu_j, C_j \}_{j=1, 2, \ldots, N}$} 
and \mbox{$\{\widebar{\mu}_j, 
	\widebar{C}_j \}_{j=1, 2, \ldots, \widebar{N}}$}, 
	which define 
$F(m)$ and $\widebar{F}(m)$ as in 
(\ref{F_form}) and (\ref{F_bar_form}). 
Because 
$F(m)$ and $\widebar{F}(m)$ 
are 
``linear" in the scattering data, 
we can consider 
a linear superposition 
of different sets of scattering data 
at this level. 
In fact, we will 
show in the next subsection that 
$F(m)$ and $\widebar{F}(m)$ 
satisfy 
linear 
evolution equations. 

\subsection{Time evolution}
\label{sTDsd}

Under the rapidly decaying boundary 
conditions (\ref{zero-bc}), 
the temporal part of the Lax representation 
(\ref{mAL-time}) 
for the Ablowitz--Ladik lattice (\ref{mALsys}) 
has the asymptotic 
behavior: 
\begin{equation}
\left[
\begin{array}{c}
 \Psi_{1, n} \\
 \Psi_{2, n} \\
\end{array}
\right]_t 
\sim 
 \left[
 \begin{array}{cc}
  (-\mu+1) b I  &  O \\
   O  &  (1-\mu^{-1}) a I \\
 \end{array}
 \right]
\left[
\begin{array}{c}
 \Psi_{1, n} \\
 \Psi_{2, n} \\
\end{array}
\right] 
\hspace{4mm}{\rm as}~~~ n \rightarrow \pm \infty. 
\nonumber
\end{equation}
%
This 
can be used to fix
the time dependence of 
the leading order 
terms 
in $\Psi_{1, n}$ and $\Psi_{2, n}$ 
(cf.~(\ref{leftJost})). 
%
Thus, 
we can 
introduce 
the explicitly 
time-dependent Jost solutions 
$\phi_n^{(t)}$ and $\widebar{\phi}_n^{\hspace{1pt}(t)}$ as 
\begin{subequations}
\label{Jost-time}
\begin{eqnarray}
\left.
\begin{array}{l}
 z^{n} \phi_n^{(t)} := \mathrm{e}^{(-\mu+1) b t} z^{n} \phi_n
 \to  \mathrm{e}^{(-\mu+1) b t}
 \left[
 \begin{array}{c}
  I \\
  O \\
 \end{array}
 \right]
\vspace{2mm}
\\
%
 z^{-n} \widebar{\phi}_n^{\hspace{1pt}(t)} 
	:=  \mathrm{e}^{(1-\mu^{-1}) a t} 
 z^{-n} \widebar{\phi}_n
 \to 
 \mathrm{e}^{(1-\mu^{-1})a t}
 \left[
 \begin{array}{c}
  O \\
  -I \\
 \end{array}
 \right]
\end{array}
\right\}
 \hspace{4mm} {\rm as}~~n \rightarrow -\infty \hspace{8mm} 
\label{Jost-time-phi}
 \end{eqnarray}
and 
$\psi_n^{(t)}$ and $\widebar{\psi}_n^{\hspace{1pt}(t)}$ as 
\begin{eqnarray}
\left.
\begin{array}{l}
 z^{-n} \psi_n^{(t)} :=  \mathrm{e}^{(1- \mu^{-1}) a t} 
 z^{-n} \psi_n
 \to 
 \mathrm{e}^{(1- \mu^{-1})a t}
 \left[
 \begin{array}{c}
  O \\
  I \\
 \end{array}
 \right]
\vspace{2mm}
\\
%
 z^{n} \widebar{\psi}_n^{\hspace{1pt}(t)} := \mathrm{e}^{(-\mu+1) b t} z^{n} \widebar{\psi}_n
 \to  \mathrm{e}^{(-\mu+1) b t}
 \left[
 \begin{array}{c}
  I \\
  O \\
 \end{array}
 \right]
\label{Jost-time-psi}
\end{array}
\right\}
 \hspace{4mm} {\rm as}~~n \rightarrow +\infty, \hspace{8mm} 
 \end{eqnarray}
\end{subequations}
respectively; they
satisfy both the eigenvalue problem 
(\ref{mAL1}) 
and 
the time-evolution equation (\ref{mAL-time}). 

To determine 
the time dependence of the scattering data, 
we 
rewrite
the 
defining relations (\ref{ref102})
as 
\begin{align}
& \phi_n^{\mathrm (t)} = \widebar{\psi}_n^{\hspace{1pt}\mathrm (t)}A
 + \psi_n^{\mathrm (t)}B
\mathrm{e}^{-[(\mu-1) b + ( 1-\mu^{-1}) a ] t},
\nonumber 
\\[1mm]
& \widebar{\phi}_n^{\hspace{1pt}\mathrm (t)} = 
 \widebar{\psi}_n^{\hspace{1pt}\mathrm (t)} \widebar{B}
\mathrm{e}^{[(\mu-1) b + ( 1-\mu^{-1}) a ] t}
- \psi_n^{\mathrm (t)}\widebar{A}.
\nonumber
\end{align}
Note that 
$\phi_n^{(t)}$, 
$\widebar{\phi}_n^{\hspace{1pt}(t)}$, 
$\widebar{\psi}_n^{\hspace{1pt}(t)}$
and ${\psi}_n^{(t)}$ satisfy 
the same 
equation (\ref{mAL-time})
and the columns of 
$\widebar{\psi}_n^{\hspace{1pt}(t)}$ and ${\psi}_n^{(t)}$ are linearly independent. 
Thus, 
the time 
dependences 
of $A$, $B$ and 
$\widebar{A}$, $\widebar{B}$
for \mbox{$|\mu|=1$} 
are given by 
\begin{align}
A(\mu,t) &= A(\mu,0), \hspace{5mm}
%
 B(\mu,t) 
=B(\mu, 0)
\mathrm{e}^{[(\mu-1) b + ( 1-\mu^{-1}) a ] t}
\label{BA_time}
\end{align}
 %
and
\begin{align}
\widebar{A}(\mu,t) &= \widebar{A}(\mu,0),\hspace{5mm}
\widebar{B}(\mu,t) 
= \widebar{B}(\mu, 0) 
 \mathrm{e}^{-[(\mu-1) b + (1 - \mu^{-1})a] t},
\label{BA_bar_time}
\end{align}
respectively. 
This implies that 
$A(\mu,t)$, $\widebar{A}(\mu,t)$,
$B(\mu, t)\mathrm{e}^{-[(\mu-1) b + ( 1-\mu^{-1}) a ] t}$
and 
$\widebar{B}(\mu, t) \mathrm{e}^{[(\mu-1) b + (1 - \mu^{-1})a] t}$ 
are generating functions 
of the 
integrals of motion. 

Because 
the ``analytic continuation" of 
$A(\mu)$ and 
$\widebar{A}(\mu)$ 
into the regions 
\mbox{$|\mu| \le 1$} and \mbox{$|\mu| \ge 1$}, 
respectively, is unique and 
remains 
time-independent 
(cf.~(\ref{AAbar-ex})), 
the positions of the simple 
poles 
of $A(\mu)^{-1}$ and $\widebar{A}(\mu)^{-1}$
and 
the corresponding residues, 
\mbox{$
\bigl\{ \mu_j, A_j^{(-1)} \bigr\}_{j=1, 2, \ldots, N }$} 
and \mbox{$\bigl\{ \widebar{\mu}_j, 
\widebar{A}_j^{\hspace{1pt}(-1)}  
\bigr\}_{j=1, 2, \ldots, \widebar{N}}$},  
are also time-independent. 
For (\ref{connect3}) and  (\ref{connect4}), 
we can apply 
a similar 
discussion 
as used to obtain 
(\ref{BA_time}) and (\ref{BA_bar_time}), 
so 
the time dependences of 
$C_j$ and 
$\widebar{C}_j$ are 
given by 
\begin{align}
 C_j(t) &= C_j(0)\mathrm{e}^{[(\mu_j-1) b + (1-\mu_j^{-1})a] t}
 \label{C_time}
\end{align}
 %
and
\begin{align}
 \widebar{C}_j(t) &= \widebar{C}_j(0)
\mathrm{e}^{-[(\widebar{\mu}_j-1) b 
 + (1-\widebar{\mu}_j^{\hspace{1pt}-1}) a] t},
\label{C_bar_time}
\end{align}
respectively. 

Substituting (\ref{BA_time})--(\ref{C_bar_time}) 
into (\ref{F_form}) and (\ref{F_bar_form}), 
we obtain 
the 
explicitly 
time-dependent forms 
of $F(n)$ and $\widebar{F}(n)$ 
as
 %
\begin{align}
F(n, t) &= \frac{1}{2\pi \mathrm{i}} \oint_{C}
 B(\mu,0)A(\mu,0)^{-1} \mu^{n} 
 \mathrm{e}^{[(\mu-1) b + (1-\mu^{-1})a]t}\mathrm{d} \mu
 \nonumber \\
 & \hphantom{=} \; \mbox{}
     - \sum_{j=1}^N C_j(0) \mu_j^{n}
 \mathrm{e}^{[(\mu_j-1) b + (1-\mu_j^{-1})a] t},
\label{F_time1}
\\[2mm]
 \widebar{F}(n,t) &= \frac{1}{2\pi \mathrm{i}} \oint_{C}
 \widebar{B}(\mu,0)\widebar{A}(\mu,0)^{-1} \mu^{-n-2} 
  \mathrm{e}^{-[(\mu-1) b + (1-\mu^{-1})a]t}
        \mathrm{d} \mu 
\nonumber \\ & 
\hphantom{=} \; \mbox{}
+ \sum_{j=1}^{\widebar{N}} \widebar{C}_j(0) 
	\widebar{\mu}_j^{\hspace{1pt}-n-2} 
       \mathrm{e}^{-[(\widebar{\mu}_j-1) b 
	+ (1-\widebar{\mu}_j^{\hspace{1pt}-1})a] t}.
\label{Fbar_time1}
\end{align}
Thus, 
it is easy to see that $F(n,t)$ and $\widebar{F}(n,t)$ 
satisfy the 
pair of 
uncoupled 
linear evolution equations:
\begin{subnumcases}{\label{AL-linear}}
{}
\frac{\partial F(n,t)}{\partial t} - b F (n+1,t) + a F(n-1,t) + (b-a) F(n,t)=O, \hspace{10mm}
\\[1mm]
\frac{\partial \widebar{F}(n,t)}{\partial t} -a \widebar{F} (n+1,t) + b \widebar{F}(n-1,t) 
	+ (a-b) \widebar{F}(n,t)=O. \hspace{10mm}
\end{subnumcases}
Note that 
these equations 
coincide with 
the linear 
part of the 
equations 
for 
$R_n$ and $Q_n$ 
(see (\ref{mALsys})). 
In addition, $F(n,t)$ and $\widebar{F}(n,t)$ 
are required to
decay 
rapidly 
as \mbox{$n \to +\infty$}
so that 
the Gel'fand--Levitan--Marchenko
equations 
(\ref{GLM_1}) and (\ref{GLM_2}) 
are 
well-posed. 

Because of the linear nature of the 
sum 
terms in 
(\ref{F_time1}) and (\ref{Fbar_time1}), 
we can 
take a coalescence limit of two or more 
simple poles of 
$A(\mu)^{-1}$ and $\widebar{A}(\mu)^{-1}$ 
directly. 
Thus, we obtain the following 
generalized expressions for 
$F(n,t)$ and $\widebar{F}(n, t)$:
\begin{align}
F(n, t) &= \frac{1}{2\pi \mathrm{i}} \oint_{C}
 B(\mu,0)A(\mu,0)^{-1} \mu^{n} 
 \mathrm{e}^{[(\mu-1) b + (1-\mu^{-1})a]t}\mathrm{d} \mu
 \nonumber \\
 & \hphantom{=} \; \mbox{} 
- \sum_{j=1}^N \sum_{k=0}^{M_j} C_j^{(k)}(0) \left( 
\frac{\partial}{\partial \mu_j} \right)^k
\mu_j^{n}
 \mathrm{e}^{[(\mu_j-1) b + (1-\mu_j^{-1})a] t},
\nonumber
\\[3mm]
 \widebar{F}(n,t) &= \frac{1}{2\pi \mathrm{i}} \oint_{C}
 \widebar{B}(\mu,0)\widebar{A}(\mu,0)^{-1} \mu^{-n-2} 
  \mathrm{e}^{-[(\mu-1) b + (1-\mu^{-1})a]t}
        \mathrm{d} \mu 
\nonumber \\ & 
\hphantom{=} \; \mbox{}
+ \sum_{j=1}^{\widebar{N}} \sum_{k=0}^{L_j} 
\widebar{C}_j^{(k)}(0)  
\left( \frac{\partial}{\partial \widebar{\mu}_j} \right)^k
	\widebar{\mu}_j^{\hspace{1pt}-n-2} 
       \mathrm{e}^{-[(\widebar{\mu}_j-1) b 
	+ (1-\widebar{\mu}_j^{\hspace{1pt}-1})a] t}.
\nonumber 
\end{align}
These expressions 
encompass the most 
general 
case 
where  
$A(\mu)^{-1}$ and $\widebar{A}(\mu)^{-1}$ 
have 
arbitrarily 
higher 
order poles. 
Moreover, 
they 
satisfy the same linear 
equations 
(\ref{AL-linear}) as 
the original expressions 
(\ref{F_time1}) and (\ref{Fbar_time1}). 

Instead of 
using
the partial 
differentiation 
with respect to 
$\mu_j$ and
$\widebar{\mu}_j$, 
one can 
consider 
the 
matrix 
functions
\mbox{$X^{n} \mathrm{e}^{bt(X-I)  + at(I-X^{-1})} $} 
and 
\mbox{$Y^{-n-2}\mathrm{e}^{-bt(Y-I)  - at(I-Y^{-1})} $}
with 
constant invertible 
matrices $X$ and $Y$ in Jordan normal 
form.  
Indeed, they satisfy the linear 
equations 
of the form 
(\ref{AL-linear}), 
so 
linear combinations of 
the independent 
elements 
of 
each matrix function can be used to 
replace 
the 
sum terms in $F(n,t)$ and $\widebar{F}(n, t)$. 
Readers interested in 
such an approach 
are referred, 
{\it e.g.}, 
to~\cite{Scieb00,Scieb04,Akto07,Akto10,DM2010,DM-SIGMA,DKM}.
%

\subsection{Exact linearization}


To reconstruct the potentials $Q_n$ and $R_n$ 
from 
the scattering data through 
(\ref{K2}) and (\ref{K_bar2}), 
we 
rewrite 
the Gel'fand--Levitan--Marchenko
equations 
(\ref{GLM_1}) and (\ref{GLM_2}) 
as 
``closed" 
linear summation equations for 
$K_1(n,m)$ and $\widebar{K}_2 (n,m)$. 
Thus, 
the 
general solution formulas 
for the matrix 
Ablowitz--Ladik lattice (\ref{mALsys}) 
can be presented 
in the form: 
%
\begin{subequations}
\label{ALlinearization}
\begin{align}
& Q_n = K_1(n,n),
\label{ALlin-1}
\\[1mm]
& R_n = \widebar{K}_2 (n,n),
\label{ALlin-2}
\\
& K_1(n,m) = \widebar{F}(m) -\sum_{i=0}^{\infty} 
\sum_{k=0}^{\infty} K_1(n,n+i) F(n+i+k+1) \widebar{F}(m+k+1),
\hspace{5mm} 
m \geq n, 
\label{K1-closed}
\\
& \widebar{K}_2 (n,m) = -F(m) -\sum_{i=0}^{\infty} 
\sum_{k=0}^{\infty} \widebar{K}_2 (n,n+i) 
 \widebar{F}(n+i+k+1) F(m+k+1), 
\hspace{5mm} 
m \geq n. 
\label{K2-closed}
\end{align}
\end{subequations}
Here, 
the time dependence of the functions 
is suppressed; 
$F(n)$ and $\widebar{F}(n)$ 
are solutions of 
the linear uncoupled system (\ref{AL-linear}) 
and 
decay rapidly 
as \mbox{$n \to +\infty$}. 
More generally, the set of formulas (\ref{ALlinearization}) 
can provide the 
solutions for any flow of 
the matrix 
Ablowitz--Ladik hierarchy
if 
$F(n)$ and $\widebar{F}(n)$ 
satisfy 
the 
linear part of the equations 
for $R_n$ and $Q_n$, 
instead of (\ref{AL-linear}). 
Hence, 
(\ref{ALlinearization}) 
realizes 
an {\it exact linearization} of 
the matrix 
Ablowitz--Ladik hierarchy 
in the sense 
of~\cite{ARS} (also see~\cite{GGKM74,Wadati76}
and Proposition~\ref{Prop.A.2}). 
As long as 
$F(n)$ and $\widebar{F}(n)$ 
decay rapidly 
as \mbox{$n \to +\infty$}, 
so do 
$Q_n$ and $R_n$ determined by (\ref{ALlinearization}). 
However, 
the requirement that $Q_n$ and $R_n$ should also 
decay 
as \mbox{$n \to -\infty$} imposes 
nontrivial 
conditions 
on $F(n)$ and $\widebar{F}(n)$,  
which will be 
touched upon 
in the 
next subsection.

\subsection{Multisoliton solutions}
\label{subsec3.6}

To construct 
exact solutions  
in explicit form, 
we consider the 
special 
case of
\mbox{$B(\mu)=\widebar{B}(\mu)=O$} on 
\mbox{$|\mu|=1$};
this is preserved under the time evolution 
(cf.~(\ref{BA_time}) and (\ref{BA_bar_time}))
and corresponds to the reflectionless potentials (cf.~(\ref{ref102})). 
Moreover, 
we
assume 
that $A(\mu)^{-1}$ and $\widebar{A}(\mu)^{-1}$ 
only have simple poles
(see~\cite{DM2010} for 
the more general case). 
Thus, 
we can set 
\begin{align}
F(n, t) = - \sum_{j=1}^N C_j(t) \mu_j^{n}, 
\hspace{7mm}
 \widebar{F}(n,t) = \sum_{j=1}^{\widebar{N}} \widebar{C}_j(t) 
	\widebar{\mu}_j^{\hspace{1pt}-n-2}, 
\label{F-reflec}
\end{align}
where 
the time dependences of $C_j$ and $\widebar{C}_j$ 
are 
given 
by 
(\ref{C_time}) and (\ref{C_bar_time}). 
We also 
set 
\begin{equation}
K_1(n, m; t) = \sum_{j=1}^{\widebar{N}} G_j(n,t) 
	\widebar{\mu}_j^{\hspace{1pt}-m-2}, 
\hspace{7mm}
 \widebar{K}_2 (n, m; t) = \sum_{j=1}^{N} H_j(n,t) 
 \mu_j^{m}, 
\label{G-H}
\end{equation}
and substitute 
all 
these 
expressions 
into (\ref{K1-closed}) and (\ref{K2-closed}); 
recalling 
that \mbox{$|\mu_j| < 1$} 
$(j=1, 2, \ldots, N)$ and \mbox{$|\widebar{\mu}_j| >1$}
$(j=1, 2, \ldots, \widebar{N})$, 
we can 
evaluate 
the infinite sum. 
Thus, we obtain 
a linear algebraic system for 
determining 
$G_j$ and that for $H_j$  
as 
\begin{subequations}
\label{GH-sys}
\begin{align}
& \left[
\begin{array}{cccc}
\! G_1 \hspace{1pt} \widebar{\mu}_1^{\hspace{1pt}-n-2} \! 
 & \! G_2 \hspace{1pt} \widebar{\mu}_2^{\hspace{1pt}-n-2} \! & \! 
\cdots \! & \! G_{\widebar{N}} \hspace{1pt} 
\widebar{\mu}_{\widebar{N}}^{\hspace{1pt}-n-2}\! 
\end{array}
\right]
\left[
\begin{array}{ccc}
 U_{11} & \cdots & U_{1\widebar{N}} \\
 \vdots &  \ddots & \vdots \\
 U_{\widebar{N}1} & \cdots & U_{\widebar{N}\widebar{N}} \\
\end{array}
\right]
\nonumber \\[1mm]
&= \left[
\begin{array}{cccc}
\! \widebar{C}_1 \hspace{1pt} \widebar{\mu}_1^{\hspace{1pt}-n-2}
 \! & \! \widebar{C}_2 \hspace{1pt} 
	\widebar{\mu}_2^{\hspace{1pt}-n-2}\! 
	& \! \cdots \! & \! \widebar{C}_{\widebar{N}}\hspace{1pt} 
\widebar{\mu}_{\widebar{N}}^{\hspace{1pt}-n-2} \! 
\end{array}
\right]
\label{GH-sys1}
\end{align}
and 
\begin{align}
&
\left[
\begin{array}{cccc}
\! H_1 \hspace{1pt} \mu_1^{n}
\! & \! H_2 \hspace{1pt} \mu_2^{n} 
\! & \! \cdots \! & \! H_{N} \hspace{1pt} \mu_N^{n} \! 
\end{array}
\right]
\left[
\begin{array}{ccc}
 V_{11} & \cdots & V_{1N} \\
 \vdots &  \ddots & \vdots \\
 V_{N1} & \cdots & V_{NN} \\
\end{array}
\right]
= \left[
\begin{array}{cccc}
\! C_1 \hspace{1pt} \mu_1^{n}
\! & \! C_2 \hspace{1pt} \mu_2^{n} 
\! & \! \cdots \! & \! C_{N} \hspace{1pt} \mu_N^{n} \! 
\end{array}
\right].
\label{GH-sys2}
\end{align}
\end{subequations}
%
Here, all the entries in 
(\ref{GH-sys}) 
are \mbox{$l \times l$} matrices;
the block matrices 
\mbox{$U=(U_{jk})_{1 \le j,k \le \widebar{N}}$} and 
\mbox{$V=(V_{jk})_{1 \le j,k \le N}$} are 
defined as 
\begin{align}
U_{jk} &:= \delta_{jk} I - \sum_{i=1}^{N} 
\frac{\mu_i^{n+1}\widebar{\mu}_k^{\hspace{1pt}-n-3}}{
\displaystyle 
\left( 1-\frac{\mu_i}{\widebar{\mu}_j}\right) 
\left( 1-\frac{\mu_i}{\widebar{\mu}_k} \right)} 
C_i (t)\widebar{C}_k(t) 
\nonumber 
\end{align}
and
\begin{align}
V_{jk} &:= \delta_{jk} I - \sum_{i=1}^{\widebar{N}} 
\frac{\widebar{\mu}_i^{\hspace{1pt} -n-3}  \mu_k^{n+1}}{
\displaystyle
\left( 1-\frac{\mu_j}{\widebar{\mu}_i}\right)
\left( 1- \frac{\mu_k}{\widebar{\mu}_i} \right)} \widebar{C}_i(t) C_k(t), 
 \nonumber
\end{align}
respectively. 
Here, 
$\delta_{jk}$ denotes the Kronecker delta. 
Thus, using (\ref{ALlin-1}),
(\ref{ALlin-2}) and (\ref{G-H}),  
we obtain 
\begin{subequations}
\label{QR-soliton}
\begin{align}
Q_n (t) &= K_1(n,n;t)
\nonumber \\
&= \left[
\begin{array}{ccc}
\! G_1 \hspace{1pt} \widebar{\mu}_1^{\hspace{1pt}-n-2} 
\! & \! 
\cdots \! & \! G_{\widebar{N}} \hspace{1pt} 
\widebar{\mu}_{\widebar{N}}^{\hspace{1pt}-n-2}\! 
\end{array}
\right]
\left[
\begin{array}{c}
 I \\
 \vdots \\
 I \\ 
\end{array}
\right]
\nonumber \\
&= \left[
\begin{array}{ccc}
\! \widebar{C}_1(t) \hspace{1pt}\widebar{\mu}_1^{\hspace{1pt}-n-2}
\! & \! \cdots \! & \! 
\widebar{C}_{\widebar{N}}(t) \hspace{1pt}
\widebar{\mu}_{\widebar{N}}^{\hspace{1pt}-n-2} \! 
\end{array}
\right] 
\left[
\begin{array}{ccc}
 U_{11} & \cdots & U_{1\widebar{N}} \\
 \vdots &  \ddots & \vdots \\
 U_{\widebar{N}1} & \cdots & U_{\widebar{N}\widebar{N}} \\
\end{array}
\right]^{-1} 
\left[
\begin{array}{c}
 I \\
 \vdots \\
 I \\ 
\end{array}
\right], 
\label{Q-Nsol}
\\[1mm]
R_n (t) &= \widebar{K}_2 (n,n;t)
\nonumber \\
&= 
\left[
\begin{array}{ccc}
\! H_1 \hspace{1pt} \mu_1^{n}
\! & \! 
\cdots \! & \! H_{N} \hspace{1pt} \mu_N^{n} \! 
\end{array}
\right]
\left[
\begin{array}{c}
 I \\
 \vdots \\
 I \\ 
\end{array}
\right]
\nonumber \\
&= \left[
\begin{array}{ccc}
\! C_1 (t) \hspace{1pt} \mu_1^{n} \! & \! 
\cdots \! & \! C_{N} (t) \hspace{1pt} \mu_N^{n} \! 
\end{array}
\right]
\left[
\begin{array}{ccc}
 V_{11} & \cdots & V_{1N} \\
 \vdots &  \ddots & \vdots \\
 V_{N1} & \cdots & V_{NN} \\
\end{array}
\right]^{-1}
\left[
\begin{array}{c}
 I \\
 \vdots \\
 I \\ 
\end{array}
\right]. 
\end{align}
\end{subequations}
This provides the 
multisoliton solutions of the 
nonreduced 
matrix 
Ablowitz--Ladik lattice (\ref{mALsys}); 
some additional conditions on 
\mbox{$\{ \mu_j, C_j \}_{j=1, 2, \ldots, N}$} 
and \mbox{$\{\widebar{\mu}_j, 
	\widebar{C}_j \}_{j=1, 2, \ldots, \widebar{N}}$} 
need to be satisfied 
for (\ref{QR-soliton}) to exhibit 
solitonic behavior. 

In the simplest nontrivial case of 
\mbox{$N=\widebar{N}=1$}, we obtain 
the one-soliton solution of 
(\ref{mALsys}) in the form: 
\begin{subequations}
\label{one-soli}
\begin{align}
Q_n (t) &= \widebar{D}(n,t) 
\left[ I - \frac{\frac{\mu_1}{\widebar{\mu}_1}}
{\left( 1-\frac{\mu_1}{\widebar{\mu}_1} \right)^2} 
D(n,t) 
\widebar{D} (n,t) \right]^{-1}, 
\label{one-soli1} \\[2mm]
R_n (t) &= D(n,t) \left[
I - \frac{\frac{\mu_1}{\widebar{\mu}_1}}
{\left( 1-\frac{\mu_1}{\widebar{\mu}_1} \right)^2}
 \widebar{D}(n,t) D(n,t)
\right]^{-1}, 
\label{one-soli2}
\end{align}
\end{subequations}
where 
\mbox{$\widebar{D}(n,t) := \widebar{C}_1(0)
\widebar{\mu}_1^{\hspace{1pt} -n-2} 
\mathrm{e}^{-[(\widebar{\mu}_1-1) b 
 + (1-\widebar{\mu}_1^{\hspace{1pt}-1}) a] t}$}
and \mbox{$D(n,t) := C_1 (0) \mu_1^n \mathrm{e}^{[(\mu_1-1) b + 
(1-\mu_1^{-1})a] t}$}.  
For 
this solution 
to 
decay 
also 
as \mbox{$n \to -\infty$}, 
we 
require 
that 
\mbox{$\lim_{n \to -\infty} Q_n (t) \vt{l} 
=\vt{0}$} 
for any 
$n$-independent 
column vector $\vt{l}$ of dimension $l$ 
and similar for $R_n (t)$. Thus, 
considering the Maclaurin series 
for \mbox{$(I-X)^{-1}$} 
where $X$ is 
the corresponding matrix in 
(\ref{one-soli1}) or (\ref{one-soli2}), 
we 
obtain the following conditions 
on 
the kernels of 
the \mbox{$l \times l$} matrices 
$\widebar{C}_1$ 
and 
$C_1$:
\[
\mathrm{Ker} 
\left( \widebar{C}_1 C_1 \widebar{C}_1 \right)
=\mathrm{Ker} \left( \widebar{C}_1 \right)
\] 
and 
\[
\mathrm{Ker} 
\left( C_1 \widebar{C}_1 C_1 \right)
= \mathrm{Ker} \left( C_1 \right).  
\]
Note that these conditions 
remain invariant under 
the 
time evolution (cf.~(\ref{C_time}) and (\ref{C_bar_time})). 
They can also be 
written in a more 
easy-to-understand form:
\[ 
\mathrm{Ker} \left( C_1 \right) \cap
\mathrm{Im} \left( \widebar{C}_1 \right) = 
\mathrm{Ker} \left( \widebar{C}_1 \right) \cap
\mathrm{Im} \left( C_1 \right) = \{ \vt{0} \}.
\]
Consequently,
we have 
\mbox{$\mathrm{rank} 
\left(  \widebar{C}_1\right) 
= \mathrm{rank} \left( C_1 \right)$}. 

For general values of $N$ and $\widebar{N}$, 
it 
is 
rather 
difficult 
to 
grasp the condition 
that 
$Q_n$ and $R_n$ given by (\ref{QR-soliton}) 
should 
also decay 
as \mbox{$n \to -\infty$}. 
Thus, we take a different
route. 
In view of 
the first component of 
(\ref{GLM_2}) and the second component of 
(\ref{GLM_1}), 
relations (\ref{F-reflec}) and 
(\ref{G-H}) together with (\ref{psi_form})
imply that 
\begin{align}
& [z^{n} \widebar{\psi}_n ](\widebar{\mu}_j) 
 \widebar{C}_j \widebar{\mu}_j^{\hspace{1pt}-n-2}
= \left[
\begin{array}{c}
 G_j (n,t) \widebar{\mu}_j^{\hspace{1pt}-n-2} \\
 \ast \\
\end{array}
\right], \hspace{5mm}
j=1, 2, \ldots, \widebar{N}, 
%
\nonumber \\[2mm]
& [z^{-n} \psi_n ](\mu_j) C_j \mu_j^{n} 
 =\left[
\begin{array}{c}
 \ast \\
 H_j (n,t) \mu_j^{n} \\
\end{array}
\right], \hspace{5mm}
j=1, 2, \ldots, N.
\nonumber
\end{align}
Thus, 
$G_j$ and $H_j$ are closely related to 
the 
bound-state eigenfunctions. 
Owing to 
the connection formulas 
(\ref{connect4}) and (\ref{connect3})
as well as the boundary conditions (\ref{phi_bar}), 
$G_j \widebar{\mu}_j^{\hspace{1pt}-n-2}$ and 
$H_j \mu_j^{n}$ 
must 
decay 
as \mbox{$n \to -\infty$}. 
If this is satisfied, 
then 
$Q_n$ and $R_n$ in
(\ref{QR-soliton}) 
naturally 
vanish 
as \mbox{$n \to -\infty$}. 
The relations (\ref{GH-sys})
for determining 
$G_j \widebar{\mu}_j^{\hspace{1pt}-n-2}$ and 
$H_j \mu_j^{n}$  
can be rewritten as 
\begin{subequations}
\begin{align}
& \left[
\begin{array}{cccc}
\! G_1 \! & \! G_2  \! & \! \cdots \! & \! G_{\widebar{N}} \! 
\end{array}
\right]
- \left[
\begin{array}{cccc}
\! G_1 \! & \! G_2  \! & \! \cdots \! & \! G_{\widebar{N}} \! 
\end{array}
\right]
\left[
\begin{array}{ccc}
\frac{\mu_1^{n+1} \widebar{\mu}_1^{\hspace{1pt}-n-2}}
	{1-\frac{\mu_1}{\widebar{\mu}_1}} I
& \cdots & \frac{\mu_N^{n+1} \widebar{\mu}_1^{\hspace{1pt}-n-2}}
	{1-\frac{\mu_N}{\widebar{\mu}_1}} I \\
 \vdots &  \ddots & \vdots \\
 \frac{\mu_1^{n+1} \widebar{\mu}_{\widebar{N}}^{\hspace{1pt}-n-2}}
	{1-\frac{\mu_1}{\widebar{\mu}_{\widebar{N}}}} 
 I & \cdots & 
\frac{\mu_N^{n+1} \widebar{\mu}_{\widebar{N}}^{\hspace{1pt}-n-2}}
	{1-\frac{\mu_N}{\widebar{\mu}_{\widebar{N}}}} I \\
\end{array}
\right]
\nonumber \\[2mm]
& \times 
\left[
\begin{array}{ccc}
\frac{1}{\widebar{\mu}_1-\mu_1} C_1 \widebar{C}_1 
& \cdots & \frac{1}{\widebar{\mu}_{\widebar{N}} - \mu_1 } 
 C_1 \widebar{C}_{\widebar{N}} \\
 \vdots &  \ddots & \vdots \\
 \frac{1}{\widebar{\mu}_1 -\mu_N} 
C_N \widebar{C}_1 & \cdots & 
\frac{1}{\widebar{\mu}_{\widebar{N}}-\mu_N} 
C_N \widebar{C}_{\widebar{N}} \\
\end{array}
\right]
= \left[
\begin{array}{cccc}
\! \widebar{C}_1 \! & \! \widebar{C}_2 \! 
& \! \cdots \! & \! \widebar{C}_{\widebar{N}} \! 
\end{array}
\right]
\label{GH-sys3}
\end{align}
and 
\begin{align}
&
\left[
\begin{array}{cccc}
\! H_1 \! & \! H_2 \! & \! \cdots \! & \! H_{N} \! 
\end{array}
\right]
-
\left[
\begin{array}{cccc}
\! H_1 \! & \! H_2 \! & \! \cdots \! & \! H_{N} \! 
\end{array}
\right]
\left[
\begin{array}{ccc}
\frac{\mu_1^n \widebar{\mu}_1^{\hspace{1pt}-n-3}}
	{1-\frac{\mu_1}{\widebar{\mu}_1}} I
& \cdots & \frac{\mu_1^n 
	\widebar{\mu}_{\widebar{N}}^{\hspace{1pt}-n-3}}
	{1-\frac{\mu_1}{\widebar{\mu}_{\widebar{N}}}} I \\
 \vdots &  \ddots & \vdots \\
 \frac{\mu_N^n \widebar{\mu}_1^{\hspace{1pt}-n-3}}
	{1-\frac{\mu_N}{\widebar{\mu}_1}} 
 I & \cdots & 
\frac{\mu_N^n \widebar{\mu}_{\widebar{N}}^{\hspace{1pt}-n-3}}
	{1-\frac{\mu_N}{\widebar{\mu}_{\widebar{N}}}} I \\
\end{array}
\right]
\nonumber \\[2mm]
& \times 
\left[
\begin{array}{ccc}
\frac{1}{\frac{1}{\mu_1} -\frac{1}{\widebar{\mu}_1}}  \widebar{C}_1 C_1
& \cdots & \frac{1}{\frac{1}{\mu_N} - \frac{1}{\widebar{\mu}_1}} 
 \widebar{C}_1 C_N \\
 \vdots &  \ddots & \vdots \\
 \frac{1}{\frac{1}{\mu_1} -\frac{1}{\widebar{\mu}_{\widebar{N}}}} 
  \widebar{C}_{\widebar{N}} C_1 & \cdots & 
\frac{1}{\frac{1}{\mu_N} -\frac{1}{\widebar{\mu}_{\widebar{N}}}} 
 \widebar{C}_{\widebar{N}} C_N \\
\end{array}
\right]
= \left[
\begin{array}{cccc}
\! C_1 \! & \! C_2 \! & \! \cdots \! & \! C_{N} \! 
\end{array}
\right].
\label{GH-sys4}
\end{align}
\end{subequations}
Then, 
we multiply both sides of 
(\ref{GH-sys3}) from the right 
by 
an $n$-independent 
column vector of dimension $l \times \widebar{N}$
and 
consider the limit \mbox{$n \to -\infty$}. 
Thus, we obtain the condition
\begin{subequations}
\label{necess}
\begin{equation}
\mathrm{Ker}
\left[
\begin{array}{ccc}
\frac{1}{ \widebar{\mu}_1 - \mu_1} C_1 \widebar{C}_1 
& \cdots & \frac{1}{\widebar{\mu}_{\widebar{N}} - \mu_1} 
 C_1 \widebar{C}_{\widebar{N}} \\
 \vdots &  \ddots & \vdots \\
 \frac{1}{\widebar{\mu}_1 - \mu_N} 
C_N \widebar{C}_1 & \cdots & 
\frac{1}{\widebar{\mu}_{\widebar{N}}- \mu_N} 
C_N \widebar{C}_{\widebar{N}} \\
\end{array}
\right]
\subseteq
\mathrm{Ker} 
\left[
\begin{array}{cccc}
\! \widebar{C}_1 \! & \! \widebar{C}_2 \! 
& \! \cdots \! & \! \widebar{C}_{\widebar{N}} \! 
\end{array}
\right].
\label{nece1}
\end{equation}
Similarly, from 
(\ref{GH-sys4}),  we 
obtain 
\begin{equation}
\mathrm{Ker}
\left[
\begin{array}{ccc}
\frac{1}{\frac{1}{\mu_1} -\frac{1}{\widebar{\mu}_1}}  \widebar{C}_1 C_1
& \cdots & \frac{1}{\frac{1}{\mu_N} -\frac{1}{\widebar{\mu}_1}} 
 \widebar{C}_1 C_N \\
 \vdots &  \ddots & \vdots \\
 \frac{1}{\frac{1}{\mu_1} - \frac{1}{\widebar{\mu}_{\widebar{N}}}} 
  \widebar{C}_{\widebar{N}} C_1 & \cdots & 
\frac{1}{\frac{1}{\mu_N} - \frac{1}{\widebar{\mu}_{\widebar{N}}}} 
 \widebar{C}_{\widebar{N}} C_N \\
\end{array}
\right]
\subseteq
\mathrm{Ker} 
\left[
\begin{array}{cccc}
\! C_1 \! & \! C_2 \! & \! \cdots \! & \! C_{N} \! 
\end{array}
\right].
\label{nece2}
\end{equation}
\end{subequations}
The two conditions (\ref{nece1}) and (\ref{nece2}) can 
be combined and simplified 
to provide more easy-to-understand conditions 
on the scattering data in the reflectionless case. 
For this purpose, we need the following lemma. 
%
\begin{lemma}
\label{L3.1}
Define 
the \mbox{$\left( M-1 \right) \times M$} matrix 
elements \mbox{$d_{ij} \in {\mathbb C}$} 
as
\[
d_{ij} := 
\frac{ \langle \vt{a}_i, \vt{b}_j \rangle}
{\lambda_i - \nu_j}, \hspace{5mm} 
i \in \{ 
i_1, i_2, \ldots, i_{M-1} \}, 
\hspace{5mm}
j \in \{ 
j_1, j_2, \ldots, j_M \}.
\]
Here, 
$\lambda_i$ and $\nu_j$ are 
parameters, 
$\vt{a}_i$ and $\vt{b}_j$ are nonzero 
vectors of 
dimension $l$ 
and 
\mbox{$\langle \, \cdot \, , \, \cdot \, \rangle$} 
stands for the scalar product. 
For $d_{ij}$ to be 
well-defined, 
we assume 
\mbox{$\lambda_i \neq \nu_j$} for all $i$ and $j$, 
but we do 
not require 
\mbox{$\lambda_{i_\alpha} \neq \lambda_{i_\beta}$} 
or \mbox{$\nu_{j_\alpha} \neq \nu_{j_\beta}$} 
for \mbox{$\alpha \neq \beta$}. 
Instead, 
we 
assume the following condition: 
for any subset 
\mbox{$\{ k_1, k_2, \ldots, k_\gamma \} 
\subseteq \{ j_1, j_2, \ldots, j_M \}$} 
such that 
\mbox{$\nu_{k_1} = \nu_{k_2} =\cdots=\nu_{k_\gamma}$}, 
the vectors
\[
\vt{b}_{k_1}, \vt{b}_{k_2}, \ldots, \vt{b}_{k_\gamma}
\]
are 
linearly independent. 
Then, 
if 
the equality 
\[
\sum_{\alpha=1}^M 
(-1)^{\alpha -1} 
\left|
\begin{array}{cccccc}
d_{i_1 j_1} & \cdots & d_{i_1 j_{\alpha-1}} & 
d_{i_1 j_{\alpha+1}} & \cdots & d_{i_1 j_M} \\
\vdots & & \vdots & \vdots &  & \vdots \\
d_{i_{M-1} j_1} & \cdots & d_{i_{M-1} j_{\alpha-1}} & 
d_{i_{M-1} j_{\alpha+1}} & \cdots & d_{i_{M-1} j_M} \\
\end{array}
\right| \vt{b}_{j_\alpha} = \vt{0} 
\]
is valid, 
all the 
scalar 
coefficients 
must be 
zero, 
where 
\mbox{$| \cdot |$} 
stands for 
the 
determinant. 
In other words, 
the above vector equation 
holds 
true 
only in the trivial case; 
note that this equation 
can be 
written 
compactly 
as 
\[
\left|
\begin{array}{ccc}
\vt{b}_{j_1} & \cdots & \vt{b}_{j_M} \\
d_{i_1 j_1} & \cdots & d_{i_1 j_M} \\
\vdots &  & \vdots \\
d_{i_{M-1} j_1} & \cdots & d_{i_{M-1} j_M} \\
\end{array}
\right| = \vt{0}, 
\]
using 
the 
Laplace expansion 
formally. 
\\
\end{lemma}
We omit the proof 
of this lemma.  
To obtain 
useful 
information from
the conditions (\ref{necess}), we 
first 
remove 
the 
trivial subspace 
of the kernels 
commonly 
contained 
on
both sides.
From 
a given \mbox{$l \times l$} matrix $W$, 
we extract 
the 
maximum number of 
linearly independent column 
vectors 
to 
form 
an 
\mbox{$l \times \mathrm{rank}(W)$} matrix $W^{(\mathrm{c})}$. 
Similarly, we extract 
the maximum number of linearly independent row 
vectors from $W$ 
to form 
a \mbox{$\mathrm{rank}(W) \times l$} matrix $W^{(\mathrm{r})}$. 
With this notation, 
(\ref{nece1}) and (\ref{nece2}) 
can be rewritten 
in 
more 
compact 
forms 
as
\begin{subequations}
\label{necess2}
\begin{equation}
\mathrm{Ker}
\left[
\begin{array}{ccc}
\frac{1}{\widebar{\mu}_1 - \mu_1} C_1^{(\mathrm{r})}
	\widebar{C}_1^{(\mathrm{c})} 
& \cdots & \frac{1}{\widebar{\mu}_{\widebar{N}}- \mu_1} 
 C_1^{(\mathrm{r})} \widebar{C}_{\widebar{N}}^{(\mathrm{c})} \\
 \vdots &  \ddots & \vdots \\
 \frac{1}{\widebar{\mu}_1 - \mu_N} 
C_N^{(\mathrm{r})} \widebar{C}_1^{(\mathrm{c})} & \cdots & 
\frac{1}{\widebar{\mu}_{\widebar{N}} - \mu_N} 
C_N^{(\mathrm{r})} \widebar{C}_{\widebar{N}}^{(\mathrm{c})} \\
\end{array}
\right]
\subseteq
\mathrm{Ker} 
\left[
\begin{array}{cccc}
\! \widebar{C}_1^{(\mathrm{c})} \! & \! 
	\widebar{C}_2^{(\mathrm{c})} \! 
& \! \cdots \! & \! \widebar{C}_{\widebar{N}}^{(\mathrm{c})} \! 
\end{array}
\right]
\label{nece3}
\end{equation}
and 
\begin{equation}
\mathrm{Ker}
\left[
\begin{array}{ccc}
\frac{1}{\frac{1}{\mu_1} -\frac{1}{\widebar{\mu}_1}}  \widebar{C}_1^{(\mathrm{r})}
	C_1^{(\mathrm{c})}
& \cdots & \frac{1}{\frac{1}{\mu_N} - \frac{1}{\widebar{\mu}_1}} 
 \widebar{C}_1^{(\mathrm{r})} C_N^{(\mathrm{c})} \\
 \vdots &  \ddots & \vdots \\
 \frac{1}{\frac{1}{\mu_1} - \frac{1}{\widebar{\mu}_{\widebar{N}}}} 
  \widebar{C}_{\widebar{N}}^{(\mathrm{r})}
	C_1^{(\mathrm{c})} & \cdots & 
\frac{1}{\frac{1}{\mu_N} -\frac{1}{\widebar{\mu}_{\widebar{N}}}} 
 \widebar{C}_{\widebar{N}}^{(\mathrm{r})} C_N^{(\mathrm{c})} \\
\end{array}
\right]
\subseteq
\mathrm{Ker} 
\left[
\begin{array}{cccc}
\! C_1^{(\mathrm{c})} \! & \! C_2^{(\mathrm{c})}
  \! & \! \cdots \! & \! C_{N}^{(\mathrm{c})} \! 
\end{array}
\right].
\label{nece4}
\end{equation}
\end{subequations}

With the aid of 
Lemma~\ref{L3.1}, 
we can 
prove 
the following two propositions. 

\begin{proposition}
\label{prop3.2}
Assume that 
\[
\sum_{j=1}^{N} \mathrm{rank}\left( C_j \right) \le
\sum_{j=1}^{\widebar{N}}
\mathrm{rank}\left( \widebar{C}_j \right),
\]
and the condition (\ref{nece3}) is satisfied. 
Then, the above inequality becomes 
an equality, 
\begin{equation}
\sum_{j=1}^{N} \mathrm{rank}\left( C_j \right) =
\sum_{j=1}^{\widebar{N}}
\mathrm{rank}\left( \widebar{C}_j \right),
\label{sca-con1}
\end{equation}
and 
the matrix on the left-hand 
side of (\ref{nece3}) 
must be 
invertible, i.e. 
\begin{equation}
\left|
\begin{array}{ccc}
\frac{1}{\widebar{\mu}_1 - \mu_1} C_1^{(\mathrm{r})}
	\widebar{C}_1^{(\mathrm{c})} 
& \cdots & \frac{1}{\widebar{\mu}_{\widebar{N}}-\mu_1} 
 C_1^{(\mathrm{r})} \widebar{C}_{\widebar{N}}^{(\mathrm{c})} \\
 \vdots &  \ddots & \vdots \\
 \frac{1}{\widebar{\mu}_1 - \mu_N} 
C_N^{(\mathrm{r})} \widebar{C}_1^{(\mathrm{c})} & \cdots & 
\frac{1}{\widebar{\mu}_{\widebar{N}} - \mu_N} 
C_N^{(\mathrm{r})} \widebar{C}_{\widebar{N}}^{(\mathrm{c})} \\
\end{array}
\right| \neq 0.
\label{sca-con2}
\end{equation}
\hphantom{aa}
\\
\end{proposition}
\begin{proposition}
\label{prop3.3}
Assume that 
\[
\sum_{j=1}^{N} \mathrm{rank}\left( C_j \right) \ge
\sum_{j=1}^{\widebar{N}}
\mathrm{rank}\left( \widebar{C}_j \right),
\]
and the condition (\ref{nece4}) is satisfied. 
Then, the above inequality becomes 
an equality, 
\[
\sum_{j=1}^{N} \mathrm{rank}\left( C_j \right) =
\sum_{j=1}^{\widebar{N}}
\mathrm{rank}\left( \widebar{C}_j \right),
\]
and 
the matrix on the left-hand 
side of (\ref{nece4}) 
must be 
invertible, i.e. 
\begin{equation}
\left|
\begin{array}{ccc}
\frac{1}{\frac{1}{\mu_1} -\frac{1}{\widebar{\mu}_1}}  \widebar{C}_1^{(\mathrm{r})}
	C_1^{(\mathrm{c})}
& \cdots & \frac{1}{\frac{1}{\mu_N} - \frac{1}{\widebar{\mu}_1}} 
 \widebar{C}_1^{(\mathrm{r})} C_N^{(\mathrm{c})} \\
 \vdots &  \ddots & \vdots \\
 \frac{1}{\frac{1}{\mu_1} - \frac{1}{\widebar{\mu}_{\widebar{N}}}} 
  \widebar{C}_{\widebar{N}}^{(\mathrm{r})}
	C_1^{(\mathrm{c})} & \cdots & 
\frac{1}{\frac{1}{\mu_N} - \frac{1}{\widebar{\mu}_{\widebar{N}}}} 
 \widebar{C}_{\widebar{N}}^{(\mathrm{r})} C_N^{(\mathrm{c})} \\
\end{array}
\right| \neq 0.
\label{sca-con3}
\end{equation}
\hphantom{aa}
\\
\end{proposition}

By combining 
Propositions \ref{prop3.2} and \ref{prop3.3}, 
we 
arrive at 
the following theorem. 
\begin{theorem}
\label{theorem3.4}
Assume that 
(\ref{QR-soliton}) 
provides 
the multisoliton solutions
of the 
matrix 
Ablowitz--Ladik lattice (\ref{mALsys}), 
which 
decay 
as \mbox{$n \to \pm \infty$} 
and 
produce 
the 
bound states 
of 
the 
associated 
eigenvalue 
problem (\ref{mAL0}).  
Then, the 
scattering data 
must 
satisfy 
the three conditions 
(\ref{sca-con1})--(\ref{sca-con3}). 
\\
\end{theorem}
\noindent
Note that the time evolution does not 
change 
these 
conditions 
(cf.~(\ref{C_time}) and (\ref{C_bar_time})). 
\\
\\
{\it Remark.} 
Here, we 
only 
considered the case 
where 
the time variable 
$t$ 
is 
fixed at 
some 
finite value. 
In the limits \mbox{$t \to \pm \infty$},  
the solitons generally 
separate from one another 
and restore their original shapes. 
Thus, for 
(\ref{QR-soliton}) to 
describe 
proper 
multisoliton collisions 
through the 
passage of time, 
we 
have to impose 
additional conditions, i.e., 
the 
above 
conditions 
for 
subsets of 
\mbox{$\{ \mu_j, C_j \}_{j=1, 2, \ldots, N}$} 
and \mbox{$\{\widebar{\mu}_j, 
	\widebar{C}_j \}_{j=1, 2, \ldots, \widebar{N}}$}. 
Some relevant results were obtained independently in~\cite{DM2010}.

\subsection{Complex conjugation reduction}

When \mbox{$b=a^\ast$}, 
the 
matrix 
Ablowitz--Ladik lattice (\ref{mALsys}) 
allows the 
complex conjugation reduction
\mbox{$R_n = \sigma Q_n^\ast$} 
with 
a real 
constant $\sigma$ (cf.~\cite{GI82}).   
In particular, 
the simplest 
reduction \mbox{$R_n = -Q_n^\ast$} can be 
realized in formulas (\ref{ALlinearization}) 
by identifying 
$\widebar{F} (n)$ 
with the complex conjugate of $F(n)$, i.e.
\begin{equation}
\widebar{F} (n) = 
\left\{ F(n) \right\}^\ast, 
\label{F-Fbar-rel}
\end{equation}
which is naturally 
preserved under the 
time evolution (\ref{AL-linear}) with \mbox{$b=a^\ast$}. 
Indeed, 
this relation can 
be 
derived 
by exploiting 
the symmetry 
of 
the eigenvalue problem (\ref{mAL0}) with \mbox{$R_n = -Q_n^\ast$}; 
that is, if 
\[
\left[
\begin{array}{c}
 \Psi_{1, n} (z)\\
 \Psi_{2, n} (z)\\
\end{array}
\right] 
\]
is 
an eigenfunction, 
then 
\[
\pm 
\left[
\begin{array}{c}
 - \Psi_{2, n}(1/z^\ast) \\
 \Psi_{1, n} (1/z^\ast) \\
\end{array}
\right]^\ast
\]
gives another eigenfunction 
of the same 
problem. 
Thus, we 
can reflect 
this symmetry 
in the 
Jost solutions and 
the 
scattering data
to confirm (\ref{F-Fbar-rel}). 
%

In particular, the $N$-soliton solution 
of the matrix 
Ablowitz--Ladik equation, 
\[
 Q_{n,t} - a Q_{n+1} + a^\ast Q_{n-1} + (a-a^\ast) Q_n 
	-a Q_n Q_n^\ast Q_{n+1} +a^\ast Q_{n-1} Q_n^\ast Q_n = O, 
\]
is obtained by setting \mbox{$b=a^\ast$}, 
\mbox{$\widebar{N}=N$} and 
\[
\widebar{\mu}_j= \frac{1}{\mu_j^\ast}, 
\hspace{5mm}
\widebar{C}_j(t)  \widebar{\mu}_j^{\hspace{1pt}-2} 
	= -\left\{ C_j (t) \right\}^\ast, 
\hspace{5mm} j=1, 2, \ldots, N
\]
in 
formula 
(\ref{Q-Nsol}) with (\ref{C_time}). 
The reduced set of scattering data 
is required to satisfy 
the 
conditions 
(\ref{sca-con2}) and (\ref{sca-con3}); in fact, 
they 
are 
equivalent 
under 
this 
reduction. 
In addition, (\ref{sca-con2}) (or (\ref{sca-con3})) 
for 
subsets of 
\mbox{$\{ \mu_j, C_j \}_{j=1, 2, \ldots, N}$} 
should 
also be satisfied. 
%
Throughout 
this paper, 
we do not 
discuss 
the issue of 
regularity of 
solutions
and the term ``soliton solution" is used in a 
broad sense. 
That is, 
it 
may have 
singularities at some
values of the independent 
variables $n$ and $t$. 

\section{Solution formulas for 
the derivative NLS lattices}


\subsection{Solutions of the space-discrete Gerdjikov--Ivanov system} 
\label{subs4.1}

In this subsection, we 
solve the space-discrete Gerdjikov--Ivanov system 
derived 
in subsection~\ref{subs2.2} 
by using 
the 
results 
in section~\ref{sec3}. 
%
Note that 
the nonzero parameter \mbox{$\mu 
$} in 
the space-discrete Gerdjikov--Ivanov system (\ref{sdGI}) 
is 
nonessential;  
it 
can be fixed 
at any 
nonzero 
value, say $1$, 
using a simple point 
transformation and 
rescalings of the parameters $a$ and $b$ 
(cf.~\cite{Tsuchi02,TsuJMP11}). 
In addition, as is clear from 
the defining relation 
of the Miura map (\ref{AL-R1}), 
the limit \mbox{$\mu \to 0$} is trivial 
and need not be considered separately. 
%
%
%
%
Thus, in the following, 
we consider 
the space-discrete 
Gerdjikov--Ivanov system (\ref{sdGI}) 
with 
\mbox{$\mu =1$}: 
\begin{subnumcases}{\label{sdGI2}}
{}
 Q_{n,t}- a Q_{n+1} + b Q_{n-1} + (a-b)Q_n 
	+ a Q_{n} 
\left( P_{n} - P_{n+1} \right) Q_{n+1} 
\nonumber \\
\mbox{}- b Q_{n-1} \left( P_{n} 
 - P_{n+1} \right) Q_{n}
+ a Q_{n} P_{n}Q_n P_{n+1} Q_{n+1} 
- b Q_{n-1}  P_{n}Q_n P_{n+1} Q_{n}
=O, \hspace{12mm}
\label{sdGIeq}
\\[2mm]
P_{n,t} -b P_{n+1} + a P_{n-1} +(b-a)P_n -b P_{n}
\left( Q_{n-1} - Q_n \right) P_{n+1}
\nonumber \\
\mbox{}+a P_{n-1}\left( Q_{n-1} - Q_n \right) P_{n}
+ b P_{n} Q_{n-1} P_{n}Q_{n} P_{n+1} 
- a P_{n-1} Q_{n-1} P_{n}Q_{n} P_{n}
=O.
\label{}
\end{subnumcases}

In section~\ref{sec3}, 
we developed the inverse scattering method 
associated with the matrix Ablowitz--Ladik eigenvalue 
problem (\ref{mAL0}) 
under the 
vanishing 
boundary conditions 
on the potentials $Q_n$ and $R_n$ 
(cf.\ (\ref{zero-bc})). 
%
The remaining unknown 
$P_n$ 
in (\ref{sdGI2}) 
can be determined 
from 
a linear eigenfunction 
through the simple 
formula  
\begin{equation}
P_n = 
\left. \hspace{-1pt}
\Psi_{2,n} \Psi_{1,n}^{-1} \right|_{\mu
\hspace{1pt}(=z^2)
=1}.
\label{P-formu1}
\end{equation}
Because 
there is some arbitrariness in 
choosing 
the 
linear eigenfunction, 
we need 
to specify 
boundary conditions 
to 
determine 
$P_n$ 
uniquely. 
Thus, 
we 
assume that 
not only $Q_n$ but also 
$P_n$ 
decays 
rapidly as \mbox{$n \to \pm \infty$}:
\begin{equation}
\label{zero-bc2}
\lim_{n \to \pm \infty} Q_n = \lim_{n \to \pm \infty} P_n =O.
\end{equation}
This is consistent if we 
choose 
the linear eigenfunction 
appearing in 
the right-hand side of (\ref{P-formu1}) 
as
\[
\left[
\begin{array}{c}
 \Psi_{1, n} \\
 \Psi_{2, n} \\
\end{array}
\right] 
:= \widebar{\psi}_n, 
\]
and assume that the scattering data $B(\mu)$ 
vanishes at \mbox{$\mu=1$}, 
i.e., 
\mbox{$B(1)=O$}
(see (\ref{leftJost}) and 
(\ref{phi_relation})). 
In fact, 
the Jost solution $\widebar{\psi}_n$ 
as well as $\phi_n$ 
does not satisfy the time part of the Lax representation 
(\ref{mAL-time}), 
so 
it is more appropriate to 
use 
the explicitly time-dependent 
Jost solutions introduced 
in subsection~\ref{sTDsd}.
However, the overall 
multiplicative factor \mbox{$\mathrm{e}^{(-\mu+1) b t}$} 
as introduced in (\ref{Jost-time}) 
plays no
role in formula (\ref{P-formu1}), 
so 
in view of (\ref{5Kdef}), we 
can express $P_n$ as 
\[
P_n = 
\left\{ \sum_{k=0}^{\infty} \widebar{K}_2 (n,n+k) \right\}
\left\{
I + \sum_{k=0}^{\infty} \widebar{K}_1 (n,n+k)
\right\}^{-1}. 
\]
This expression  
enables us 
to 
determine $P_n$ from 
the set of 
scattering data 
with the aid of 
the Gel'fand--Levitan--Marchenko equations 
(\ref{GLM_1}) and (\ref{GLM_2}); 
however, 
for later convenience, 
we take an alternative approach.

Because the space-discrete 
Gerdjikov--Ivanov system (\ref{sdGI2}) 
is ``symmetric" with respect to 
$Q_n$ and $P_n$, 
there 
must be a formula 
for expressing 
$Q_n$
in a manner 
similar to (\ref{P-formu1}). 
Such a 
formula can be 
established 
by identifying 
an appropriate 
Ablowitz--Ladik 
eigenvalue problem that is 
gauge equivalent to the original problem 
(\ref{mAL0}); 
the corresponding 
gauge transformation is 
often referred to as a B\"acklund--Darboux transformation.  
Then, in the 
new gauge, 
the roles of $Q_n$ and $P_n$ are 
swapped 
and 
$P_n$ appears directly 
as a potential 
in the Ablowitz--Ladik 
eigenvalue problem.
In other words, there 
exists 
another Miura map from the space-discrete 
Gerdjikov--Ivanov system (\ref{sdGI2}) 
to the Ablowitz--Ladik lattice 
in the form \mbox{$(Q_n,P_n) \mapsto (\widetilde{Q}_n,P_n)$}. 

Let us 
consider the original Ablowitz--Ladik 
eigenvalue problem 
%
%
(\ref{mAL0}) 
with \mbox{$R_n = P_{n} - P_{n+1} + P_{n}Q_n P_{n+1}$} 
(cf.~(\ref{AL-R1})): 
\begin{flalign}
\mathrm{(AL1):} \hspace{5mm}
\left[
\begin{array}{c}
 \Psi_{1, n} \\
 \Psi_{2, n} \\
\end{array}
\right]
&= 
 \left[
 \begin{array}{cc}
  z I & z Q_n \\
 z^{-1} \left( P_{n} - P_{n+1} + P_{n}Q_n P_{n+1} \right)
 & z^{-1} I \\
 \end{array}
 \right]
\left[
\begin{array}{c}
 \Psi_{1, n+1} \\
 \Psi_{2, n+1} \\     
\end{array}
\right].&
\label{ALsca1}
\end{flalign}
Indeed, 
this 
can be 
rewritten 
as 
another 
Ablowitz--Ladik 
eigenvalue problem:  
\begin{flalign}
\mathrm{(AL2):} \hspace{5mm}
\left[
\begin{array}{c}
 \Phi_{1, n} \\
 \Phi_{2, n} \\
\end{array}
\right]
&= 
 \left[
 \begin{array}{cc}
  z I & z \left( -Q_n +Q_{n+1} + Q_n P_{n+1} Q_{n+1} \right) \\
 z^{-1} P_{n+1} & z^{-1} I \\
 \end{array}
 \right]
\left[
\begin{array}{c}
 \Phi_{1, n+1} \\
 \Phi_{2, n+1} \\     
\end{array}
\right],&
\label{ALsca2}
\end{flalign}
using 
the gauge transformation 
defined as 
\begin{align}
\hspace{-3mm}
\left[
\begin{array}{c}
 \Phi_{1, n} \\
 \Phi_{2, n} \\
\end{array}
\right]
& := \left[
\begin{array}{c}
\left( z^{-2}-1 \right) \Psi_{1, n} -Q_n \left( I-P_n Q_n \right)^{-1}
 \left( \Psi_{2,n} -z^{-2} P_n \Psi_{1,n} \right)	\\
 \left( I-P_n Q_n \right)^{-1}
 \left( \Psi_{2,n} -z^{-2} P_n \Psi_{1,n} \right) \\
\end{array}
\right]
\label{gauge-def1} \\
& \hphantom{:} =: g_n 
\left[
\begin{array}{c}
 \Psi_{1, n} \\
 \Psi_{2, n} \\
\end{array}
\right]. 
\nonumber
\end{align}
The explicit form of the 
transformation matrix $g_n$ 
is 
unimportant;
only 
its
asymptotic behavior as \mbox{$n \to \pm \infty$} 
is needed: 
\[
\lim_{n \to \pm \infty} g_n = 
\left[
\begin{array}{cc}
\left( z^{-2}-1 \right) I & O \\
 O & I \\
\end{array}
\right]. 
\]
For 
the original Jost solutions 
for (AL1) 
defined as
(\ref{leftJost}), 
the Jost solutions 
for (AL2) are given 
as 
\begin{subequations}
\label{Jost-AL1-2}
\begin{align}
& \phi_n^{\mathrm{(AL2)}} (z) = \frac{1}{z^{-2}-1} g_n
	\phi_n^{\mathrm{(AL1)}} (z), \hspace{5mm} 
 \widebar{\phi}_n^{\mathrm{\hspace{1pt}(AL2)}} (z) = 
 g_n \widebar{\phi}_n^{\mathrm{\hspace{1pt}(AL1)}} (z), 
\label{Jost-AL-phi}
\\[2mm]
& \psi_n^{\mathrm{(AL2)}} (z) 
	= g_n \psi_n^{\mathrm{(AL1)}} (z), \hspace{5mm}
 \widebar{\psi}_n^{\mathrm{\hspace{1pt}(AL2)}} (z)	= 
   \frac{1}{z^{-2}-1} g_n
 \widebar{\psi}_n^{\mathrm{\hspace{1pt}(AL1)}} (z). 
\label{Jost-AL-psi}
\end{align}
\end{subequations}
Indeed, they satisfy both 
the eigenvalue problem (\ref{ALsca2}) and 
the boundary conditions 
(\ref{leftJost}). 
The 
case 
of \mbox{$z^2=1$}
can be understood 
in the corresponding 
limit. 

In view of 
the aforementioned condition \mbox{$B(1)=O$}, 
it is natural to 
modify 
the 
defining relations (\ref{ref102})
of 
the scattering data 
for (AL1) 
on \mbox{$|\mu|=1$} 
as
\begin{subequations}\label{AL1-scat}
 \begin{align}
 \phi_n^{\mathrm{(AL1)}} 
&= \widebar{\psi}_n^{\hspace{1pt}\mathrm{(AL1)}} A(\mu)
        + \psi_n^{\mathrm{(AL1)}}  (\mu^{-1}-1) B(\mu),
 \\[0.5mm]
 \widebar{\phi}_n^{\hspace{1pt}\mathrm{(AL1)}} 
&= \widebar{\psi}_n^{\hspace{1pt}\mathrm{(AL1)}} \widebar{B}(\mu)
        - \psi_n^{\mathrm{(AL1)}} \widebar{A}(\mu).
 \end{align}
\end{subequations}
Then, relations 
(\ref{Jost-AL1-2}) between 
the Jost solutions for (AL1) and 
those for (AL2)  
imply 
that 
the defining relations of 
the scattering data 
for (AL2) on \mbox{$|\mu|=1$} 
become 
\begin{subequations}\label{AL2-scat}
 \begin{align}
 \phi_n^{\mathrm{(AL2)}} 
&= \widebar{\psi}_n^{\hspace{1pt}\mathrm{(AL2)}} A(\mu)
        + \psi_n^{\mathrm{(AL2)}} B(\mu),
 \\[0.5mm]
 \widebar{\phi}_n^{\hspace{1pt}\mathrm{(AL2)}} 
&= \widebar{\psi}_n^{\hspace{1pt}\mathrm{(AL2)}} 
	(\mu^{-1}-1)\widebar{B}(\mu)
        - \psi_n^{\mathrm{(AL2)}} \widebar{A}(\mu).
\end{align}
\end{subequations}
The bound-state eigenvalues are 
determined 
by the positions of 
the simple poles of $A(\mu)^{-1}$ in \mbox{$|\mu| <1$}
and $\widebar{A}(\mu)^{-1}$ in \mbox{$|\mu| >1$}; 
the more general case of higher
order poles can be recovered by taking a 
suitable 
coalescence limit. 
Because of 
the uniqueness of the ``analytic continuation", 
the bound-state eigenvalues 
are common to 
(AL1) and (AL2):
\begin{align}
& \mu_j^{\mathrm{(AL1)}}
=\mu_j^{\mathrm{(AL2)}}= \mu_j, \hspace{5mm} j=1, 2, \ldots, N, 
\nonumber \\[2mm]
& \widebar{\mu}_j^{\hspace{1pt}\mathrm{(AL1)}}
=\widebar{\mu}_j^{\hspace{1pt}\mathrm{(AL2)}}
= \widebar{\mu}_j, \hspace{5mm} j=1, 2, \ldots, \widebar{N}.
\nonumber 
\end{align}
Owing to 
(\ref{Jost-AL1-2}), 
the corresponding 
matrices 
$C_j$ and $\widebar{C}_j$ 
(cf.~(\ref{connect3}) and (\ref{connect4}))
for (AL1) and (AL2) 
can be expressed as 
\begin{subequations}
\label{C-AL-GI}
\begin{align}
& C_j^{\mathrm{(AL1)}}= \left( \mu_j^{-1}-1 \right) C_j, \hspace{5mm}
C_j^{\mathrm{(AL2)}}= C_j,\hspace{5mm} j=1, 2, \ldots, N, 
\\[2mm]
& \widebar{C}_j^{\hspace{1pt}\mathrm{(AL1)}}= \widebar{C}_j, \hspace{5mm}
 \widebar{C}_j^{\hspace{1pt}\mathrm{(AL2)}} 
 = \left( \widebar{\mu}_j^{\hspace{1pt}-1}-1 \right) \widebar{C}_j,
 \hspace{5mm} j=1, 2, \ldots, \widebar{N}.
\end{align}
\end{subequations}
By 
combining the above relations, 
the functions $F(m)$ and $\widebar{F}(m)$ 
for (AL1) and (AL2) 
(cf.\ (\ref{F_form}) and (\ref{F_bar_form}))
can be written 
as 
\begin{subequations}
\label{F-GI}
\begin{align}
& F^{\mathrm{(AL1)}}(m)= F(m-1)-F(m), \hspace{5mm}
 F^{\mathrm{(AL2)}}(m)= F(m), 
\label{F-GI1}
\\[2mm]
& \widebar{F}^{\hspace{1pt}\mathrm{(AL1)}}(m)= \widebar{F}(m), 
\hspace{5mm} \widebar{F}^{\hspace{1pt}\mathrm{(AL2)}}(m)
= \widebar{F}(m+1) - \widebar{F}(m),
\label{F-GI2}
\end{align}
\end{subequations}
in terms of the original 
$F(m)$ and $\widebar{F}(m)$ 
defined as 
(\ref{F_form}) and (\ref{F_bar_form}). 
Clearly, 
the 
modification of the scattering 
data for (AL1) and (AL2) as described above 
does not change 
their 
time dependences 
given by (\ref{BA_time})--(\ref{C_bar_time}). 
Thus, 
each of the 
pairs \mbox{$( F, \widebar{F} \hspace{1pt} )$}, 
\mbox{$( F^{\mathrm{(AL1)}}, \widebar{F}^{\hspace{1pt}\mathrm{(AL1)}} )$}
and 
\mbox{$( F^{\mathrm{(AL2)}}, \widebar{F}^{\hspace{1pt}\mathrm{(AL2)}} )$} 
satisfies the same 
linear 
evolutionary system 
(\ref{AL-linear}).  

We can construct 
the Gel'fand--Levitan--Marchenko equations 
for the space-discrete 
Gerdjikov--Ivanov system (\ref{sdGI2}) 
by combining 
(\ref{ALlin-1}) and (\ref{K1-closed}) 
for (AL1) 
and 
(\ref{ALlin-2}) and (\ref{K2-closed})
for (AL2), 
i.e.
%
\begin{align}
& Q_n = K_1^{\mathrm{(AL1)}} (n,n),
\nonumber
\\[1mm]
& P_{n+1} = \widebar{K}^{\hspace{1pt}\mathrm{(AL2)}}_2 (n,n),
\nonumber \\
& K_1^{\mathrm{(AL1)}}(n,m) 
 = \widebar{F}^{\hspace{1pt}\mathrm{(AL1)}}(m) -\sum_{i=0}^{\infty} 
\sum_{k=0}^{\infty} K_1^{\mathrm{(AL1)}}(n,n+i) 
F^{\mathrm{(AL1)}}(n+i+k+1) 
\nonumber \\
& \hspace{28mm} \times 
 \widebar{F}^{\hspace{1pt}\mathrm{(AL1)}}(m+k+1),
\hspace{5mm} 
m \geq n, 
\nonumber \\[2mm]
& \widebar{K}^{\hspace{1pt}\mathrm{(AL2)}}_2 (n,m) 
= -F^{\mathrm{(AL2)}}(m) -\sum_{i=0}^{\infty} 
\sum_{k=0}^{\infty} \widebar{K}^{\hspace{1pt}\mathrm{(AL2)}}_2 (n,n+i) 
 \widebar{F}^{\hspace{1pt}\mathrm{(AL2)}}(n+i+k+1) 
\nonumber \\
& \hspace{28mm} \times 
 F^{\mathrm{(AL2)}}(m+k+1), 
\hspace{5mm} 
m \geq n. 
\nonumber 
\end{align}
Substituting 
(\ref{F-GI}) 
and 
changing 
the notation 
slightly, 
we 
obtain 
\begin{subequations}
\label{GIlinearization2}
\begin{align}
& Q_n = {\mathscr K} (n,n),
\label{GIlin-1}
\\[1mm]
& P_n = \widebar{\mathscr K} (n,n),
\label{GIlin-2}
\\
& {\mathscr K} (n,m) 
 = \widebar{F}(m) +\sum_{i=0}^{\infty} 
\sum_{k=0}^{\infty} {\mathscr K} (n,n+i) 
\left\{ F(n+i+k+1) -F(n+i+k) \right\}
\nonumber \\
& \hspace{21.5mm} \times 
 \widebar{F}(m+k+1),
\hspace{5mm} 
m \geq n, 
\label{K11-closed}
\\[2mm]
& \widebar{\mathscr K} (n,m) 
= -F(m-1) -\sum_{i=0}^{\infty} 
\sum_{k=0}^{\infty} \widebar{\mathscr K} (n,n+i) 
\left\{ \widebar{F}(n+i+k+1) - \widebar{F}(n+i+k) \right\}
\nonumber \\
& \hspace{21.5mm} \times 
 F(m+k), 
\hspace{5mm} 
m \geq n. 
\label{K22-closed}
\end{align}
\end{subequations}
Note that $F$ and $\widebar{F}$ satisfy 
the linear 
part of the equations 
for 
$P_n$ and $Q_n$, 
which is 
(\ref{AL-linear}) for the space-discrete 
Gerdjikov--Ivanov system (\ref{sdGI2}).   

In the 
case of 
\mbox{$B(\mu)=\widebar{B}(\mu)=O$} on 
\mbox{$|\mu|=1$}, 
which 
corresponds 
to the reflectionless potentials 
for both (AL1) and (AL2), we can
solve 
the Gel'fand--Levitan--Marchenko equations (\ref{GIlinearization2}) 
to obtain the soliton solutions 
in closed form. 
The derivation 
is essentially the same as 
in 
the Ablowitz--Ladik case described 
in subsection~\ref{subsec3.6}; 
naturally, 
the multisoliton solutions 
of the space-discrete 
Gerdjikov--Ivanov system (\ref{sdGI2}) can be obtained directly 
by applying 
the correspondence relations 
(\ref{C-AL-GI}) 
to (\ref{QR-soliton}), i.e. 
%
\begin{subequations}
\label{QP-soliton}
\begin{align}
 Q_n (t) &= \left[
\begin{array}{ccc}
\! \widebar{C}_1(t) \hspace{1pt}\widebar{\mu}_1^{\hspace{1pt}-n-2}
\! & \! \cdots \! & \! 
\widebar{C}_{\widebar{N}}(t) \hspace{1pt}
\widebar{\mu}_{\widebar{N}}^{\hspace{1pt}-n-2} \! 
\end{array}
\right] 
\left[
\begin{array}{ccc}
 \mathscr{U}_{11} & \cdots & \mathscr{U}_{1\widebar{N}} \\
 \vdots &  \ddots & \vdots \\
 \mathscr{U}_{\widebar{N}1} & \cdots 
	& \mathscr{U}_{\widebar{N}\widebar{N}} \\
\end{array}
\right]^{-1} 
\left[
\begin{array}{c}
 I \\
 \vdots \\
 I \\ 
\end{array}
\right], 
\label{Q-Nsol2}
\\[1mm]
 P_n (t) &= \left[
\begin{array}{ccc}
\! C_1 (t) \hspace{1pt} \mu_1^{n-1} \! & \! 
\cdots \! & \! C_{N} (t) \hspace{1pt} \mu_N^{n-1} \! 
\end{array}
\right]
\left[
\begin{array}{ccc}
 \mathscr{V}_{11} & \cdots & \mathscr{V}_{1N} \\
 \vdots &  \ddots & \vdots \\
 \mathscr{V}_{N1} & \cdots & \mathscr{V}_{NN} \\
\end{array}
\right]^{-1}
\left[
\begin{array}{c}
 I \\
 \vdots \\
 I \\ 
\end{array}
\right]. 
\end{align}
\end{subequations}
%
%
Here, all the entries in (\ref{QP-soliton}) 
are \mbox{$l \times l$} matrices;
the block matrices 
\mbox{$\mathscr{U}=(\mathscr{U}_{jk})_{1 \le j,k \le \widebar{N}}$} and 
\mbox{$\mathscr{V}=(\mathscr{V}_{jk})_{1 \le j,k \le N}$} are 
defined as 
\begin{align}
& \mathscr{U}_{jk} := \delta_{jk} I - \sum_{i=1}^{N} 
\frac{\left( 1- \mu_i \right) \mu_{i}^n \widebar{\mu}_k^{\hspace{1pt}-n-3}}
{\displaystyle \left(1-\frac{\mu_i}{\widebar{\mu}_j}\right) 
	\left(1-\frac{\mu_i}{\widebar{\mu}_k}\right)} 
C_i(t)\widebar{C}_k(t),
\nonumber \\[2mm]
& \mathscr{V}_{jk} := \delta_{jk} I + \sum_{i=1}^{\widebar{N}} 
\frac{ \left( 1- \widebar{\mu}_i^{\hspace{1pt}-1} \right) 
 \widebar{\mu}_i^{\hspace{1pt} -n-2}  \mu_k^{n} }
{\displaystyle \left( 1-\frac{\mu_j}{\widebar{\mu}_i} \right) 
 \left( 1-\frac{\mu_k}{\widebar{\mu}_i} \right) }
 \widebar{C}_i(t) C_k(t), 
 \nonumber
\end{align}
and
the time dependences of $C_j$ and $\widebar{C}_j$ 
are 
given by 
(\ref{C_time}) and (\ref{C_bar_time}). 
Note that 
the three conditions 
(\ref{sca-con1})--(\ref{sca-con3}) 
must be satisfied 
for (\ref{QP-soliton}) 
to describe proper 
multisoliton solutions 
decaying as \mbox{$n \to \pm \infty$} 
(cf.~Theorem \ref{theorem3.4}). 
In addition, we 
need to impose 
similar conditions 
for subsets of the soliton parameters 
so that the solitons interact 
with 
each other 
properly 
throughout the time evolution. 
%
%

When \mbox{$b=a^\ast$}, 
the space-discrete 
Gerdjikov--Ivanov system (\ref{sdGI2}) 
allows the 
complex conjugation reduction
\mbox{$P_n = \mathrm{i} \sigma Q_{n-
1/2
}^{\,\ast}$} 
with 
a real constant $\sigma$~\cite{Tsuchi02}. 
That is, 
two originally uncoupled systems, 
(\ref{sdGI2}) with \mbox{$n \in \mathbb{Z}$}
and (\ref{sdGI2}) with \mbox{$n \in \mathbb{Z}+1/2$}, 
can be related 
by 
this reduction
to give
a single equation
with 
\mbox{$n \in \mathbb{Z}/2$}. 
Clearly, 
the value of $\sigma$ is nonessential, so 
we set
\mbox{$\sigma=1$} 
and consider the reduction \mbox{$P_n = \mathrm{i} Q_{n-
1/2
}^{\,\ast}$}. 
This reduction 
can be realized 
at the level of 
formulas (\ref{GIlinearization2}) 
by setting
\begin{equation}
\widebar{F} (n) 
 = -\mathrm{i} 
\left\{ F \left( n-\mbox{$\tiny{\frac{1}{2}}$} \right) \right\}^\ast, 
\nonumber 
\end{equation}
which is 
consistent with 
the 
time evolution (\ref{AL-linear}) 
with \mbox{$b=a^\ast$}. 
In particular, the 
$N$-soliton solution 
of the space-discrete Gerdjikov--Ivanov equation,  
\begin{align}
& Q_{n,t} - a Q_{n+1} + a^\ast Q_{n-1} + (a-a^\ast )Q_n 
	+ \mathrm{i}a Q_{n} 
\left( Q_{n-\frac{1}{2}}^{\,\ast}
 - Q_{n+\frac{1}{2}}^{\,\ast} \right) Q_{n+1} 
\nonumber \\[1mm]
& \mbox{}
- \mathrm{i} a^\ast Q_{n-1} \left( Q_{n-\frac{1}{2}}^{\,\ast}
 - Q_{n+\frac{1}{2}}^{\,\ast} \right) Q_{n}
- a Q_{n} Q_{n-\frac{1}{2}}^{\,\ast} 
	Q_n Q_{n+\frac{1}{2}}^{\,\ast} Q_{n+1} 
+ a^\ast Q_{n-1} Q_{n-\frac{1}{2}}^{\,\ast} Q_n 
	Q_{n+\frac{1}{2}}^{\,\ast} Q_{n} =O, 
\nonumber
\end{align}
is obtained by setting \mbox{$b=a^\ast$}, 
\mbox{$\widebar{N}=N$} and 
\[
\widebar{\mu}_j= \frac{1}{\mu_j^\ast}, \hspace{5mm}
\widebar{C}_j(t)  \widebar{\mu}_j^{\hspace{1pt}-2} 
	= \mathrm{i} \left\{ C_j (t) \mu_j^{-\frac{1}{2}} \right\}^\ast, 
\hspace{5mm} j=1, 2, \ldots, N
\]
in 
formula 
(\ref{Q-Nsol2}) with (\ref{C_time}). 
The imaginary unit (roman $\mathrm{i}$) 
should not be 
confused with the index of summation 
(italic $i$) in the definition of $\mathscr{U}_{jk}$.  
The reduced set of scattering data 
is required to satisfy 
the condition (\ref{sca-con2}) 
(or 
(\ref{sca-con3})) 
and its 
smaller versions 
corresponding to 
subsets of the 
solitons. 

\subsection{Solutions of the space-discrete  
Kaup--Newell system}
\label{subs4.2}

In this subsection, we 
solve the space-discrete Kaup--Newell system 
(\ref{sdKN}) 
derived in subsection~\ref{subs2.3} by 
applying the results 
in section~\ref{sec3}.
Because the parameter $\mu
$ is 
nonessential 
in (\ref{sdKN}), 
we 
set 
\mbox{$\mu\hspace{1pt}(=z^2)
=1$} 
and consider 
the space-discrete Kaup--Newell system 
in the form: 
\begin{subnumcases}{\label{sdKN2}}
{}
 q_{n,t} - \boldsymbol{\Delta}_n^+ 
 \left[ a \left( I - q_{n}r_{n} \right)^{-1} q_{n}
 + b \left( I + q_{n-1} r_{n} \right)^{-1} q_{n-1} \right] = O,
\\[1mm]
 r_{n,t} - \boldsymbol{\Delta}_n^+ 
 \left[ b  \left( I + r_{n} q_{n-1} \right)^{-1} r_{n}
 + a \left( I - r_{n-1} q_{n-1} \right)^{-1} r_{n-1} \right] 
 = O. 
\hspace{15mm} 
\end{subnumcases}
Recall that 
$\boldsymbol{\Delta}_n^+$ 
denotes the forward difference operator: 
\mbox{$\boldsymbol{\Delta}_n^+ f_n := f_{n+1} - f_{n} $}. 
We assume 
vanishing boundary conditions at spatial infinity: 
\begin{equation}
\label{zero-bc3}
\lim_{n \to \pm \infty} q_n = \lim_{n \to \pm \infty} r_n =O.
\end{equation}
Propositions~\ref{prop1} and \ref{prop2}
imply 
that 
the 
solution of (\ref{sdKN2}) 
can be 
obtained 
as
\begin{equation}
q_n = \boldsymbol{\Delta}_n^+ \left. \hspace{-1pt}
\left( \Psi_{1,n}^{(1)\,-1} \Psi_{1,n}^{(2)} 
	\right)\right|_{\mu
=1},
 \hspace{5mm}
r_n = \left. \hspace{-1pt}
 \left( \Psi_{2,n}^{(1)\,-1} \Psi_{2,n}^{(2)} 
	- \Psi_{1,n}^{(1)\,-1} \Psi_{1,n}^{(2)} \right)^{-1}
 \right|_{\mu
=1},  
\label{qr-formu}
\end{equation}
using 
two 
linearly 
independent 
eigenfunctions 
of 
the linear problem, 
(\ref{mAL1}) and (\ref{mAL-time}), 
associated with 
the Ablowitz--Ladik lattice 
(cf.~(\ref{gen-Lax})); 
this is 
in contrast to the space-discrete Gerdjikov--Ivanov system, 
which can 
be 
derived 
using only one linear 
eigenfunction as described in subsection~\ref{subs2.2}.  
Proposition~\ref{prop3} implies 
that 
$r_n$ in (\ref{qr-formu}) 
can 
be 
rewritten 
in the difference form 
as $q_n$; 
to 
see 
this explicitly, 
we need to identify 
an appropriate 
linear problem 
for 
the Ablowitz--Ladik lattice, which is 
gauge 
equivalent 
to the original 
problem. 
%
It turns out that 
the gauge transformation 
connecting (AL1) 
and 
(AL2) considered 
in subsection~\ref{subs4.1} 
plays the desired role. 

Suppose 
that 
the two 
eigenfunctions 
appearing in (\ref{qr-formu}) 
satisfy 
(AL1).  
%
Moreover, 
we 
choose 
the 
linear 
eigenfunction appearing in (\ref{P-formu1}) 
as
\[
P_n = 
\left. \hspace{-1pt}
\Psi_{2,n}^{(1)} \Psi_{1,n}^{(1)\, -1} \right|_{\mu
\hspace{1pt}(=z^2)
=1}.
\]
Recall that (AL1) and (AL2) involve this $P_n$ and 
are connected through 
the gauge transformation (\ref{gauge-def1}). 
Thus, 
the two 
linear eigenfunctions 
for (AL2) can be introduced 
as 
\begin{align}
\left[
\begin{array}{c}
 \Phi_{1, n}^{(1)} \\
 \Phi_{2, n}^{(1)} \\
\end{array}
\right]
 := \frac{1}{z^{-2}-1}
g_n 
\left[
\begin{array}{c}
 \Psi_{1, n}^{(1)} \\
 \Psi_{2, n}^{(1)} \\
\end{array}
\right],
\hspace{5mm}
\left[
\begin{array}{c}
 \Phi_{1, n}^{(2)} \\
 \Phi_{2, n}^{(2)} \\
\end{array}
\right]
 := g_n 
\left[
\begin{array}{c}
 \Psi_{1, n}^{(2)} \\
 \Psi_{2, n}^{(2)} \\
\end{array}
\right], 
\label{redef-Jost}
\end{align}
so that both of them 
become 
nontrivial 
in the limit \mbox{$\mu\hspace{1pt}(=z^2) \to 1$}. 
Note that (\ref{ALsca2}) and 
(\ref{gauge-def1}) imply that 
\begin{align}
\Phi_{2,n}^{(1)} &= z^{-1} P_{n+1} 
 \Psi_{1,n+1}^{(1)} 
	+ z^{-1} \left( I-P_{n+1} Q_{n+1} \right) \Phi_{2,n+1}^{(1)},
\nonumber \\[0.5mm]
\Phi_{2,n}^{(2)} &= z^{-1} \left( z^{-2}-1 \right) P_{n+1} 
 \Psi_{1,n+1}^{(2)} 
	+ z^{-1} \left( I-P_{n+1} Q_{n+1} \right) \Phi_{2,n+1}^{(2)}.
\nonumber
\end{align}
%
%
Thus, 
the ratio 
between 
these quantities 
in the limit \mbox{$\mu\hspace{1pt}(=z^2) \to 1$} 
satisfies 
\begin{align}
& \left. \hspace{-1pt}
\Phi_{2,n-1}^{(2)\,-1} \Phi_{2,n-1}^{(1)} \right|_{\mu
\to 1}
\nonumber \\
=\; & \left. \hspace{-1pt}
\left[ \left( I-P_{n} Q_{n} \right) \Phi_{2,n}^{(2)} 
\right]^{-1}
\left[ P_{n} \Psi_{1,n}^{(1)} 
	+ \left( I-P_{n} Q_{n} \right) \Phi_{2,n}^{(1)}
\right] \right|_{\mu
\to 1}
\nonumber \\
=\; & \left. \hspace{-1pt}
 \left( \Psi_{2,n}^{(2)} -z^{-2} P_n \Psi_{1,n}^{(2)} \right)^{-1} 
 P_{n} \Psi_{1,n}^{(1)} 
+ 
\Phi_{2,n}^{(2)\,-1} \Phi_{2,n}^{(1)} \right|_{\mu
\to 1}
\nonumber \\
=\; & \left. \hspace{-1pt}
 \left( \Psi_{2,n}^{(1)\, -1} \Psi_{2,n}^{(2)} -  \Psi_{1,n}^{(1)\, -1}
 \Psi_{1,n}^{(2)} \right)^{-1} 
+ 
\Phi_{2,n}^{(2)\,-1} \Phi_{2,n}^{(1)} \right|_{\mu
\to 1}. 
\nonumber
\end{align}
Therefore, 
the formula for determining $r_n$ in 
(\ref{qr-formu}) can be replaced with 
\begin{equation}
r_n = - \boldsymbol{\Delta}_n^+ \left. \hspace{-1pt}
\left( \Phi_{2,n-1}^{(2)\,-1} \Phi_{2,n-1}^{(1)} \right) \right|_{\mu
\to 1},
\label{qr-formu2}
\end{equation}
which uses the 
two linear eigenfunctions for (AL2). 

%
We set  
\begin{align}
\left[
\begin{array}{c}
 \Psi_{1, n}^{(1)} \\
 \Psi_{2, n}^{(1)} \\
\end{array}
\right] &:=
\widebar{\psi}_n^{\hspace{1pt}\mathrm{(AL1)}}, 
\hspace{5mm}
\left[
\begin{array}{c}
 \Psi_{1, n}^{(2)} \\
 \Psi_{2, n}^{(2)} \\
\end{array}
\right] :=
 \psi_n^{\mathrm{(AL1)}}, 
\end{align}
so that (cf.~(\ref{redef-Jost}) and (\ref{Jost-AL-psi})) 
\begin{align}
\left[
\begin{array}{c}
 \Phi_{1, n}^{(1)} \\
 \Phi_{2, n}^{(1)} \\
\end{array}
\right] =
\widebar{\psi}_n^{\hspace{1pt}\mathrm{(AL2)}},
\hspace{5mm}
\left[
\begin{array}{c}
 \Phi_{1, n}^{(2)} \\
 \Phi_{2, n}^{(2)} \\
\end{array}
\right] &=
\psi_n^{\mathrm{(AL2)}}. 
\end{align}
In view of 
(\ref{leftJost}), 
(\ref{AL1-scat}) and (\ref{AL2-scat}), 
this choice is indeed 
consistent with 
the boundary conditions (\ref{zero-bc3}). 
To be 
precise, we should use 
the explicitly time-dependent 
Jost solutions introduced 
in subsection~\ref{sTDsd}, 
which satisfy 
not only the Ablowitz--Ladik eigenvalue problem (\ref{mAL0}) but also 
the time-evolution equation (\ref{mAL-time}); 
however, this 
makes no difference 
in the limit \mbox{$\mu \to 1$} (cf.~(\ref{Jost-time-psi})),
so the above choice is valid 
in formulas (\ref{qr-formu}) and (\ref{qr-formu2}). 
%
%
Thus, 
with the aid of (\ref{psi_form}), 
the solution of the space-discrete Kaup--Newell system 
(\ref{sdKN2}) can be expressed as 
\begin{subequations}
\label{qr-formu3}
\begin{align}
q_n &= 
\boldsymbol{\Delta}_n^+ \left\{
\left[ I+ \sum_{k=0}^{\infty} 
 \widebar{K}^{\hspace{1pt}\mathrm{(AL1)}}_1 (n,n+k) \right]^{-1}
\sum_{k=0}^{\infty} K^{\mathrm{(AL1)}}_1 (n, n+k)
\right\}, 
\\[2mm]
r_n &= - \boldsymbol{\Delta}_n^+ \left\{
\left[ I+ \sum_{k=0}^{\infty} K^{\mathrm{(AL2)}}_2 (n-1, n-1+k)
\right]^{-1} \sum_{k=0}^{\infty} 
\widebar{K}^{\hspace{1pt}\mathrm{(AL2)}}_2 (n-1,n-1+k)
\right\}.
\end{align}
\end{subequations}
Recall that the infinite 
sums in (\ref{qr-formu3}) 
are assumed to be convergent; 
this is satisfied 
if the potentials in (AL1) and (AL2) decay sufficiently 
rapidly as \mbox{$n \to \pm \infty$} (cf.~(\ref{psi_form})). 
To realize an exact linearization 
of the space-discrete Kaup--Newell system 
(\ref{sdKN2}), 
we introduce new quantities 
${\cal K}(n,m)$ and $\widebar{\cal K}(n,m)$ for 
\mbox{$m \ge n$} as
\begin{align}
{\cal K}(n,m) &:=
\left[ I+ \sum_{k=0}^{\infty} 
 \widebar{K}^{\hspace{1pt}\mathrm{(AL1)}}_1 (n,n+k) \right]^{-1}
\sum_{s=m}^{\infty} K^{\mathrm{(AL1)}}_1 (n, s), 
\nonumber \\[2mm]
\widebar{\cal K}(n,m) &:= 
- \left[ I+ \sum_{k=0}^{\infty} K^{\mathrm{(AL2)}}_2 (n-1, n-1+k)
\right]^{-1} \sum_{s=m-1}^{\infty} 
\widebar{K}^{\hspace{1pt}\mathrm{(AL2)}}_2 (n-1,s),
\nonumber
\end{align}
so that 
\mbox{$q_n = \boldsymbol{\Delta}_n^+ {\cal K}(n,n)$} 
and \mbox{$r_n=\boldsymbol{\Delta}_n^+ \widebar{\cal K}(n,n)$}. 
Let us derive the linear summation equation 
for ${\cal K}(n,m)$ 
from the Gel'fand--Levitan--Marchenko equations 
(\ref{GLM_1}) and (\ref{GLM_2}) for (AL1). 
From (\ref{GLM_1}), we have
\[
 \widebar{K}^{\hspace{1pt}\mathrm{(AL1)}}_1 (n,p) +
\sum_{j=0}^{\infty} K^{\mathrm{(AL1)}}_1 (n,n+j) 
 F^{\mathrm{(AL1)}}(p+j+1) = O,
\hspace{5mm} p \geq n.
\]
Thus, taking the sum with respect to $p$, 
we obtain the relation
\begin{align}
\sum_{p=n+k}^\infty
\widebar{K}^{\hspace{1pt}\mathrm{(AL1)}}_1 (n,p) 
&=- \sum_{j=0}^{\infty} \left[
\sum_{s=n+j}^\infty K^{\mathrm{(AL1)}}_1 (n,s) -
\sum_{s=n+j+1}^\infty K^{\mathrm{(AL1)}}_1 (n,s)
\right] 
\nonumber \\
& \hphantom{=} \; 
\mbox{}\times 
\sum_{p=n+k}^\infty
 F^{\mathrm{(AL1)}}(p+j+1). 
\label{4.22}
\end{align}
From (\ref{GLM_2}), we have
\begin{align}
 K^{\mathrm{(AL1)}}_1 (n,s) 
&= \widebar{F}^{\hspace{1pt}\mathrm{(AL1)}}(s) 
 + \sum_{k=0}^{\infty} \widebar{K}^{\hspace{1pt}\mathrm{(AL1)}}_1 (n,n+k) 
 \widebar{F}^{\hspace{1pt}\mathrm{(AL1)}}(s)
\nonumber \\ & 
\hphantom{=}\; 
- \sum_{k=0}^{\infty} \left[ \sum_{p=n+k}^\infty
\widebar{K}^{\hspace{1pt}\mathrm{(AL1)}}_1 (n,p) 
\right] \left[
 \widebar{F}^{\hspace{1pt}\mathrm{(AL1)}}(s+k) 
 - \widebar{F}^{\hspace{1pt}\mathrm{(AL1)}}(s+k+1)
\right], 
\hspace{5mm} s \geq n,
\nonumber
\end{align}
where (a variant of) 
the summation by parts formula 
is 
used. 
Thus, 
taking the sum with respect to $s$ 
and 
using the fact that 
\mbox{$\widebar{F}^{\hspace{1pt}\mathrm{(AL1)}}(n)$} 
decays rapidly as \mbox{$n \to +\infty$}, 
we obtain 
\begin{align}
\sum_{s=m}^\infty 
 K^{\mathrm{(AL1)}}_1 (n,s) & = 
\left[ I + \sum_{k=0}^{\infty} 
 \widebar{K}^{\hspace{1pt}\mathrm{(AL1)}}_1 (n,n+k) \right] 
\sum_{s=m}^\infty  \widebar{F}^{\hspace{1pt}\mathrm{(AL1)}}(s)
\nonumber \\
& \hphantom{=}\; 
- \sum_{k=0}^{\infty} \left[ \sum_{p=n+k}^\infty
\widebar{K}^{\hspace{1pt}\mathrm{(AL1)}}_1 (n,p) 
\right] 
 \widebar{F}^{\hspace{1pt}\mathrm{(AL1)}}(m+k), 
\hspace{5mm} m \geq n.
\nonumber
\end{align}
Substituting (\ref{4.22}) 
into the last term 
and 
multiplying both sides 
from the left by 
\mbox{$\left[ 
I + \sum_{k=0}^{\infty} 
 \widebar{K}^{\hspace{1pt}\mathrm{(AL1)}}_1 (n,n+k) 
\right]^{-1}$}, 
we arrive at 
the linear summation equation 
for ${\cal K}(n,m)$:  
\begin{align}
{\cal K} (n,m) & = \sum_{s=m}^{\infty
} \widebar{F}^{\hspace{1pt}\mathrm{(AL1)}}(s) 
+ \sum_{j=0}^{\infty} \sum_{k=0}^{\infty}
\left[
{\cal K}(n,n+j) - {\cal K}(n,n+j+1)\right] 
\nonumber \\
& \hphantom{=}\; \mbox{}\times
\sum_{p=n+k}^\infty F^{\mathrm{(AL1)}}(p+j+1)
\widebar{F}^{\hspace{1pt}\mathrm{(AL1)}}(m+k), \hspace{5mm} m \geq n,
\nonumber 
\end{align}
which can be 
rewritten using (\ref{F-GI}) as 
\begin{align}
{\cal K} (n,m) = \sum_{s=m}^{\infty} \widebar{F}(s) 
+ \sum_{j=0}^{\infty} \sum_{k=0}^{\infty}
\left[
{\cal K}(n,n+j) - {\cal K}(n,n+j+1)\right] F(n+j+k) \widebar{F}(m+k), 
\hspace{5mm} m \geq n. 
\nonumber
\end{align}

In a similar way, we can 
derive the linear summation equation 
for $\widebar{\cal K}(n,m)$ 
from the Gel'fand--Levitan--Marchenko equations 
(\ref{GLM_1}) and (\ref{GLM_2}) for (AL2) as 
%
%
%
\begin{align}
\widebar{\cal K} (n,m)
&= \sum_{s=m-1}^{\infty
} F^{\mathrm{(AL2)}}(s) +
\sum_{j=0}^{\infty}
\sum_{k=0}^{\infty}
\left[ \widebar{\cal K}(n,n+j) - \widebar{\cal K}(n,n+j+1) 
\right] 
\nonumber \\
& \hphantom{=}\; \mbox{}\times
\sum_{p=n+k-1}^\infty
\widebar{F}^{\hspace{1pt}\mathrm{(AL2)}}(p+j+1) F^{\mathrm{(AL2)}}(m+k-1)
\nonumber \\[2mm]
&= 
\sum_{s=m-1}^{\infty
} F(s) -
\sum_{j=0}^{\infty}
\sum_{k=0}^{\infty}
\left[ \widebar{\cal K}(n,n+j) - \widebar{\cal K}(n,n+j+1) 
\right] \widebar{F}(n+j+k) F(m+k-1), 
\nonumber \\
& \hspace{110mm} m \geq n,
\nonumber
\end{align}
where (\ref{F-GI}) is used. 
Combining the above results, 
we obtain 
a set of 
formulas for 
the solutions of 
the space-discrete 
Kaup--Newell 
system (\ref{sdKN2}), 
which tend to zero as \mbox{$n \to + \infty$}, 
in the 
form~\cite{TsuJMP10}:
\begin{subequations}
\label{sdKNsol}
\begin{align}
q_n 
&=
\boldsymbol{\Delta}_n^+ {\cal K}(n,n
),
\label{KN-lin1}
\\[4pt]
r_n 
&=
\boldsymbol{\Delta}_n^+ {\cal \widebar{K}}(n,n
),
\label{KN-lin2}
\\[4pt]
{\cal K} (n,m) 
&= 
 {\cal \widebar{F}} (m) +\sum_{j=0}^{\infty} \sum_{k=0}^{\infty}
\left[ {\cal K}(n,n+j) - {\cal K}(n,n+j+1)\right]
\nonumber \\
& \hphantom{=} \;
\mbox{}\times 
\left[ {\cal F} (n+j+k+1) - {\cal F}(n+j+k+2) \right]
\left[ {\cal \widebar{F}}(m+k) - {\cal \widebar{F}}(m+k+1) \right],
\hspace{5mm} m \geq n,
\label{calK-KN}
\\[4pt]
{\cal \widebar{K}} (n,m)
&=
 {\cal F}(m) -
\sum_{j=0}^{\infty}
\sum_{k=0}^{\infty}
\left[ {\cal \widebar{K}}(n,n+j) - {\cal \widebar{K}}(n,n+j+1) \right]
\nonumber \\
& \hphantom{=} \;
\mbox{} \times
\left[ {\cal \widebar{F}}(n+j+k) -{\cal \widebar{F}}(n+j+k+1) \right]
\left[ {\cal F}(m+k) - {\cal F}(m+k+1) \right],
\hspace{5mm} m \geq n.
\label{calKbar-KN}
\end{align}
\end{subequations}
Here, the functions ${\cal F}(n)$ and ${\cal \widebar{F}}(n)$  
are defined as \mbox{${\cal F}(n) :=\sum_{s=n-1}^\infty F(s)$} and
\mbox{${\cal \widebar{F}} (n) :=\sum_{s=n}^\infty \widebar{F}(s)$}, 
respectively; 
they 
satisfy the same 
linear 
evolutionary system as 
$F(n)$ and $\widebar{F}(n)$ 
(cf.~(\ref{AL-linear}))
and decay rapidly as \mbox{$n \to + \infty$}. 
Note that 
the set of formulas (\ref{sdKNsol}) 
can 
provide the solutions 
for 
any flow 
of the space-discrete 
Kaup--Newell hierarchy 
if 
${\cal F}(n)$ and ${\cal \widebar{F}}(n)$ 
satisfy 
the 
corresponding linear 
system. 

In the same way as for the 
other 
lattice 
systems, 
we can 
derive 
the multisoliton solutions of
the space-discrete 
Kaup--Newell system (\ref{sdKN2}) 
from the set of formulas 
(\ref{sdKNsol}); 
this corresponds to 
the special case of 
reflectionless potentials 
for 
both (AL1) and (AL2).
For simplicity, 
we
assume 
that $A(\mu)^{-1}$ and $\widebar{A}(\mu)^{-1}$ 
only have simple poles
and set 
\begin{align}
{\cal F}(n, t) = \sum_{j=1}^N 
{\cal C}_j(t) \mu_j^{n}, 
\hspace{7mm}
{\cal \widebar{F}}(n,t) 
= \sum_{j=1}^{\widebar{N}} \widebar{\cal C}_j(t) 
	\widebar{\mu}_j^{\hspace{1pt}-n}, 
\label{F-Fbar-KN}
\end{align}
where 
the time dependences of ${\cal C}_j$ and 
$\widebar{\cal C}_j$ 
are 
given as 
(\ref{C_time}) and (\ref{C_bar_time}), i.e. 
\[
{\cal C}_j(t) = {\cal C}_j(0)
\mathrm{e}^{[(\mu_j-1) b + (1-\mu_j^{-1})a] t}, \hspace{5mm} 
\widebar{\cal C}_j(t) = \widebar{\cal C}_j(0)
\mathrm{e}^{-[(\widebar{\mu}_j-1) b 
 + (1-\widebar{\mu}_j^{\hspace{1pt}-1}) a] t}.
\]
We also 
set 
\begin{equation}
{\cal K}(n, m; t) = \sum_{j=1}^{\widebar{N}} {\cal G}_j(n,t) 
	\widebar{\mu}_j^{\hspace{1pt}-m}, 
\hspace{7mm}
 \widebar{\cal K} (n, m; t) = \sum_{j=1}^{N} {\cal H}_j(n,t) 
 \mu_j^{m}, 
\label{G-H-KN}
\end{equation}
and substitute 
the expressions (\ref{F-Fbar-KN}) and (\ref{G-H-KN}) 
into (\ref{calK-KN}) and (\ref{calKbar-KN}). 
Because 
\mbox{$|\mu_j| < 1$} 
$(j=1, 2, \ldots, N)$ and \mbox{$|\widebar{\mu}_j| >1$}
$(j=1, 2, \ldots, \widebar{N})$, 
we can 
evaluate the infinite sum 
to 
obtain 
linear algebraic systems for 
determining 
${\cal G}_j$ and 
${\cal H}_j$, respectively. 
They 
can be 
written as 
\begin{subequations}
\label{GH-sys-KN}
\begin{align}
& \left[
\begin{array}{cccc}
\! {\cal G}_1 \hspace{1pt} \widebar{\mu}_1^{\hspace{1pt}-n} \! 
 & \! {\cal G}_2 \hspace{1pt} \widebar{\mu}_2^{\hspace{1pt}-n} \! & \! 
\cdots \! & \! {\cal G}_{\widebar{N}} \hspace{1pt} 
\widebar{\mu}_{\widebar{N}}^{\hspace{1pt}-n}\! 
\end{array}
\right]
\left[
\begin{array}{ccc}
 {\cal U}_{11} & \cdots & {\cal U}_{1\widebar{N}} \\
 \vdots &  \ddots & \vdots \\
 {\cal U}_{\widebar{N}1} & \cdots & {\cal U}_{\widebar{N}\widebar{N}} \\
\end{array}
\right]
= \left[
\begin{array}{cccc}
\! \widebar{\cal C}_1 \hspace{1pt} \widebar{\mu}_1^{\hspace{1pt}-n}
 \! & \! \widebar{\cal C}_2 \hspace{1pt} 
	\widebar{\mu}_2^{\hspace{1pt}-n}\! 
	& \! \cdots \! & \! \widebar{\cal C}_{\widebar{N}}\hspace{1pt} 
\widebar{\mu}_{\widebar{N}}^{\hspace{1pt}-n} \! 
\end{array}
\right]
\label{GH-sys1-KN}
\end{align}
and 
\begin{align}
&
\left[
\begin{array}{cccc}
\! {\cal H}_1 \hspace{1pt} \mu_1^{n}
\! & \! {\cal H}_2 \hspace{1pt} \mu_2^{n} 
\! & \! \cdots \! & \! {\cal H}_{N} \hspace{1pt} \mu_N^{n} \! 
\end{array}
\right]
\left[
\begin{array}{ccc}
 {\cal V}_{11} & \cdots & {\cal V}_{1N} \\
 \vdots &  \ddots & \vdots \\
 {\cal V}_{N1} & \cdots & {\cal V}_{NN} \\
\end{array}
\right]
= \left[
\begin{array}{cccc}
\! {\cal C}_1 \hspace{1pt} \mu_1^{n}
\! & \! {\cal C}_2 \hspace{1pt} \mu_2^{n} 
\! & \! \cdots \! & \! {\cal C}_{N} \hspace{1pt} \mu_N^{n} \! 
\end{array}
\right].
\label{GH-sys2-KN}
\end{align}
\end{subequations}
%
Here, all the entries in 
(\ref{GH-sys-KN}) 
are \mbox{$l \times l$} matrices;
the block matrices 
\mbox{${\cal U}=({\cal U}_{jk})_{1 \le j,k \le \widebar{N}}$} and 
\mbox{${\cal V}=({\cal V}_{jk})_{1 \le j,k \le N}$} are 
defined as 
\begin{align}
{\cal U}_{jk} &:= \delta_{jk} I - \sum_{i=1}^{N} 
\frac{\displaystyle \left( 1-\widebar{\mu}_j^{\hspace{1pt}-1}\right) \left(1-\mu_i\right)
 \left(1-\widebar{\mu}_k^{\hspace{1pt}-1} \right) }
{\displaystyle \left( 1-\frac{\mu_i}{\widebar{\mu}_j}\right) 
 \left( 1-\frac{\mu_i}{\widebar{\mu}_k} \right)} 
\mu_i^{n+1} \widebar{\mu}_k^{\hspace{1pt}-n}\hspace{1pt}
{\cal C}_i (t) \hspace{1pt} \widebar{\cal C}_k(t) 
\nonumber 
\end{align}
and
\begin{align}
{\cal V}_{jk} &:= \delta_{jk} I + \sum_{i=1}^{\widebar{N}} 
\frac{\left( 1- \mu_j\right) \left(1-\widebar{\mu}_i^{\hspace{1pt}-1} \right) 
	\left(1-\mu_k\right)}
{\displaystyle \left( 1-\frac{\mu_j}{\widebar{\mu}_i}\right) 
 \left( 1- \frac{\mu_k}{\widebar{\mu}_i} \right)} 
 \widebar{\mu}_i^{\hspace{1pt} -n}  \mu_k^{n} \hspace{1pt}
 \widebar{\cal C}_i(t) \hspace{1pt} {\cal C}_k(t), 
 \nonumber
\end{align}
respectively. 
Thus, 
with the aid of 
(\ref{KN-lin1}),
(\ref{KN-lin2}) and (\ref{G-H-KN}),  
we obtain 
the 
multisoliton solutions of the 
space-discrete 
Kaup--Newell 
system (\ref{sdKN2}) 
in the difference 
form:
\begin{subequations}
\label{KN-soliton}
\begin{align}
q_n (t) 
&= \boldsymbol{\Delta}_n^+ \left\{ 
\left[
\begin{array}{ccc}
\! \widebar{\cal C}_1(t) \hspace{1pt}\widebar{\mu}_1^{\hspace{1pt}-n}
\! & \! \cdots \! & \! 
\widebar{\cal C}_{\widebar{N}}(t) \hspace{1pt}
\widebar{\mu}_{\widebar{N}}^{\hspace{1pt}-n} \! 
\end{array}
\right] 
\left[
\begin{array}{ccc}
 {\cal U}_{11} & \cdots & {\cal U}_{1\widebar{N}} \\
 \vdots &  \ddots & \vdots \\
 {\cal U}_{\widebar{N}1} & \cdots & {\cal U}_{\widebar{N}\widebar{N}} \\
\end{array}
\right]^{-1} 
\left[
\begin{array}{c}
 I \\
 \vdots \\
 I \\ 
\end{array}
\right] \right\}, 
\label{Q-Nsol-KN}
\\[1mm]
r_n (t) 
&= \boldsymbol{\Delta}_n^+ \left\{ 
\left[
\begin{array}{ccc}
\! {\cal C}_1 (t) \hspace{1pt} \mu_1^{n} \! & \! 
\cdots \! & \! {\cal C}_{N} (t) \hspace{1pt} \mu_N^{n} \! 
\end{array}
\right]
\left[
\begin{array}{ccc}
 {\cal V}_{11} & \cdots & {\cal V}_{1N} \\
 \vdots &  \ddots & \vdots \\
 {\cal V}_{N1} & \cdots & {\cal V}_{NN} \\
\end{array}
\right]^{-1}
\left[
\begin{array}{c}
 I \\
 \vdots \\
 I \\ 
\end{array}
\right] \right\}. 
\end{align}
\end{subequations}
%

When \mbox{$b=a^\ast$}, 
the space-discrete 
Kaup--Newell system (\ref{sdKN2})
allows not only 
the complex conjugation reduction
\mbox{$r_n = \mathrm{i} \sigma q_{n-
1/2
}^{\,\ast}$} 
but also 
the Hermitian conjugation reduction \mbox{$r_n = \mathrm{i} 
\sigma q_{n-
1/2
}^{\,\dagger}$}, 
where 
$\sigma$ is a real constant~\cite{Tsuchi02}. 
Each reduction relates 
two originally 
uncoupled 
systems, 
(\ref{sdKN2}) with \mbox{$n \in \mathbb{Z}$}
and (\ref{sdKN2}) with \mbox{$n \in \mathbb{Z}+1/2$}, 
to provide 
a single equation
with 
\mbox{$n \in \mathbb{Z}/2$}. 
Clearly, 
the value of $\sigma$ is nonessential, so 
we set
\mbox{$\sigma=1$} 
and consider the Hermitian conjugation
reduction \mbox{$r_n = \mathrm{i} q_{n-
1/2
}^{\,\dagger}$}. 
This reduction 
can be realized 
at the level of 
formulas (\ref{sdKNsol}) 
by setting
\begin{equation}
{\cal \widebar{F}} (n) 
 = \mathrm{i} 
\left\{ {\cal F} 
 \left( n+\mbox{$\tiny{\frac{1}{2}}$} \right) \right\}^\dagger, 
\nonumber 
\end{equation}
which is 
consistent with 
the 
time evolution 
(cf.~(\ref{AL-linear}) with \mbox{$b=a^\ast$}). 
In particular, the 
$N$-soliton solution 
of the space-discrete 
Kaup--Newell equation, 
\begin{equation}
 q_{n,t} - \boldsymbol{\Delta}_n^+ 
 \left[ a \left( I -  \mathrm{i}q_{n} q_{n-\frac{1}{2}}^{\,\dagger}
 \right)^{-1} q_{n}
 + a^\ast \left( I +  \mathrm{i}q_{n-1} q_{n-\frac{1}{2}}^{\,\dagger}
 \right)^{-1} q_{n-1} \right] = O,
\label{matrixKNeq}
\end{equation}
is obtained by setting \mbox{$b=a^\ast$}, 
\mbox{$\widebar{N}=N$} and 
\[
\widebar{\mu}_j= \frac{1}{\mu_j^\ast}, \hspace{5mm}
\widebar{\cal C}_j(t)  
	= \mathrm{i} \left\{ {\cal C}_j (t) 
	\mu_j^{\frac{1}{2}} \right\}^\dagger, 
\hspace{5mm} j=1, 2, \ldots, N
\]
in 
formula 
(\ref{Q-Nsol-KN}), where 
\mbox{${\cal C}_j(t) = {\cal C}_j(0)
\mathrm{e}^{[(\mu_j-1) a^\ast + (1-\mu_j^{-1})a] t}$}. 
The imaginary unit (roman $\mathrm{i}$) 
should not be 
confused with the index of summation 
(italic $i$) in the definitions of ${\cal U}_{jk}$ 
and ${\cal V}_{jk}$.  
The $N$-soliton solution thus obtained 
is a space-discrete analog 
of the $N$-soliton solution 
of the continuous 
matrix Kaup--Newell equation
reported 
in~\cite{talk08}. 
Note that 
the square 
matrix 
equation (\ref{matrixKNeq}) 
can be 
further 
reduced 
to a vector equation 
by setting all but one of the columns (or rows) 
of $q_n$ to zero; 
its $N$-soliton solution can
be obtained 
in the same way as described in~\cite{Tsuchi04}. 

\section{Concluding remarks}

In this paper, 
we have developed 
the inverse scattering method 
associated with 
the matrix Ablowitz--Ladik 
eigenvalue problem 
and its applications to 
space-discrete analogs of 
derivative NLS systems. 
In particular, 
the most streamlined version of 
the inverse scattering method on a lattice 
is formulated, 
which 
can avoid redundant processes
present in the existing literature. 
Thus, 
we 
are now able to understand the 
inverse scattering method for the Ablowitz--Ladik lattice 
as a
direct discrete 
analog of 
the inverse scattering method 
for the 
continuous 
NLS system;   
%
in essence, 
the discrete case 
is no longer more complicated 
than the continuous case. 
Moreover, we can characterize 
the space-discrete 
derivative NLS 
systems 
using the potentials 
and linear eigenfunctions 
appearing 
in the Lax representation for 
the Ablowitz--Ladik 
lattice. 
On the basis of this characterization, 
we can solve 
the space-discrete 
derivative NLS 
systems 
by 
preparing 
two 
relevant 
copies
of the 
inverse scattering formulas for the Ablowitz--Ladik lattice
and 
considering 
a 
B\"acklund--Darboux transformation 
between them. 
This provides 
a unification of 
the inverse scattering method 
for the NLS system and that for the 
derivative NLS systems in the 
discrete setting; 
such a unification is also possible 
in 
the continuous case (see~\cite{talk07,talk08}). 
The 
multisoliton solutions 
of the space-discrete 
derivative NLS systems 
can be obtained 
in a straightforward manner
within this unified framework; 
they 
reduce to 
the multisoliton solutions 
of the 
derivative NLS systems 
in the continuous space 
limit. 
%
Note that the space-discrete Kaup--Newell system 
allows the introduction of the potential 
variables with respect to the discrete spatial coordinate
and can be 
rewritten locally 
in terms of these 
variables; our solution formulas 
reflect this property accurately, 
that is, by construction 
any solution 
is written 
in the difference form using the forward difference operator.  

We assumed 
vanishing 
boundary conditions 
at spatial infinity, 
namely, 
as 
\mbox{$n \to + \infty$} 
and 
\mbox{$n \to - \infty$}.  
In the case of matrix-valued dependent variables,
this assumption 
imposes 
highly 
nontrivial 
conditions 
on 
the scattering data, 
which 
become almost 
trivial in the scalar case 
(cf.~(\ref{sca-con1})). 
For simplicity, we derived such conditions 
in the 
reflectionless 
case of 
the 
potentials; 
however, 
they are expected to be 
valid in the 
general 
case, 
because 
in the limits \mbox{$t \to \pm \infty$} 
the 
contribution of
the continuous 
spectrum 
would 
become negligible (cf.~(\ref{F_time1}) and 
(\ref{Fbar_time1}) with \mbox{$b=a^\ast$}). 
Note also that 
our 
approach 
of solving the 
derivative NLS systems 
using the 
NLS 
eigenvalue problem 
is 
applicable 
under 
other 
boundary conditions 
that 
are 
amenable to 
the inverse scattering method 
or its 
generalizations. 
In addition, although 
we mainly 
considered 
the first nontrivial 
flows of the 
integrable hierarchies, 
our approach 
can be applied, 
with minor amendments, 
to 
the higher 
flows 
as well as 
the 
negative flows of the hierarchies. 
%

%
%
%

\appendix
\section
{Reduction to the vector modified Volterra lattice}
\label{app2}

In this appendix, we 
consider
the reduction of 
the matrix Ablowitz--Ladik 
lattice
(\ref{mALsys}) 
to the vector modified Volterra lattice: 
\begin{equation}
\vt{q}_{n,t} = \bigl(1+ \sca{\vt{q}_n}{\vt{q}_n}\bigr)
(\vt{q}_{n+1} -\vt{q}_{n-1}). 
\label{vector-mV}
\end{equation}
Here, $\vt{q}_n$
is a row vector 
of 
arbitrary dimension 
and 
\mbox{$\sca{\,\cdot\,}{\,\cdot\,}$} 
stands 
for the scalar product. 
The scalar (i.e., one-component) 
modified Volterra lattice was introduced 
by Hirota~\cite{Hiro73}
and the two-component case 
was studied 
by Ablowitz and Ladik~\cite{AL76}. 
The vector generalization (\ref{vector-mV}) 
was recognized as an integrable system 
in the late 1990s (see, {\it e.g.}, 
references in~\cite{Adler08});
it can also be  
considered as 
the simplest 
space-discrete analog of 
the vector modified KdV equation~\cite{Svi93,SviSok94}: 
\begin{align}
& \vt{q}_{t} =
\vt{q}_{xxx} + 6 \sca{\vt{q}}{\vt{q}}\vt{q}_x. 
\label{vmKdV1}
\end{align}

As was shown in our previous paper~\cite{TUW98} 
(also see~\cite{TUW99}), 
the inverse scattering 
method 
formulated 
for the matrix Ablowitz--Ladik 
lattice
(\ref{mALsys}) can be 
applied to 
the vector modified Volterra lattice (\ref{vector-mV}) 
by imposing suitable 
reduction conditions 
on the 
scattering data. 
Here, we revisit this problem and 
derive the reduction conditions 
in a more convincing 
manner.

We introduce a set of \mbox{$2^{M-1} \times 2^{M-1}$} matrices 
\mbox{$\{ e_1, e_2, 
\ldots, e_{2M-1} \}$} 
that are linearly independent and satisfy 
the anticommutation relations:
\begin{equation}
\{ e_j , e_k \}_+
:=e_j e_k + e_k e_j = -2 \delta_{jk} I. 
\label{a-commute}
\end{equation}
Note that 
they 
are 
the 
generators of 
the Clifford algebra. 
We set the matrix-valued 
dependent variables 
$Q_n$ and $R_n$ 
as 
\begin{subequations}
\label{reduct}
\begin{align}
Q_n &= q_n^{(1)} I + \sum_{j=1}^{2M-1} q_n^{(j+1)} e_j, 
\label{reduct-Q}
\\[1mm]
R_n &= -q_n^{(1)} I+ \sum_{j=1}^{2M-1} q_n^{(j+1)} e_j. 
\label{reduct-R}
\end{align}
\end{subequations}
%
Then,
because of 
(\ref{a-commute}), 
they 
satisfy 
the important relation: 
\[
Q_n R_n = R_n Q_n = - \sca{\vt{q}_n}{\vt{q}_n} I. 
\]
Here, 
\mbox{$\vt{q}_n = ( q_n^{(1)}, \ldots, q_n^{(2M)} )$}.  
Using this relation, 
it is easy to see that 
the choice (\ref{reduct}) reduces 
the matrix Ablowitz--Ladik 
lattice
(\ref{mALsys}) with \mbox{$a=b=1$}
to the vector modified Volterra lattice (\ref{vector-mV}). 
Lax representations involving the 
generators of
the Clifford algebra 
were 
introduced 
in the 
pioneering 
papers~\cite{EP79,KuSk81}, 
but our choice (\ref{reduct}) 
is more 
efficient and useful 
because 
it 
contains 
the unit matrix $I$. 
As a natural 
analog 
of 
the complex 
conjugate, 
we 
define 
the Clifford 
conjugate 
for the linear span of 
\mbox{$\{ I, e_1, e_2, \ldots, e_{2M-1} \}$} 
as 
\[
\widehat{Q}_n = q_n^{(1)} I - \sum_{j=1}^{2M-1} q_n^{(j+1)} e_j,
\quad 
\widehat{R}_n = -q_n^{(1)} I - \sum_{j=1}^{2M-1} q_n^{(j+1)} e_j.
\]
Note that \mbox{$\widehat{\widehat{Q}
}_n = Q_n$}, 
\mbox{$R_n = -\widehat{Q}_n$}, 
\mbox{$Q_n \widehat{Q}_n = \widehat{Q}_n Q_n 
 = \sca{\vt{q}_n}{\vt{q}_n} I$}, etc. 
In short, the Clifford 
conjugate 
denoted 
by 
\mbox{$\widehat{\;\;}$} changes the sign of 
the coefficients of \mbox{$\{e_1, e_2, \ldots, e_{2M-1} \}$}. 
This definition of  
the 
Clifford conjugate is very useful 
in the following 
discussion. 

Let us discuss 
how the reduction (\ref{reduct}) constrains 
the 
scattering data for the 
matrix 
Ablowitz--Ladik 
lattice. 
%
The main role is played by the quantity 
\mbox{$P_n = \Psi_{2,n} \Psi_{1,n}^{-1}$} 
defined from 
the Lax representation, 
(\ref{mAL1}) and (\ref{mAL-time}), which
satisfies 
the pair of discrete and continuous 
matrix Riccati equations 
(\ref{AL-R}). 
We can show in an inductive manner that
under the reduction 
(\ref{reduct}), 
$P_n$ 
also 
takes its values in 
the linear span of 
\mbox{$\{ I, e_1, e_2, \ldots, e_{2M-1} \}$}. 
We assume that the expression, 
\begin{equation}
P_n = -p_n^{(1)} I + \sum_{j=1}^{2M-1} p_n^{(j+1)} e_j,
\label{P-exp}
\end{equation}
is valid at 
some 
value of $n$ 
and write 
its coefficients 
as \mbox{$
\vt{p}_n := 
( p_n^{(1)}, \ldots, p_n^{(2M)} )
$}.  
Then, 
noting 
the 
relations 
derivable from 
the anticommutation relations (\ref{a-commute}), 
\begin{align}
& P_{n} \widehat{P}_{n} = \widehat{P}_{n} P_{n} 
	= \sca{\vt{p}_n}{\vt{p}_n} I, 
\nonumber 
\\[1mm]
& \widehat{Q}_n \widehat{P}_{n} + P_{n}Q_n 
 =  -2 \hspace{1pt} \sca{\vt{p}_n}{\vt{q}_n} I,
\label{conj-scalar}
\\[1mm]
& \widehat{Q}_n \widehat{P}_{n} \widehat{Q}_n =
\left( \widehat{Q}_n \widehat{P}_{n} + P_{n}Q_n \right) \widehat{Q}_n
- P_{n}Q_n \widehat{Q}_n 
= -2 \hspace{1pt} \sca{\vt{p}_n}{\vt{q}_n} \widehat{Q}_n 
  -\sca{\vt{q}_n}{\vt{q}_n} P_{n},
\nonumber 
\\[1mm]
& \left( I - \mu \widehat{Q}_n \widehat{P}_{n} \right) 
\left( I - \mu P_{n}Q_n \right) 
= \left( 1+ 2\mu \sca{\vt{p}_n}{\vt{q}_n} 
 + \mu^2 \sca{\vt{p}_n}{\vt{p}_n} \sca{\vt{q}_n}{\vt{q}_n} \right) I, 
\nonumber 
\end{align}
we can rewrite (\ref{AL-R1}) as 
\begin{align}
& P_{n+1} 
= \left( I - \mu P_{n}Q_n \right)^{-1} 
 \left( \mu P_{n} + \widehat{Q}_n \right)
\nonumber 
\\ &\! = \frac{1}{1+ 2\mu \sca{\vt{p}_n}{\vt{q}_n} 
 + \mu^2 \sca{\vt{p}_n}{\vt{p}_n} \sca{\vt{q}_n}{\vt{q}_n}}
\left( \mu P_{n} + \widehat{Q}_n 
 - \mu^2 \widehat{Q}_n \widehat{P}_{n} P_n
 - \mu \widehat{Q}_n \widehat{P}_{n} \widehat{Q}_n \right)
\nonumber 
\\ &\! = \frac{1}{1+ 2\mu \sca{\vt{p}_n}{\vt{q}_n} 
 + \mu^2 \sca{\vt{p}_n}{\vt{p}_n} \sca{\vt{q}_n}{\vt{q}_n}}
\left[ \mu \left( 1 + \sca{\vt{q}_n}{\vt{q}_n} \right) P_{n} 
 + \left( 1  +2 \mu \sca{\vt{p}_n}{\vt{q}_n}
 - \mu^2  \sca{\vt{p}_n}{\vt{p}_n} \right) \widehat{Q}_n 
\right]. 
\nonumber 
\end{align}
Thus, the expression (\ref{P-exp}) is also valid for $P_{n+1}$,  
and 
the coefficients 
satisfy the recursion relation 
written in the 
vector form:
\begin{align}
& \vt{p}_{n+1}
= \frac{ \mu \left( 1 + \sca{\vt{q}_n}{\vt{q}_n} \right) \vt{p}_n
 - \left( 1  +2 \mu \sca{\vt{p}_n}{\vt{q}_n}
 - \mu^2  \sca{\vt{p}_n}{\vt{p}_n} \right) \vt{q}_n}
 {1+ 2\mu \sca{\vt{p}_n}{\vt{q}_n} 
 + \mu^2 \sca{\vt{p}_n}{\vt{p}_n} \sca{\vt{q}_n}{\vt{q}_n}}. 
\nonumber 
\end{align}
In a similar way, 
we can show using (\ref{AL-R1}) 
that if 
the expression (\ref{P-exp}) is 
valid for $P_{n+1}$, 
then it is also valid for $P_n$.
Therefore, 
under a suitable boundary condition
on 
$P_n$, 
such as 
\mbox{$\lim_{n \to - \infty} P_n =O$}
or \mbox{$\lim_{n \to +\infty} P_n =O$}, 
so that the 
expression (\ref{P-exp}) is valid at the boundary, 
$P_n$ indeed 
takes its values in 
the linear span of 
\mbox{$\{ I, e_1, e_2, \ldots, e_{2M-1} \}$} 
for all \mbox{$n \in \mathbb{Z}$}.
Moreover, 
using (\ref{AL-R2}), 
we can easily check 
that 
this property 
is 
preserved under 
the time evolution. 

Substituting the expressions
\begin{align}
R_n &= a_n P_n + b_n P_{n+1}, 
\nonumber \\
Q_n &= -\widehat{R}_n = -a_n \widehat{P}_n - b_n \widehat{P}_{n+1}
\nonumber
\end{align}
into (\ref{AL-R1}), we can 
express 
the 
scalar coefficients 
$a_n$ and $b_n$ 
in terms of 
$\vt{p}_n$ and $\vt{p}_{n+1}$. 
Thus, 
$\vt{q}_n$ can be 
written 
explicitly 
using 
$\vt{p}_n$ and $\vt{p}_{n+1}$, which defines 
a Miura map to the vector modified Volterra lattice (\ref{vector-mV}).  
The corresponding modified system, 
i.e., 
the closed differential-difference 
equation 
for $\vt{p}_n$
can be obtained 
from (\ref{AL-R2}), but we do not present 
it here (cf.~\cite{Yang94,Adler08}). 

Because 
(\ref{AL-R1}) 
can be rewritten 
as 
\[
\mu^{-n-1} R_n = \mu^{-n} P_{n} - 
\mu^{-n-1} P_{n+1} + \left( \mu^{-n} P_{n} \right) 
\left( \mu^{n+1} Q_n \right) \left( \mu^{-n-1} P_{n+1} \right), 
\]
it is possible 
that 
\mbox{$ \mu^{-n} P_{n} 
$} 
for \mbox{$|\mu|=1$} 
has 
finite 
limits
as 
\mbox{$n \to \pm \infty$} 
under the 
decaying boundary 
conditions on $Q_n$ and $R_n$ 
(cf.~(\ref{zero-bc})). 
In particular, 
the above difference equation 
with the boundary condition 
\mbox{$\lim_{n \to - \infty} \mu^{-n}P_n =O$} 
defines 
the other boundary value 
\mbox{$\lim_{n \to + \infty} \mu^{-n}P_n$}, which 
belongs 
to the linear span of 
\mbox{$\{ I, e_1, e_2, \ldots, e_{2M-1} \}$}. 
The defining relation 
(\ref{phi_relation}) 
of the scattering data 
together 
with 
(\ref{leftJost}) 
implies that 
on the unit circle \mbox{$|\mu|=1$}, 
\begin{align}
\left[
\begin{array}{cc}
 z^n I & O \\
 O & z^{-n} I \\
\end{array}
\right] 
\phi_n & = 
\left[
\begin{array}{cc}
 I & O \\
 O & \mu^{-n} I \\
\end{array}
\right] z^n
\widebar{\psi}_n A + 
\left[
\begin{array}{cc}
 \mu^n I & O \\
 O & I \\
\end{array}
\right] z^{-n} 
\psi_n B
\nonumber \\[2mm]
& \to 
\left\{ 
\begin{array}{l}
\left[
\begin{array}{c}
 I \\
 O \\
\end{array}
\right]
\hspace{4mm}
 {\rm as}~~~ n \rightarrow -\infty, 
\vspace{2mm} 
\\
 \left[
\begin{array}{c}
 A (\mu) \\
 B(\mu) \\
\end{array}
\right]
\hspace{4mm}
 {\rm as}~~~ n \rightarrow +\infty.
\end{array} \right.
\nonumber 
\end{align}
%
Thus, 
if we set 
\mbox{$ \mu^{-n} P_{n}$} as  
\mbox{$ \mu^{-n} P_{n} 
=( z^{-n} \phi_{2,n} )( z^n \phi_{1,n} )^{-1}$} 
using the 
above linear eigenfunction, 
we 
find that 
\mbox{$\lim_{n \to + \infty} \mu^{-n}P_n 
 = B(\mu) A (\mu)^{-1}$} 
takes its values in 
the linear span of 
\mbox{$\{ I, e_1, e_2, \ldots, e_{2M-1} \}$}. 
%
More precisely, we should 
use 
the explicitly time-dependent 
Jost 
solution 
$\phi_n^{(t)}$ as introduced in 
(\ref{Jost-time-phi}),  
but this 
makes no difference 
in the above discussion. 
In addition, 
the established 
property of \mbox{$B(\mu) A (\mu)^{-1}$} 
is preserved 
under the time evolution. 

For \mbox{$|\mu|<1$}, 
we 
can 
still 
consider 
the 
difference equation 
(\ref{AL-R1}) for 
\mbox{$P_n 
$} 
with 
the boundary condition 
\mbox{$\lim_{n \to - \infty} 
P_n =O$} 
by setting \mbox{$P_n 
=( z^n \phi_{2,n} ) ( z^n \phi_{1,n} )^{-1}
$};
however, 
in this case 
the other boundary value 
\mbox{$\lim_{n \to + \infty} \mu^{-n}P_n$} 
does not
exist in general. 
Thus, we need to 
take 
a more delicate limit for 
$n$
and $\mu$
to 
extract 
meaningful information from $P_n$. 
Let us evaluate 
the \mbox{$n \to + \infty$} behavior of the Jost solution 
\mbox{$ [z^{n} \phi_n](\mu) 
$} 
in the neighborhood of 
\mbox{$\mu = \mu_j$}
(\mbox{$|\mu_j| < 1$})
at which 
$ A(\mu)^{-1}$ has a simple pole. 
Recalling the results 
in subsections~\ref{subs3.2} and \ref{GLM_eq}, 
we obtain that 
\[
 [z^{n} \phi_n](\mu) \times 
(\mu-\mu_j) A(\mu)^{-1} = 
\left[
\begin{array}{c}
 (\mu-\mu_j) \left\{ I + \mathrm{o}(1) \right\}
 + \mathrm{o} (\mu_j^{n}) \\
 (\mu-\mu_j) \hspace{2pt} \mathrm{o}(1) 
 + C_j \mu_j^{n} + \mathrm{o} (\mu_j^{n}) \\
\end{array}
\right].
\]
Thus, if we take 
the limit of 
\mbox{$n \to + \infty$} and \mbox{$\mu \to \mu_j$} 
while maintaining the 
balance 
condition 
\mbox{$
\left| \mu-\mu_j \right| \sim \left| \mu_j^{n} \right|
$}, 
we arrive at 
the 
desired formula:
\[
P_n 
= \frac{1}{\mu-\mu_j} C_j \mu_j^{n} + \mathrm{o}(1).
\]
Hence, 
$C_j$ 
also 
takes its values in 
the linear span of 
\mbox{$\{ I, e_1, e_2, \ldots, e_{2M-1} \}$}; 
naturally, 
this property is preserved 
under the time evolution. 

Combining the above results, 
we 
conclude
that $F(n, t)$ as given in 
(\ref{F_time1}) with \mbox{$a=b=1$} 
(or its generalization 
corresponding to the case of higher order poles of $A(\mu)^{-1}$) 
can be 
expressed in the form: 
\begin{align}
F(n, t) &= f^{(1)} (n, t) I - \sum_{j=1}^{2M-1} f^{(j+1)} (n, t) e_j. 
\label{F-Cliff}
\end{align}
%
We remark that 
the quantity 
\mbox{${\cal P}_n := \Psi_{1,n} \Psi_{2,n}^{-1}$} 
defined 
from 
another linear eigenfunction 
of 
the Ablowitz--Ladik eigenvalue problem 
(\ref{mAL1}) 
satisfies 
\[
Q_n = \mu^{-1} {\cal P}_{n} - {\cal P}_{n+1} 
 + \mu^{-1} {\cal P}_{n}R_n {\cal P}_{n+1},
\]
or equivalently, 
\[
{\cal P}_{n+1} = \left( I 
 - \mu^{-1} {\cal P}_{n} R_n \right)^{-1} 
 \left( \mu^{-1} {\cal P}_{n} - Q_n \right).
\]
%
Because $Q_n$ and $R_n$ 
in (\ref{reduct}) 
are related 
as
\mbox{$R_n = -\widehat{Q}_n$}, 
we can 
identify ${\cal P}_n$ 
with the Clifford conjugate of
$P_n$  
as
\[
{\cal P}_n (\mu) = -\widehat{P}_n (\mu^{-1}). 
\]
Here, 
we assume that the boundary conditions 
for ${\cal P}_n$ and $P_n$ are compatible 
in 
the above 
identification, 
{\it e.g.}, 
\mbox{$\lim_{n \to - \infty} 
{\cal P}_n = \lim_{n \to - \infty} P_n = O$}. 
Thus, if we set 
\mbox{${\cal P}_n =
( z^{-n} \widebar{\phi}_{1,n})(z^{-n} \widebar{\phi}_{2,n})^{-1}
$} 
using the Jost solution $\widebar{\phi}_n 
$ 
as introduced in (\ref{phi_bar}) 
(or 
the explicitly time-dependent one
$\widebar{\phi}_n^{\hspace{1pt}(t)}$ 
in (\ref{Jost-time-phi})) 
and compare it with \mbox{$P_n =
 ( z^n \phi_{2,n}) ( z^n \phi_{1,n} )^{-1}$}, 
we obtain 
the following relations: 
\begin{itemize}
\item 
\mbox{$\widebar{B}(\mu) \widebar{A} (\mu)^{-1}$} on
\mbox{$|\mu|=1$} 
is the Clifford conjugate of \mbox{$B(\mu^{-1}) A (\mu^{-1})^{-1}$},   
\item 
\mbox{$\displaystyle \widebar{\mu}_j= \frac{1}{\mu_j}, 
\hspace{3mm}
\widebar{C}_j  
 = - \frac{1}{\mu_j^2}
 \widehat{C}_j, 
\hspace{3mm} j=1, 2, \ldots, N ( =\widebar{N} )$}, 
up to renumbering. 
\end{itemize}
Therefore, $\widebar{F}(n, t)$ as given in 
(\ref{Fbar_time1}) with \mbox{$a=b=1$}
is 
equal to 
the Clifford conjugate of $F(n, t)$: 
\begin{align}
\widebar{F}(n, t) &= \widehat{F}(n, t)
\nonumber \\ 
&= f^{(1)} (n, t) I + \sum_{j=1}^{2M-1} f^{(j+1)} (n, t) e_j. 
\label{Fbar-Cliff}
\end{align}
%
Here, 
the vector 
\mbox{$\vt{f}_n (t) := ( f^{(1)}(n,t), \ldots, f^{(2M)}(n,t) )$} 
satisfies the linear evolution equation 
\mbox{$\vt{f}_{n,t}=  \vt{f}_{n+1} - \vt{f}_{n-1}$}
(cf.~(\ref{AL-linear})) 
and 
decays rapidly 
as \mbox{$n \to +\infty$}. 
%

In fact, 
(\ref{F-Cliff}) and (\ref{Fbar-Cliff}) 
provide 
not only 
necessary but also 
sufficient 
conditions 
for the corresponding 
potentials $Q_n$ and $R_n$ 
to be 
expressed 
as 
(\ref{reduct})
without using 
$e_j e_k$, 
$e_i e_j e_k$, etc. 
\begin{proposition}
\label{Prop.A.1}
If 
$F(n, t)$ and $\widebar{F}(n, t)$ 
are given 
as 
(\ref{F-Cliff}) and (\ref{Fbar-Cliff}), 
then 
the potentials $Q_n$ and $R_n$ 
reconstructed from 
the 
set of 
exact linearization formulas (\ref{ALlinearization}) 
can 
be expressed 
in the form 
(\ref{reduct}). 
\end{proposition}
%
Before 
proving 
Proposition~\ref{Prop.A.1}, 
we need to state one proposition and 
two lemmas. 
We 
first 
show that the set of 
formulas (\ref{ALlinearization}) 
realizes an exact linearization 
of the matrix Ablowitz--Ladik lattice. 
\begin{proposition}
\label{Prop.A.2}
Suppose that 
$F(n,t)$ and $\widebar{F}(n,t)$ 
satisfy 
the 
pair of 
linear evolution equations (\ref{AL-linear}) 
and decay sufficiently 
rapidly 
as \mbox{$n \to +\infty$}.  
Then, 
the Liouville--Neumann-type series 
for the 
potentials $Q_n$ and $R_n$
defined by
formulas (\ref{ALlinearization}) 
solve
the matrix Ablowitz--Ladik lattice 
(\ref{mALsys}) exactly. 
\end{proposition}

\noindent
{\it Remark.} 
Only 
the 
case 
\mbox{$a=b=1$}
is relevant to the vector modified Volterra 
lattice (\ref{vector-mV}), but 
we 
find it more 
convenient 
to 
prove 
Proposition~\ref{Prop.A.2} for 
general values 
of $a$ and $b$. 
\\
\\
{\it Proof of Proposition~\ref{Prop.A.2}.} 
Using (\ref{ALlinearization}), 
we obtain 
the Liouville--Neumann-type series for 
$Q_n$ and $R_n$
in the form: 
\begin{subequations}
\label{AL-LN}
\begin{align}
Q_n &= \widebar{F}(n) -\sum_{i_1,i_2=0}^{\infty} 
  \widebar{F}(n+i_1) F(n+i_1+i_2+1) \widebar{F}(n+i_2+1)
\nonumber \\ &
\mbox{}+ \sum_{i_1,i_2,i_3,i_4=0}^{\infty} 
 \widebar{F}(n+i_1) F(n+i_1+i_2+1) \widebar{F}(n+i_2+i_3+1)
 F(n+i_3+i_4+1) \widebar{F}(n+i_4+1)
\nonumber \\ & \mbox{}- \cdots 
\nonumber \\ &=
\sum_{k=0}^\infty (-1)^k
{\cal Q}_n^{(k)},
\label{Q-LN}
\\[2mm]
R_n & = - F(n) + \sum_{i_1,i_2=0}^{\infty} 
F(n+i_1) \widebar{F}(n+i_1+i_2+1) F(n+i_2+1)
\nonumber \\ & 
\mbox{}- \sum_{i_1,i_2,i_3,i_4=0}^{\infty} 
F(n+i_1) \widebar{F}(n+i_1+i_2+1) F(n+i_2+i_3+1) 
\widebar{F}(n+i_3+i_4+1) F(n+i_4+1)
\nonumber \\ & \mbox{}+ \cdots 
\nonumber \\ 
&= \sum_{k=0}^\infty (-1)^{k+1}
{\cal R}_n^{(k)}. 
\label{R-LN}
\end{align}
\end{subequations}
Here, 
\begin{subequations}
\label{Qk-Rk-def}
\begin{align}
{\cal Q}_n^{(0)}:=\widebar{F}(n), \hspace{5mm} 
{\cal R}_n^{(0)}:=F(n), 
\label{Q0-R0-def}
\end{align} 
and 
${\cal Q}_n^{(k)}$ and ${\cal R}_n^{(k)}$ 
for \mbox{$k \ge 1$} are 
defined as 
\begin{align}
{\cal Q}_n^{(k)} &:= 
 \sum_{i_1,i_2,\ldots, 
 i_{2k}=0}^{\infty} 
 \widebar{F}(n+i_1) F(n+i_1+i_2+1) 
 \widebar{F}(n+i_2+i_3+1) F(n+i_3+i_4+1)
\nonumber \\ & \hspace{13mm}
\mbox{} \times \cdots 
 \widebar{F}(n+i_{2k-2}+i_{2k-1}+1)
 F(n+i_{2k-1}+i_{2k}+1) \widebar{F}(n+i_{2k}+1),
\label{Qk-def}
\\
{\cal R}_n^{(k)} &:= 
\sum_{i_1,i_2,\ldots,i_{2k}=0}^{\infty} 
F(n+i_1) \widebar{F}(n+i_1+i_2+1) F(n+i_2+i_3+1) \widebar{F}(n+i_3+i_4+1)
\nonumber \\ & \hspace{13mm}
\mbox{} \times \cdots F(n+i_{2k-2}+i_{2k-1}+1)
\widebar{F}(n+i_{2k-1}+i_{2k}+1) F(n+i_{2k}+1). 
\end{align}
\end{subequations}
Note that ${\cal Q}_n^{(k)}$ and ${\cal R}_n^{(k)}$ 
are ``polynomials" 
of degree \mbox{$2k+1$} in $F$ and $\widebar{F}$
with their arguments 
bounded below 
but 
unbounded above. 
We only consider 
the region of \mbox{$(n,t)$} 
where 
the series in (\ref{AL-LN}) 
are 
absolutely convergent 
and 
admit termwise differentiation by $t$.  

In the following, we use 
the shift operator $E_n$ 
as well as its inverse 
$E_n^{-1}$
defined as 
\begin{align}
E_n f (n) := f (n+1), \hspace{5mm} 
E_n^{-1} f (n) := f (n-1), 
\nonumber 
\end{align}
and the forward difference operator $\boldsymbol{\Delta}_{i_\alpha}^+$ 
in each index of summation $i_\alpha$, 
\[
\boldsymbol{\Delta}_{i_\alpha}^+ 
 f (\ldots, i_\alpha, \ldots) 
	:= f (\ldots, i_\alpha +1, \ldots ) 
 - f (\ldots, i_\alpha, \ldots). 
\]
Then, using (\ref{AL-linear}), 
we can compute the time derivative of ${\cal Q}_n^{(k)}$ 
for \mbox{$k \ge 1$}
to obtain 
\begin{align}
& \;
\left[ \partial_t -a E_n + b E_n^{-1} + (a-b)I
	\right] {\cal Q}_n^{(k)}
\nonumber \\[2mm]
=& \;\, a \sum_{i_1,i_2,\ldots, 
 i_{2k}=0}^{\infty} \left\{
 \widebar{F}\uwave{(n+i_1+1)} F(n+i_1+i_2+1) 
 \widebar{F}(n+i_2+i_3+1) \cdots 
 \widebar{F}(n+i_{2k}+1) \right.
\nonumber \\
& \hspace{10mm} \mbox{} - \widebar{F}(n+i_1) F\uwave{(n+i_1+i_2)} 
 \widebar{F}(n+i_2+i_3+1) \cdots \widebar{F}(n+i_{2k}+1)
\nonumber \\
& \hspace{10mm} \mbox{}+ \cdots 
\nonumber \\
&  \hspace{10mm} \mbox{}- \widebar{F}(n+i_1) F(n+i_1+i_2+1) \cdots 
 F\uwave{(n+i_{2k-1}+i_{2k})} \widebar{F}(n+i_{2k}+1)
\nonumber \\
& \hspace{10mm} \mbox{} +  \widebar{F}(n+i_1) F(n+i_1+i_2+1) \cdots 
 F(n+i_{2k-1}+i_{2k}+1) \widebar{F}\uwave{(n+i_{2k}+2)}
\nonumber \\  
& \hspace{9mm}  \left. \mbox{}  - 
\widebar{F}\uwave{(n+i_1+1)} F\uwave{(n+i_1+i_2+2)} \cdots  
 F\uwave{(n+i_{2k-1}+i_{2k}+2)}
 \widebar{F}\uwave{(n+i_{2k}+2)} \right\}
\nonumber \\[1mm]
& \, \mbox{}+ b \sum_{i_1,i_2,\ldots, i_{2k}=0}^{\infty} \left\{
 -\widebar{F}\uwave{(n+i_1-1)} F(n+i_1+i_2+1) 
 \widebar{F}(n+i_2+i_3+1)  \cdots 
 \widebar{F}(n+i_{2k}+1)  \right.
\nonumber \\
&  \hspace{10mm} \mbox{} +
\widebar{F}(n+i_1) F\uwave{(n+i_1+i_2+2)} 
 \widebar{F}(n+i_2+i_3+1)  \cdots 
 \widebar{F}(n+i_{2k}+1)
\nonumber \\
&  \hspace{10mm} \mbox{} -
\widebar{F}(n+i_1) F(n+i_1+i_2+1)
 \widebar{F}\uwave{(n+i_2+i_3)}  \cdots 
 \widebar{F}(n+i_{2k}+1)
\nonumber \\
& \hspace{10mm} \mbox{}+ \cdots 
\nonumber \\
&  \hspace{10mm} \mbox{} +
 \widebar{F}(n+i_1) F(n+i_1+i_2+1) \cdots 
 F\uwave{(n+i_{2k-1}+i_{2k}+2)} \widebar{F}(n+i_{2k}+1)
\nonumber \\
&  \hspace{10mm} \mbox{} -
 \widebar{F}(n+i_1) F(n+i_1+i_2+1) \cdots 
 F(n+i_{2k-1}+i_{2k}+1) \widebar{F}\uwave{(n+i_{2k})}
\nonumber \\
&  \hspace{9mm} \left. \mbox{} +
 \widebar{F}\uwave{(n+i_1-1)} F \uwave{(n+i_1+i_2)} \cdots 
 F \uwave{(n+i_{2k-1}+i_{2k})} \widebar{F} \uwave{(n+i_{2k})}
\right\},
\nonumber
\end{align} 
where 
the 
arguments 
shifted from 
the original ones in 
${\cal Q}_n^{(k)}$ 
are underscored with a wavy line. 
We can further rewrite it as 
\begin{align}
&  \;\, a \sum_{i_1,i_2,\ldots, 
 i_{2k}=0}^{\infty} \left\{
 \boldsymbol{\Delta}_{i_1}^+ \left[
 \widebar{F}(n+i_1) F \uwave{(n+i_1+i_2)} 
 \widebar{F}(n+i_2+i_3+1) \cdots 
 \widebar{F}(n+i_{2k}+1) \right] \right.
\nonumber \\
& \hspace{10mm} \mbox{}+
 \boldsymbol{\Delta}_{i_3}^+ \left[
 \widebar{F}(n+i_1) F(n+i_1+i_2+1) 
 \widebar{F}(n+i_2+i_3+1)  F \uwave{(n+i_3+i_4)} \cdots 
 \widebar{F}(n+i_{2k}+1) \right]
\nonumber \\
& \hspace{10mm} \mbox{}+ \cdots 
\nonumber \\
&  \hspace{10mm} \mbox{}
+\boldsymbol{\Delta}_{i_{2k-1}}^+ \left[ 
 \widebar{F}(n+i_1) 
 \cdots \widebar{F}(n+i_{2k-2}+i_{2k-1}+1)
 F\uwave{(n+i_{2k-1}+i_{2k})} \widebar{F}(n+i_{2k}+1) \right]
\nonumber \\
&  \hspace{10mm} \mbox{}
- \sum_{\alpha=1}^k \boldsymbol{\Delta}_{i_{2\alpha -1}}^+ \left[ 
 \widebar{F}(n+i_1) F(n+i_1+i_2+1)
 \cdots \widebar{F}(n+i_{2\alpha-2}+i_{2\alpha-1}+1) \right.
\nonumber \\
&  \hspace{33mm} \left. \left. \mbox{} \times
 F(n+i_{2\alpha-1}+i_{2\alpha}+1)
 \widebar{F}\uwave{(n+i_{2\alpha}+i_{2\alpha +1}+2)} \cdots 
 \widebar{F}\uwave{(n+i_{2k}+2)} \right] \right\}
\nonumber \\[1mm]
& \, \mbox{}+ b \sum_{i_1,i_2,\ldots, i_{2k}=0}^{\infty} \left\{
 \boldsymbol{\Delta}_{i_2}^+ \left[
 \widebar{F}(n+i_1) F(n+i_1+i_2+1) 
 \widebar{F}\uwave{(n+i_2+i_3)}  \cdots 
 \widebar{F}(n+i_{2k}+1)  \right. \right]
\nonumber \\
& \hspace{10mm} \mbox{}+
 \boldsymbol{\Delta}_{i_4}^+ \left[
 \widebar{F}(n+i_1) \cdots
 F (n+i_3+i_4+1) \widebar{F}\uwave{(n+i_4+i_5)} \cdots 
 \widebar{F}(n+i_{2k}+1) \right]
\nonumber \\
& \hspace{10mm} \mbox{}+ \cdots 
\nonumber \\
&  \hspace{10mm} \mbox{}
+\boldsymbol{\Delta}_{i_{2k}}^+ \left[ 
 \widebar{F}(n+i_1) F(n+i_1+i_2+1) \cdots 
 F (n+i_{2k-1}+i_{2k}+1) \widebar{F} \uwave{(n+i_{2k})} \right]
\nonumber \\
&  \hspace{10mm} \mbox{}
- \sum_{\beta=1}^k \boldsymbol{\Delta}_{i_{2\beta}}^+ \left[ 
 \widebar{F}\uwave{(n+i_1-1)} F \uwave{(n+i_1+i_2)}
\cdots 
 F \uwave{(n+i_{2\beta-1}+i_{2\beta})} 
 \widebar{F} \uwave{(n+i_{2\beta}+i_{2\beta +1})}
\right.
\nonumber \\
&  \hspace{33mm} \left. \left. \mbox{} \times
 F (n+i_{2\beta +1}+i_{2\beta +2}+1) \cdots 
 \widebar{F}(n+i_{2k}+1) \right] \right\}
\nonumber \\[2mm]
= & \;\, a \sum_{i_1,i_2,\ldots, 
 i_{2k}=0}^{\infty} \,
 \sum_{1 \le \alpha \le \beta \le k}
 \boldsymbol{\Delta}_{i_{2\alpha -1}}^+
 \boldsymbol{\Delta}_{i_{2\beta}}^+
 \left[
  -\widebar{F}(n+i_1) F(n+i_1+i_2+1)
 \cdots \widebar{F}(n+i_{2\alpha-2}+i_{2\alpha-1}+1) \right.
\nonumber \\
&  \hspace{33mm} \left. \mbox{} \times
 F\uwave{(n+i_{2\alpha-1}+i_{2\alpha})}
 \widebar{F}(n+i_{2\alpha}+i_{2\alpha +1}+1) \cdots 
 F (n+i_{2\beta-1}+i_{2\beta}+1) \right. 
\nonumber \\
&  \hspace{33mm} \left. \mbox{} \times
 \widebar{F} (n+i_{2\beta}+i_{2\beta +1}+1)
 F \uwave{(n+i_{2\beta+1}+i_{2\beta+2}+2)} \cdots 
 \widebar{F}\uwave{(n+i_{2k}+2)}
 \right]  
\nonumber \\[1mm]
& \, \mbox{}+ b \sum_{i_1,i_2,\ldots, i_{2k}=0}^{\infty} \,
 \sum_{1 \le \alpha \le \beta \le k}
 \boldsymbol{\Delta}_{i_{2\alpha -1}}^+
 \boldsymbol{\Delta}_{i_{2\beta}}^+
 \left[ \widebar{F}\uwave{(n+i_1-1)} F \uwave{(n+i_1+i_2)}
 \cdots \widebar{F} \uwave{(n+i_{2\alpha-2}+i_{2\alpha-1})} \right.
\nonumber \\
&  \hspace{33mm} \left. \mbox{} \times
 F\uwave{(n+i_{2\alpha-1}+i_{2\alpha})}
 \widebar{F}(n+i_{2\alpha}+i_{2\alpha +1}+1) \cdots 
 F (n+i_{2\beta-1}+i_{2\beta}+1) \right. 
\nonumber \\
&  \hspace{33mm} \left. \mbox{} \times
 \widebar{F} \uwave{(n+i_{2\beta}+i_{2\beta+1})}
 F (n+i_{2\beta+1}+i_{2\beta+2}+1) \cdots 
 \widebar{F}(n+i_{2k}+1) \right],  
\nonumber
\end{align}
which is equal to 
\begin{align}
&  \;\, a 
\sum_{1 \le \alpha \le \beta \le k} 
 \left[ - \sum_{i_1,\ldots, i_{2\alpha-2}=0}^{\infty} 
 \widebar{F}(n+i_1) F(n+i_1+i_2+1)
 \cdots \widebar{F}(n+i_{2\alpha-2}+1) \right.
\nonumber \\ 
& \hspace{20mm} \left. \mbox{} \times
\sum_{i_{2\alpha}, \ldots, i_{2\beta-1}=0}^{\infty} 
 F( n+i_{2\alpha}) 
 \widebar{F} (n+i_{2\alpha}+i_{2\alpha+1}+1) \cdots 
 F (n+i_{2\beta-1}+1) \right.
\nonumber \\
&  \hspace{20mm} \left. \mbox{} \times
\sum_{ i_{2\beta +1}, \ldots, i_{2k}=0}^{\infty} 
 \widebar{F} (n+i_{2\beta +1}+1)
 F (n+i_{2\beta+1}+i_{2\beta+2}+2) \cdots 
 \widebar{F} (n+i_{2k}+2)
 \right]  
\nonumber \\[1mm]
& \, \mbox{}+ b \sum_{1 \le \alpha \le \beta \le k} 
 \left[ \sum_{i_1,\ldots, i_{2\alpha-2}=0}^{\infty} 
 \widebar{F}(n+i_1-1) F (n+i_1+i_2)
 \cdots \widebar{F}(n+i_{2\alpha-2}) \right.
\nonumber \\
&  \hspace{20mm} \left. \mbox{} \times 
 \sum_{i_{2\alpha}, \ldots, i_{2\beta-1}=0}^{\infty} F (n+i_{2\alpha})
 \widebar{F}(n+i_{2\alpha}+i_{2\alpha +1}+1) \cdots 
 F (n+i_{2\beta-1}+1) \right.
\nonumber \\
&  \hspace{20mm} \left. \mbox{} \times
 \sum_{ i_{2\beta +1}, \ldots, i_{2k}=0}^{\infty} 
 \widebar{F} (n+i_{2\beta+1})
 F (n+i_{2\beta+1}+i_{2\beta+2}+1) \cdots 
 \widebar{F}(n+i_{2k}+1) \right]
\nonumber \\[2mm]
=& \,
 -a \sum_{1 \le \alpha \le \beta \le k} 
 {\cal Q}_n^{(\alpha-1)} {\cal R}_n^{(\beta-\alpha)} 
	{\cal Q}_{n+1}^{(k-\beta)}
 + b \sum_{1 \le \alpha \le \beta \le k} 
 {\cal Q}_{n-1}^{(\alpha-1)} {\cal R}_n^{(\beta-\alpha)} 
	{\cal Q}_n^{(k-\beta)}.
\nonumber 
\end{align}
Thus, we obtain 
\begin{subequations}
\label{AL-rec}
\begin{align}
& \;
\left[ \partial_t -a E_n + b E_n^{-1} + (a-b)I
	\right] {\cal Q}_n^{(k)}
\nonumber \\[2mm]
=& \,
 -a \sum_{1 \le \alpha \le \beta \le k} 
 {\cal Q}_n^{(\alpha-1)} {\cal R}_n^{(\beta-\alpha)} 
	{\cal Q}_{n+1}^{(k-\beta)}
 + b \sum_{1 \le \alpha \le \beta \le k} 
 {\cal Q}_{n-1}^{(\alpha-1)} {\cal R}_n^{(\beta-\alpha)} 
	{\cal Q}_n^{(k-\beta)}, \hspace{5mm} k \ge 1.
\label{}
\end{align}
Similarly, 
we 
also obtain 
\begin{align}
& \;
\left[ \partial_t - b E_n + a E_n^{-1} + (b-a)I
	\right] {\cal R}_n^{(k)}
\nonumber \\[2mm]
=& \,
 -b \sum_{1 \le \alpha \le \beta \le k} 
 {\cal R}_n^{(\alpha-1)} {\cal Q}_n^{(\beta-\alpha)} 
	{\cal R}_{n+1}^{(k-\beta)}
 + a \sum_{1 \le \alpha \le \beta \le k} 
 {\cal R}_{n-1}^{(\alpha-1)} {\cal Q}_n^{(\beta-\alpha)} 
	{\cal R}_n^{(k-\beta)}, \hspace{5mm} k \ge 1.
\label{}
\end{align}
\end{subequations}
%
Because of 
(\ref{AL-LN}),  
(\ref{Q0-R0-def}) 
and (\ref{AL-linear}), 
relations 
(\ref{AL-rec}) imply 
that 
\[
\left\{
\begin{split}
& Q_{n,t} - a Q_{n+1} + b Q_{n-1} + (a-b) Q_n 
 = -a Q_n R_n Q_{n+1} +b Q_{n-1} R_n Q_n, 
\\[1mm]
& R_{n,t} -b R_{n+1} +a R_{n-1} + (b-a) R_n 
	=-b R_n Q_n R_{n+1} +a R_{n-1} Q_n R_n. 
\end{split} \right. 
\]
%
These are exactly the equations of motion for
the matrix Ablowitz--Ladik lattice (\ref{mALsys}). 
$\hfill$ $\Box$
\\

In the following, we 
set \mbox{$a=b=1$}, which 
simplifies (\ref{AL-linear}) 
to 
\begin{subnumcases}{\label{mLV-linear}}
{}
  \partial_t F(n,t) = F (n+1,t) - F(n-1,t), 
\\[1mm]
 \partial_t \widebar{F}(n,t)
 = \widebar{F} (n+1,t) - \widebar{F}(n-1,t). 
\end{subnumcases}
Here, 
$F(n,t)$ and $\widebar{F}(n,t)$ 
are required to
decay 
rapidly 
as \mbox{$n \to +\infty$}. 
For the moment, 
we 
do not impose 
the 
conditions 
(\ref{F-Cliff}) and (\ref{Fbar-Cliff}), 
and consider the general case where $F(n,t)$ and $\widebar{F}(n,t)$ 
are independent matrix 
functions.  
%
Then, 
we can rewrite 
(\ref{AL-rec}) 
as 
recurrence 
relations 
for 
${\cal Q}_n^{(k)}$ and ${\cal R}_n^{(k)}$
that 
were 
originally defined as 
(\ref{Qk-Rk-def}), i.e.\
\begin{subequations}
\label{AL-rec2}
\begin{align}
& {\cal Q}_n^{(k)}
=\left( I + E_n \partial_t - E_n^2 \right)^{-1}
\left[ 
 -\sum_{1 \le \alpha \le \beta \le k} 
 {\cal Q}_{n+1}^{(\alpha-1)} {\cal R}_{n+1}^{(\beta-\alpha)} 
	{\cal Q}_{n+2}^{(k-\beta)}
 + \sum_{1 \le \alpha \le \beta \le k} 
 {\cal Q}_{n}^{(\alpha-1)} {\cal R}_{n+1}^{(\beta-\alpha)} 
	{\cal Q}_{n+1}^{(k-\beta)} \right], 
\label{AL-Qrec2}
\\[2mm]
& {\cal R}_n^{(k)}
= \left( I + E_n \partial_t - E_n^2 \right)^{-1}
 \left[
  - \sum_{1 \le \alpha \le \beta \le k} 
 {\cal R}_{n+1}^{(\alpha-1)} {\cal Q}_{n+1}^{(\beta-\alpha)} 
	{\cal R}_{n+2}^{(k-\beta)}
  + \sum_{1 \le \alpha \le \beta \le k} 
 {\cal R}_{n}^{(\alpha-1)} {\cal Q}_{n+1}^{(\beta-\alpha)} 
	{\cal R}_{n+1}^{(k-\beta)} \right]
\label{AL-Rrec2}
\end{align}
\end{subequations}
for \mbox{$k \ge 1$}, 
and 
\mbox{${\cal Q}_n^{(0)}=\widebar{F}(n,t)$} and 
\mbox{${\cal R}_n^{(0)}=F(n,t)$}. 
%
Here, 
the inverse operator 
\mbox{$\left( I + E_n \partial_t - E_n^2 \right)^{-1}$} 
is defined using 
the 
Maclaurin series
as 
\[
\left( I + E_n \partial_t - E_n^2 \right)^{-1} 
:= \sum_{j=0}^\infty 
E_n^j \left( E_n - \partial_t \right)^j, 
\]
where the action of $\partial_t$ on 
$F(n,t)$ and $\widebar{F}(n,t)$ is 
given by 
(\ref{mLV-linear}). 

In fact, 
the linear operator \mbox{$I + E_n \partial_t -E_n^2$} 
has a nontrivial kernel 
that contains 
$F(n,t)$ and $\widebar{F}(n,t)$ 
as well as their 
spatial/temporal shifts, 
so it is not invertible in general; 
however, 
this 
does not 
cause any 
problem in 
obtaining (\ref{AL-rec2}). 
To confirm 
this, it is sufficient to note 
that 
no ``polynomials" 
of degree \mbox{$2k+1 \; (k \ge 1)$} in $F$ and $\widebar{F}$ 
can 
vanish 
by the action 
of \mbox{$I + E_n \partial_t -E_n^2$}, 
as long as 
their spatial 
arguments 
are 
bounded below.  

\begin{lemma}
\label{LemA.3}
For general \mbox{$F(n)$} and \mbox{$\widebar{F}(n)$} 
satisfying 
the linear evolution equations (\ref{mLV-linear}), 
assume that 
the following equality 
is
valid: 
\begin{align}
& \left( I + E_n \partial_t  -E_n^2 \right) 
\sum_{i_1,i_2,\ldots, i_{2k+1} =0}^{\infty}
 \gamma_{i_1 i_2 \ldots i_{2k+1}}
 \widebar{F}(n+i_1+\alpha_1) F(n+i_2+\alpha_2) 
 \widebar{F}(n+i_3+\alpha_3)
\nonumber \\ & \hspace{40mm}
\mbox{} \times \cdots 
 F(n+i_{2k}+\alpha_{2k}) 
 \widebar{F}(n+i_{2k+1}+\alpha_{2k+1}) =O. 
 \nonumber 
\end{align}
Here, \mbox{$\gamma_{i_1 i_2 \ldots i_{2k+1}} \in {\mathbb C}$} 
and  \mbox{$\alpha_j \in {\mathbb Z}_{\ge 0}$} 
are 
constants, 
and \mbox{$k \ge 1$}.   
Then, 
the 
equality 
must be 
trivial, 
i.e. 
\[
\gamma_{i_1 i_2 \ldots i_{2k+1}} = 0, 
\;\;\;
\forall i_1,i_2,\ldots, i_{2k+1} \in {\mathbb Z}_{\ge 0}. 
\]
\end{lemma}
\vspace{5mm}

This lemma can be 
easily verified 
by 
ordering the nonzero 
terms involved 
in the summation, 
{\it e.g.}, 
using 
sums of the arguments such as 
\begin{align}
(n+i_1+\alpha_1) & + (n+i_2+\alpha_2) 
 + (n+i_3+\alpha_3) 
 + \cdots 
 + (n+i_{2k+1}+\alpha_{2k+1}), 
\nonumber \\[1mm]
& \hphantom{\mbox{}+} \;\,
(n+i_2+\alpha_2) + (n+i_3+\alpha_3) 
 + \cdots 
 + (n+i_{2k+1}+\alpha_{2k+1}), \; \textrm{etc.}
\nonumber 
\end{align}
Note that 
both 
$ E_n \partial_t$ and $E_n^2$ 
increase the values of 
the arguments in
each term, 
while $I$ 
leaves them invariant. 
Thus, 
the equality 
implies 
that 
the coefficient 
of the 
first 
term in the 
ordering, 
which 
has 
the minimum 
values of 
the arguments, 
must be zero. 
Therefore, there is no first term and 
all 
$\gamma_{i_1 i_2 \ldots i_{2k+1}}$ 
must vanish. 
$\hfill$ $\Box$
\\

Lemma~\ref{LemA.3}
guarantees that the difference between 
${\cal Q}_n^{(k)}$ defined as (\ref{Qk-def}) 
and the right-hand side of 
(\ref{AL-Qrec2})
does not 
belong to the kernel 
of \mbox{$I + E_n \partial_t -E_n^2$}. 
Thus, (\ref{AL-Qrec2}) is indeed 
an exact equality; 
the same 
applies to (\ref{AL-Rrec2}). 

On the basis of the recurrence relations 
(\ref{AL-rec2}), 
we can prove Proposition~\ref{Prop.A.1} 
by induction. 
To this end, 
we need 
to 
use one lemma, which 
is a direct consequence of 
the anticommutation relations (\ref{a-commute}) 
for the generators of the Clifford algebra 
(cf.~(\ref{conj-scalar})). 
%
\begin{lemma}
\label{LemA.4}
If the 
square 
matrices  
$X$ and $Y$ 
take their values in 
the linear span of 
\mbox{$\{ I, e_1, e_2, \ldots, e_{2M-1} \}$}, 
then 
\[
X\widehat{Y}+Y\widehat{X} 
\]
is 
a scalar matrix
and coincides with 
\mbox{$\widehat{X}Y + \widehat{Y}X$}. 
Thus, 
up to an overall factor, 
it can be considered as 
the definition of 
an inner product. 
Here, \mbox{$\widehat{\;\;}$} denotes 
the Clifford conjugate, which 
changes the sign of 
the coefficients of \mbox{$\{e_1, e_2, \ldots, e_{2M-1} \}$}. 
\\
\end{lemma}
\noindent
{\it Proof of Proposition~\ref{Prop.A.1}.} 
Because of (\ref{AL-LN}), 
we need only show 
that,  
for
all 
\mbox{$k \in {\mathbb Z}_{\ge 0}$}, 
%
%
\begin{center}
\hspace{5mm}
\begin{minipage}{9.5cm}
${\cal Q}_n^{(k)}$ and ${\cal R}_n^{(k)}$ 
take their values in 
the linear span of 
\mbox{$\{ I, e_1, e_2, \ldots, e_{2M-1} \}$} and 
satisfy the  Clifford conjugation 
relation 
${\cal R}_n^{(k)} = \widehat{{\cal Q}}_n^{(k)}$. 
\end{minipage}
\hspace{5mm}
($\clubsuit$)
\end{center}
%
Note that 
this 
is true 
for 
\mbox{$k =0$} 
(cf.~(\ref{F-Cliff}), (\ref{Fbar-Cliff}) and (\ref{Q0-R0-def})). 
We assume 
that ($\clubsuit$) 
is true for 
\mbox{$
k \le L$}
and then 
%
proceed to show ($\clubsuit$) for \mbox{$k = L+1$}. 
Lemma~\ref{LemA.4} implies that 
the 
quantity appearing in 
(\ref{AL-Qrec2}) 
with \mbox{$k = L+1$}, 
\begin{align}
\sum_{\alpha=1}^{\beta} 
 {\cal Q}_{n+1}^{(\alpha-1)} {\cal R}_{n+1}^{(\beta-\alpha)} 
&= \frac{1}{2} \sum_{\alpha=1}^{\beta} \left[ 
 {\cal Q}_{n+1}^{(\alpha-1)} {\cal R}_{n+1}^{(\beta-\alpha)} 
 + {\cal Q}_{n+1}^{(\beta-\alpha)} {\cal R}_{n+1}^{(\alpha-1)} 
 \right]
\nonumber \\
&= \frac{1}{2} \sum_{\alpha=1}^{\beta} \left[ 
 {\cal Q}_{n+1}^{(\alpha-1)} \widehat{{\cal Q}}_{n+1}^{(\beta-\alpha)} 
 + {\cal Q}_{n+1}^{(\beta-\alpha)} \widehat{{\cal Q}}_{n+1}^{(\alpha-1)} 
 \right], \hspace{5mm} 1 \le \beta \le L+1, 
\nonumber
\end{align}
is a scalar matrix and coincides with the quantity appearing in (\ref{AL-Rrec2}) 
with \mbox{$k = L+1$},
\begin{align}
\sum_{\alpha=1}^{\beta} 
 {\cal R}_{n+1}^{(\alpha-1)} {\cal Q}_{n+1}^{(\beta-\alpha)} 
&= \frac{1}{2} \sum_{\alpha=1}^{\beta} \left[ 
 {\cal R}_{n+1}^{(\alpha-1)} {\cal Q}_{n+1}^{(\beta-\alpha)} 
 + {\cal R}_{n+1}^{(\beta-\alpha)} {\cal Q}_{n+1}^{(\alpha-1)} 
 \right]
\nonumber \\
&= \frac{1}{2} \sum_{\alpha=1}^{\beta} \left[ 
 \widehat{{\cal Q}}_{n+1}^{(\alpha-1)} {\cal Q}_{n+1}^{(\beta-\alpha)} 
 + \widehat{{\cal Q}}_{n+1}^{(\beta-\alpha)} {\cal Q}_{n+1}^{(\alpha-1)} 
 \right], \hspace{5mm} 1 \le \beta \le L+1. 
\nonumber
\end{align}
Similarly, we also have 
\[
\sum_{\beta=\alpha}^{L+1} {\cal R}_{n+1}^{(\beta-\alpha)} {\cal Q}_{n+1}^{(L+1-\beta)} 
= \sum_{\beta=\alpha}^{L+1} 
 {\cal Q}_{n+1}^{(\beta-\alpha)} {\cal R}_{n+1}^{(L+1-\beta)} 
= \mathrm{scalar}, 
\hspace{5mm} 1 \le \alpha \le L+1. 
\]
Thus, 
the recurrence relations 
(\ref{AL-rec2}) 
imply
that ($\clubsuit$) 
holds true 
for \mbox{$k = L+1$}.
Therefore, 
by complete induction, 
($\clubsuit$) 
is true 
for all 
\mbox{$k \in {\mathbb Z}_{\ge 0}$}. 
$\hfill$ $\Box$
\\
\\
{\it Remark.} 
By considering 
the continuous space limit of 
Proposition~\ref{Prop.A.1}, 
we can 
obtain similar results 
on the corresponding reductions  
of the continuous matrix NLS hierarchy. 
In particular, 
we can identify 
the reduction conditions 
on 
the 
scattering data
to solve 
the vector modified KdV equation (\ref{vmKdV1}) 
and the vector sine-Gordon equation~\cite{EP79}
by the inverse scattering method (see~\cite{TsuJPSJ98}). 
In addition, 
it is an easy task 
to consider 
a continuous analog of Proposition~\ref{Prop.A.2} 
that 
realizes 
an exact linearization of 
the matrix NLS system. 
\\

Proposition~\ref{Prop.A.1} demonstrates 
that 
under the restrictions (\ref{F-Cliff}) and (\ref{Fbar-Cliff}) 
on $F(n, t)$ and $\widebar{F}(n, t)$, 
we can solve 
the vector modified Volterra 
lattice (\ref{vector-mV}) by 
the inverse scattering method. 
Moreover, if 
the vector dependent variable 
$\vt{q}_n$ is 
real-valued, 
we need to 
restrict the components of the 
vector 
\mbox{$\vt{f}_n (t) = ( f^{(1)}(n,t), \ldots, f^{(2M)}(n,t) )$}
appearing 
in (\ref{F-Cliff}) and (\ref{Fbar-Cliff}) 
to be 
real-valued. 
%
Recall that 
in the 
reflectionless 
case of the 
potentials, 
if 
$A(\mu)^{-1}$ only has 
isolated 
simple poles, $F(n, t)$ is given as 
(cf.~(\ref{F_time1}) with \mbox{$a=b=1$})
\begin{align}
F(n, t) &= 
     - \sum_{j=1}^N C_j(0) \mu_j^{n}
 \mathrm{e}^{(\mu_j -\mu_j^{-1}) t}, 
\label{F-refl}
\end{align}
where \mbox{$0<|\mu_j| < 1$}. 
Thus, 
the configuration of 
\mbox{$\{\mu_1, \mu_2, \ldots, \mu_{N} \}$}  
must be symmetric 
with respect to the real $\mu$-axis,
that is, 
they 
either take real values or 
occur in complex conjugate pairs. 
Up to a reordering, 
they can be classified into the following 
three types: 
\vspace{1mm}

(I) \hspace{1pt} \mbox{$\mu_{2k-1} = \mu_{2k}^\ast=a_k \mathrm{e}^{\mathrm{i} \theta_k}, \;\hspace{1pt}
0< a_k <1, \;\hspace{1pt} 0< \theta_k <\pi, \;\hspace{1pt} k=1, 2, \ldots, N_1$},
\vspace{1mm}

(II) \hspace{1pt} \mbox{$0< \mu_j <1, \;\hspace{1pt} j=2N_1+1, 
	\ldots, 2 N_1+N_2$}, 
\vspace{1mm}

(III) \hspace{1pt} \mbox{$-1< \mu_j <0, 
	\;\hspace{1pt} j=2 N_1+N_2+1, \ldots, 2 N_1+N_2+N_3 \hspace{1pt}(=N)$}. 
\vspace{1mm}
\\
%
Each of 
the corresponding matrices \mbox{$\{C_1 (0), C_2 (0), \ldots, C_N (0) \}$} 
%
takes its 
values in the linear span of 
\mbox{$\{ I, e_1, e_2, \ldots, e_{2M-1} \}$}, 
wherein 
the coefficients are 
all real 
for type (II) and type (III). 
For type (I), 
the coefficients associated with $\mu_{2k-1}$ 
and those associated with $\mu_{2k}$ form 
a complex conjugate pair 
so that the coefficients of 
\mbox{$\{ I, e_1, e_2, \ldots, e_{2M-1} \}$}
in (\ref{F-refl}) 
become real-valued~\cite{TUW98,TUW99}. 
%
%
%

The nature of the three types (I)--(III) can be unveiled 
by constructing 
the corresponding 
one-soliton solutions. 
Type (II) 
provides the trivial 
vector 
analog of the 
one-soliton solution of the scalar modified Volterra lattice, 
\[
\vt{q}_n (t)= \frac{\sinh \alpha}
{\cosh \left[ \alpha n
  + 2 (\sinh \alpha) t + \delta \right] } \hspace{1pt} \vt{u}, 
  \hspace{5mm} \sca{\vt{u}}{\vt{u}}=1.
\]
Type (III) 
provides a very 
similar 
solution, 
\[
\vt{q}_n (t)= \frac{ (-1)^n \sinh \alpha}
{\cosh \left[ \alpha n - 2 (\sinh \alpha) t + \delta \right] } 
 \hspace{1pt} \vt{u}, \hspace{5mm} \sca{\vt{u}}{\vt{u}}=1, 
\]
which reflects the form-invariance of 
the vector modified Volterra lattice (\ref{vector-mV}) 
under the transformation 
\mbox{$\vt{q}_n 
\to (-1)^n \vt{q}_n$}, \mbox{$t 
\to -t$}. 
For type (I), we 
need to 
impose an additional condition 
to exclude 
a breather solution
so that 
a pure 
soliton solution 
with 
a time-independent 
profile 
can be obtained~\cite{TUW98,TUW99}. 
Thus, the 
one-soliton solution 
of type (I) 
is 
given by 
\[
\vt{q}_n (t)= \frac{\vt{c}\hspace{1pt} \mathrm{e}^{\mathrm{i} \beta n 
+ 2\mathrm{i} (\cosh \alpha \hspace{1pt}\sin \beta) t}
+ \vt{c}^\ast \mathrm{e}^{-\mathrm{i} \beta n -2 \mathrm{i}
 (\cosh \alpha\hspace{1pt} \sin \beta) t}}
{\cosh \left[ \alpha n
  + 2 (\sinh \alpha \hspace{1pt}\cos \beta) t + \delta \right] }
\]
with 
\mbox{$\sca{\vt{c}}{\vt{c}}
=0$} and 
\mbox{$\hspace{1pt}2 \sca{\vt{c}}{\vt{c}^\ast} = (\sinh \alpha)^2$}.
This is indeed the most general 
one-soliton solution, 
because it 
reduces to 
type (II) and type (III) in the limit 
\mbox{$\beta \to 0$} and \mbox{$\beta \to \pi$}, respectively. 
%

Proposition~\ref{Prop.A.1} 
enables us 
to formulate 
the inverse scattering method 
for the vector modified Volterra lattice (\ref{vector-mV}), 
as well as 
the relevant continuous systems, 
with satisfactory rigor.
As by-products, 
it can 
also generate 
numerous 
nontrivial identities 
for 
some 
rational functions of
\mbox{$\{\mu_1, \mu_2, \ldots, \mu_{N} \}$}. 
Indeed, 
for 
$F(n, t)$ and $\widebar{F}(n, t)$ given 
as 
(\ref{F-Cliff}) and (\ref{Fbar-Cliff}), 
the coefficients of \mbox{$e_j e_k \; (j \neq k)$}, 
\mbox{$e_i e_j e_k \; (i \neq j \neq k \neq i)$}, etc.\ 
in ${\cal Q}_n^{(k)}$ 
for \mbox{$k \ge 1$} 
defined 
as 
(\ref{Qk-def}) 
must vanish 
(cf.~($\clubsuit$)). 
For the reflectionless case 
of the potentials, 
we can 
express 
these 
coefficients explicitly in terms of the scattering data, 
which 
indeed provide 
nontrivial 
identities. 
If $A(\mu)^{-1}$ 
only has 
isolated simple poles, 
$F(n)$ and $\widebar{F}(n)$ 
are given as
 %
\begin{align}
F(n) = - \sum_{j=1}^N C_j \mu_j^{n}, \hspace{5mm} 
 \widebar{F}(n) =  - \sum_{j=1}^{N} \widehat{C}_j \mu_j^{n}. 
\label{F-gene}
\end{align}
Here, \mbox{$\{\mu_1, \mu_2, \ldots, \mu_{N} \}$} are pairwise distinct 
and satisfy 
\mbox{$0<|\mu_j| < 1$}, and \mbox{$\widehat{\;\;}$} denotes 
the Clifford conjugate. 
The time dependence 
of $C_j$ and $\widehat{C}_j$ 
is irrelevant in this context 
and thus 
is omitted. 
Substituting (\ref{F-gene}) 
into (\ref{Qk-def}) and 
computing the multiple sum with respect to \mbox{$i_1, i_2, \ldots, i_{2k}$}, 
we obtain 
\begin{align}
{\cal Q}_n^{(k)} = -\sum_{j_1,j_2,\ldots, 
 j_{2k+1}=1}^{N} \frac{\mu_{j_1}^{n} \mu_{j_2}^{n+1}\mu_{j_3}^{n+1} \mu_{j_4}^{n+1}
  \cdots \mu_{j_{2k}}^{n+1} \mu_{j_{2k+1}}^{n+1} \widehat{C}_{j_1} C_{j_2} 
 \widehat{C}_{j_3} C_{j_4} \cdots C_{j_{2k}} \widehat{C}_{j_{2k+1}}
 }
{\left( 1-\mu_{j_1}\mu_{j_2} \right) \left( 1-\mu_{j_2}\mu_{j_3} \right) 
\left( 1-\mu_{j_3}\mu_{j_4} \right) \cdots \left( 1-\mu_{j_{2k}}\mu_{j_{2k+1}} \right)}
\nonumber 
\end{align}
for \mbox{$k \ge 1$}. 
Thus, we can state the following proposition. 
\begin{proposition}
\label{A.5}
For \mbox{$\{ e_1, e_2, 
\ldots, e_{2M-1} \}$} 
satisfying 
the anticommutation relations 
\mbox{$
e_j e_k + e_k e_j = -2 \delta_{jk} I$}, 
all 
the coefficients of 
quadratic or higher terms 
such as 
\mbox{$e_j e_k \; (j \neq k)$} 
and
\mbox{$e_i e_j e_k \; (i \neq j \neq k \neq i)$}
in the quantity,
\begin{align}
 \sum_{\stackrel{\scriptstyle \{j_1,j_2,\ldots, 
 j_{2k+1}\}}{=\{ i_1, i_2, \ldots, i_{2k+1} \}}} 
 \frac{ \widehat{C}_{j_1} C_{j_2} 
 \widehat{C}_{j_3} C_{j_4} \cdots C_{j_{2k}} \widehat{C}_{j_{2k+1}}}
{\mu_{j_1} \left( 1-\mu_{j_1}\mu_{j_2} \right) \left( 1-\mu_{j_2}\mu_{j_3} \right) 
\left( 1-\mu_{j_3}\mu_{j_4} \right) \cdots \left( 1-\mu_{j_{2k}}\mu_{j_{2k+1}} \right)},
\nonumber 
\end{align}
vanish identically. 
Here, \mbox{$\{ i_1, i_2, \ldots, i_{2k+1} \} $} is 
any (repeated) combination of positive integers 
(\mbox{$1 \le i_1 \le i_2 \le \cdots \le i_{2k+1} 
$}) and 
\[
C_j = c_j^{(1)} I + \sum_{\alpha=1}^{2M-1} c_j^{(\alpha+1)} e_\alpha,
\quad 
\widehat{C}_j = c_j^{(1)} I - \sum_{\alpha=1}^{2M-1} c_j^{(\alpha+1)} e_\alpha.
\]
Because 
\mbox{$\{ c_j^{(1)}, c_j^{(2)}, \ldots, c_j^{(2M)} \}$} for 
\mbox{$j \in \{ i_1, i_2, \ldots, i_{2k+1} \} $} 
at a fixed time can be chosen arbitrarily, 
each 
identity obtained this way splits into 
as many 
identities 
as the number of different products of \mbox{$\{ c_j^{(i)} \}$} 
involved in the identity. 
\end{proposition}

For example, 
when \mbox{$\{ i_1, i_2, \ldots, i_{2k+1} \}$} are pairwise distinct, 
we can 
simply 
set 
them 
as 
\mbox{$
\{ 1, 2, \ldots, 2k+1 \}$}. 
Thus, 
the coefficients of the highest products of 
\mbox{$\{ e_j \}$} such as 
\mbox{$e_1 e_2 \cdots e_{2k+1}$} 
provide 
the identity: 
\begin{align}
 \sum_{\stackrel{\scriptstyle \{j_1,j_2,\ldots, 
 j_{2k+1}\}}{=\{ 1, 2, \ldots, 2k+1 \}}} 
 \frac{ \mathrm{sgn} \left( 
\begin{array}{cccc}
 1 \!&\! 2 \!&\! \ldots \!&\! 2k+1 \\
 j_1 \!&\! j_2 \!&\! \ldots \!&\! j_{2k+1}\\ 
\end{array}
\right)}
{\mu_{j_1} \left( 1-\mu_{j_1}\mu_{j_2} \right) \left( 1-\mu_{j_2}\mu_{j_3} \right) 
\left( 1-\mu_{j_3}\mu_{j_4} \right) \cdots \left( 1-\mu_{j_{2k}}\mu_{j_{2k+1}} \right)}=0.
\nonumber 
\end{align}
Considering the lower products 
such as \mbox{$e_1 e_2 \cdots e_{2k}$}, 
\mbox{$e_1 e_2 \cdots e_{2k-1}$}, $\ldots$, 
$e_1 e_2$, we 
obtain various extensions 
of the above identity, wherein 
the sign 
of the permutation 
is 
replaced with more elaborate 
functions 
that still take 
values in \mbox{$\{ +1, -1\}$}. 

Proposition~\ref{A.5} can be generalized in 
many different 
directions. 
First, we can consider 
the 
case 
where $A(\mu)^{-1}$ 
also has 
second or higher order poles. 
Thus, 
$\mu_j^{n}$ in (\ref{F-gene}) can be replaced 
with 
more general 
functions 
decaying 
as 
\mbox{$n \to + \infty$}. 
Second, instead of 
using 
Proposition~\ref{Prop.A.1} 
designed for solving the vector modified Volterra 
lattice (\ref{vector-mV}), 
we can consider 
corresponding results for 
other integrable systems having 
a Lax representation 
of the same type. 
That is, 
any integrable system 
with vector 
dependent variables 
can provide a similar result 
if it 
can be 
obtained as a reduction of 
a matrix-valued 
integrable system 
using 
the generators of the Clifford algebra. 
In particular, starting with 
the solution formulas for the 
matrix generalizations of 
the 
NLS, derivative NLS, modified KdV
as well as 
their 
discrete analogs~\cite{TsuJMP10}, 
we can obtain 
various interesting 
variants of 
Proposition~\ref{A.5}. 
%

\end{document}